\documentclass[twocolumn,nofootinbib,amsmath,amssymb,aps,prd,balancelastpage,superscriptaddress]{revtex4-1}

\usepackage{color}
\usepackage{xcolor}
\usepackage[active]{srcltx}
\usepackage{amsmath,amsfonts,amssymb,amsthm,amstext,amscd,eucal,srcltx}
\usepackage{epsfig,graphicx,bm}
\usepackage{epstopdf, epsf}
\usepackage{dcolumn}
\usepackage{comment}
\usepackage{hyperref}
\usepackage{tikz}
\usepackage[nointegrals]{wasysym}
\usepackage{braket}
\usepackage{etoolbox}

\newcommand{\be}{\begin{equation}}
\newcommand{\ee}{\end{equation}}

\newcommand{\bse}{\begin{subequations}}
\newcommand{\ese}{\end{subequations}}
\newcommand{\ba}{\begin{eqnarray}}
\newcommand{\ea}{\end{eqnarray}}
\newcommand{\bc}{\begin{center}}
\newcommand{\ec}{\end{center}}

\newcommand{\nn}{\nonumber}

\def\bra#1{\langle#1|}
\def\ket#1{|#1\rangle}
\def\braket#1{\langle#1\rangle}

\begin{document}
\preprint{IPM/P-2012/009}  
\vspace*{3mm}

\title{The scrambling power of gravity in black hole radiation}%

\author{Xuan-Lin Su}
\email{19110190015@fudan.edu.cn}
\affiliation{Center for Field Theory and Particle Physics \& Department of Physics, Fudan University, 200433 Shanghai, China}

\author{Alioscia Hamma}
\email{alioscia.hamma@unina.it}
\affiliation{Dipartimento di Fisica Ettore Pancini, Universit\`a degli Studi di Napoli Federico II,
Via Cintia, 80126 Napoli NA}
\affiliation{INFN, Sezione di Napoli, Italy}

\author{Antonino Marcian\`o}
\email{marciano@fudan.edu.cn}
\affiliation{Center for Field Theory and Particle Physics \& Department of Physics, Fudan University, 200433 Shanghai, China}
\affiliation{Laboratori Nazionali di Frascati INFN, Frascati (Rome), Italy, EU}

\begin{abstract}
\noindent
The black hole information paradox remains a profound challenge in theoretical physics. Among the proposed resolutions, the soft-hair approach stands out for its independence from any specific quantum gravity model. In this paper, we investigate how the inclusion of soft degrees of freedom in the unitary evolution of quantum electrodynamics, within a spacetime collapsing into a Reissner–Nordstrom black hole, leads to information scrambling. By evaluating the tripartite mutual information of this unitary evolution, we estimate the degree of information scrambling in the corresponding quantum channel. Our results show that the presence of soft degrees of freedom induces scrambling of information initially encoded in hard degrees of freedom, driven by quantum electrodynamics interactions and the nontrivial transformations arising from the non-uniqueness of the vacuum in the collapsing spacetime. This lays the groundwork for a deeper understanding of the black hole information paradox, particularly the mechanisms behind information scrambling.
\end{abstract}

\maketitle

\section{Introduction}
The black hole information paradox is one of the most profound puzzles in modern theoretical physics \cite{Harlow2016Jerusalem,Marolf2017black,Polchinski2016Black,Chen2015Black,Balasubramanian2011Quantitativ}. According to Hawking's seminal calculation, black holes are not entirely ``black'' but instead emit thermal radiation due to quantum effects—now known as Hawking radiation \cite{Hawking1975Particle,Hawking1976Breakdown}. Over time, this radiation causes the black hole to gradually evaporate and eventually disappear. However, this process appears to erase all information about the initial state beyond the black hole's mass, charge, and angular momentum, effectively transforming a pure quantum state into a thermal mixed state, hence violating the fundamental unitarity of quantum mechanics. This apparent contradiction is what we refer to as the black hole information paradox. 

To resolve this puzzle, Page proposed a perspective suggesting that the paradox may arise from an incomplete understanding of quantum evolution in such systems \cite{Page1993Information,Page1993Average,Almheiri2020entropy}. By considering the black hole and its Hawking radiation as parts of a single entangled quantum system evolving unitarily over the black hole’s lifetime, Page argued that the entanglement entropy of the radiation would initially increase but eventually decrease as the black hole shrinks, following what is now known as the Page curve --- a behavior consistent with information being preserved rather than lost.

Building on Page's insights, subsequent researches have proposed various solutions to the paradox, most of which focus on the idea that information hidden within the black hole is released through correlations between two subsystems. Notable examples include the black hole complementarity principle, which establishes complementary relationships between observers inside and outside black hole horizon \cite{Susskind1993stretched}; the AMPS paradox (and the associated firewall paradox), which considers the entanglement between early and late Hawking radiation\cite{Almheiri2013Black,Almheiri2013apologia}; the ER=EPR conjectured resolution, as an extrapolation of the AdS/CFT correspondence, wherein AdS black holes are described holographically through two entangled copies of CFT \cite{Maldacena1997Large,VanRaamsdonk2010Building,Maldacena2013Cool,Herman2020ER=EPR}; the quantum extremal island solution, which employs the Ryu-Takayanagi formula and the AdS/CFT correspondence to model entanglement between extremal islands inside the black hole and the Hawking radiation\cite{Almheiri2020Page,Almheiri2019entropy}; and the soft-hair resolution, which relies on correlations between soft and hard degrees of freedom \cite{Hawking2016Soft,Strominger2017Black}.

Despite these developments, the precise mechanism of information flow and recovery remains unclear. However, the soft hair resolution, which builds on the established understanding of both soft and hard degrees of freedom and their correlations, provides an opportunity to explicitly investigate how information flows between these degrees of freedom.

The soft hair resolution, proposed by Strominger, is based on soft hair --- additional quantum degrees of freedom induced by asymptotic symmetries beyond the classical conserved quantities of mass, angular momentum, and charge described by the no-hair theorem \cite{Hawking2016Soft}. It suggests that the correlations between the soft and hard degrees of freedom cause an observer who cannot access the former to observe the latter as being in a thermal mixed state \cite{Strominger2017Black}.

Following the idea of soft hair resolution, Carney, Chaurette, Neuenfeld, and Semenoff carried out detailed studies analyzing how a pure state describing the entire system reduces to a mixed state when partial trace is taken over the soft degrees of freedom \cite{Carney2017Infrared,Carney2018Dressed}. Their work turned the soft hair resolution from a theoretical proposal to a tangible potential solution to the black hole information paradox, paving the way for researchers to explore information flow based on the established understanding of soft hair.

Inspired by these pioneering studies, our previous work explored how the presence of soft degrees of freedom in flat spacetime affects the information carried by incoming hard degrees of freedom \cite{Su2024Scrambling}. Within the framework of quantum electrodynamics (QED), we found that the information initially encoded in the hard sector becomes delocalized across the entire system, with the source of this scrambling originating from the QED interaction. In this paper, we extend the analysis to a collapsing spacetime that forms a Reissner–Nordstrom (RN) black hole, aiming to understand how the nontrivial spacetime structure influences the degree of information scrambling. By computing the tripartite mutual information, we find that the information initially encoded in the hard sector is still scrambled, consistent with our previous findings in flat spacetime. This scrambling originates from the QED interaction, while the nontrivial structure of spacetime further modulates the degree of scrambling.

The structure of this paper is as follows. In Sec.~\ref{information-scrambling}, we introduce the general framework for estimating the degree of information scrambling, which establishes the basic methodology of our study. In Sec.~\ref{field-expansion}, we present the field-theoretic setup in collapsing spacetime, identifying the input and output states together with the corresponding quantum channel required for analyzing information scrambling. In Sec.~\ref{TMI-collapsing}, following the method outlined in Sec.~\ref{information-scrambling}, we evaluate within the QED framework how the presence of soft degrees of freedom in collapsing spacetime affects the information carried by hard degrees of freedom. Finally, in Sec.~\ref{conclusion}, we summarize our main findings.

\section{Information scrambling and tripartite mutual information}\label{information-scrambling}
Information scrambling is a phenomenon characterized by the delocalization of information \cite{Hosur2016Chaos,Harrow2021Separation,Leone2021Isospectral,Gharibyan2018Onset,Lashkari2013Towards,Maldacena2016bound,Leone2021Quantum,Hayden2007Black,Leone2022Retrieving}. A quantum channel exhibits scrambling power when the information initially carried by certain degrees of freedom of inputs becomes delocalized, such that it cannot be retrieved by local measurements of the outputs. 

To explain this, consider a quantum channel described by a bipartite unitary operator $U=U_{AB{\to}CD}$. The indices $AB$ and $CD$ label the bipartite parts of the input and output spaces. The input space, a bipartite Hilbert space $\mathcal{H}_A \otimes \mathcal{H}_B$, of this quantum channel is spanned by a set of orthonormal and complete basis $\{\ket{ab}_{\rm in}\}$, where $a = 0, \dots, d_A - 1$, $b = 0, \dots, d_B - 1$, and the composite index $ab$ runs from $0$ to $D-1$, with $D = d_A d_B$. The output space, a bipartite Hilbert space $\mathcal{H}_C \otimes \mathcal{H}_D$, is spanned by a set of orthonormal and complete basis $\{\ket{cd}_{\rm out}\}$, where $c = 0, \dots, d_C - 1$, $d = 0, \dots, d_D - 1$, and the composite index $cd$ runs from $0$ to $D-1$, with $D = d_C d_D$. 

The scrambling power of this quantum channel can be characterized by the tripartite mutual information $I_3(A:C:D)$ of the parts $A$, $C$ and $D$, which is defined as \cite{Roberts2017Chaos,Oliviero2021Random,Ding2016Conditional}
\begin{equation}
\label{TMI-definition}
I_3(A:C:D)=I(A:C)+I(A:D)-I(A:CD),
\end{equation}
where $I(X:Y)$ represents the mutual information between part $X$ and part $Y$, while $I(X:YZ)$ represents the mutual information between part $X$ and the joint part $YZ$. It quantifies how much information about part $A$ is inaccessible from either part $C$ or part $D$ individually, but is only accessible from their joint part $CD$. In other words, it characterizes the extent to which the information initially contained in part $A$ becomes non-locally distributed across the combined part $CD$, such that it cannot be retrieved from any single part $C$ or $D$. Therefore, the tripartite mutual information becomes a valuable tool for revealing the non-additive structure of information and the degree of information scrambling. In such cases, the mutual information between part $A$ and the joint part $CD$ exceeds or equals the sum of the mutual information between part $A$ and part $C$, and between part $A$ and part $D$. So that the tripartite mutual information is always non-positive, and the more negative it is, the stronger the scrambling power of the corresponding unitary operator is.

To calculate the tripartite mutual information, one requires a density matrix that fully captures the dynamics of the quantum channel. Such a density matrix can be constructed by first mapping the unitary operator of this quantum channel to its Choi state via the Choi isomorphism, and then using this state to build the corresponding density matrix. For the bipartite unitary operator $U=U_{AB{\to}CD}$, the corresponding Choi state, which is usually denoted as $\ket{U}$, is a state defined on a doubled Hilbert space $\mathcal{H}_{A}\otimes\mathcal{H}_{B}\otimes\mathcal{H}_{C}\otimes\mathcal{H}_{D}$ \cite{Hosur2016Chaos,Su2024Scrambling}, namely,
\begin{equation}
\ket{U} =\frac{1}{\sqrt{D}}\sum_{ab=0}^{D-1}\sum_{cd=0}^{D-1} u_{cdab} \ket{ab}_{\text{in}}\otimes \ket{cd}_{\text{out}}.
\end{equation}
The Choi isomorphism, which maps the bipartite unitary operator to its corresponding Choi state, can be realized by applying the operator $\mathbb{I}_{\text{in}}\otimes U_{\text{out}}$ to a $D$-dimensional maximally entangled state $\ket{I}$
\begin{equation}
\label{Choi-state-definition}
\ket{U} = (\mathbb{I}_{\text{in}}\otimes U_{\text{out}})\ket{I},
\end{equation}
where $\mathbb{I}_{\text{in}}$ represents the identity operator of the input space, $U_{\text{out}}$ is the representation of $U$ in term of the output basis $\{\ket{cd}_{\text{out}}\}$
\begin{equation}
U_{\text{out}}=\sum_{cd=0}^{D-1}\sum_{gh=0}^{D-1}u_{cdgh}\ket{cd}_{\rm out \ out} \bra{gh},
\end{equation}
while $\ket{I}$ is a normalized entangled state composed of the input basis $\{\ket{ab}_{\text{in}}\}$ and output basis $\{\ket{cd}_{\text{out}}\}$, 
\begin{equation}
\ket{I} =\frac{1}{\sqrt{D}}\sum_{ij=0}^{D-1}\ket{ij}_{\text{in}}\otimes \ket{ij}_{\text{out}}.
\end{equation}
Once the Choi state is known, the corresponding density matrix $\rho_U=\ket{U}\bra{U}$ follows immediately.

With the density matrix $\rho_U$, the tripartite mutual information can be evaluated by calculating the von Neumann entropy $S(\rho)$ of the corresponding reduced density matrix $\rho$,
\begin{equation}
S(\rho) = - {\rm tr}(\rho\log\rho),
\end{equation}
which results into \cite{Ding2016Conditional}
\begin{align}
\label{TMI-von-Neumann}
\nn
I_3 &= S(C) + S(D) - S(AC) - S(AD)\\
&= \log D - S(AC) - S(AD).
\end{align}
Here, $S(C)$ represents the von Neumann entropy of the reduced density matrix $\rho_C$, which is obtained by tracing out the $A$, $B$ and $D$ parts of the full density matrix $\rho_U$, i.e., $\rho_C={\rm tr}_{ABD}\rho_{U}$. Similarly, $S(D)$, $S(AC)$ and $S(AD)$ represent the von Neumann entropies of the reduced density matrix $\rho_D$, $\rho_{AC}$ and $\rho_{AD}$, respectively. Moreover, the tripartite mutual information $I_3$ is bounded by the inequality \cite{Ding2016Conditional}
\begin{equation}
-2 \log d_A \le I_3 \le 0.
\end{equation}
The upper bound $I_3=0$ corresponds to the situation where no scrambling occurs, while the lower bound $I_3=-2 \log d_A$ characterizes the situation that the information is maximal scrambled. In other words, the more negative $I_3$ is, the greater the degree of information scrambling.

However, in practice, the calculation of the von Neumann entropy is not always straightforward. In such cases, the 2-Renyi entropy,
\begin{equation}
\label{2-renyi-entropy}
S_{2}(\rho) = -\log{\rm tr}\rho^2,
\end{equation}
can be used to estimate the tripartite mutual information \cite{Ding2016Conditional}
\begin{align}
\label{TMI-2-Renyi}
\nn
I_{3(2)} &= S_2(C) + S_2(D) - S_2(AC) - S_2(AD)\\
&= \log D - S_2(AC) - S_2(AD).
\end{align}
Here, $S_2(C)$ denotes the 2-Renyi entropy of the reduced density matrix $\rho_C$, and $S_2(D)$, $S_2(AC)$ and $S_2(AD)$ denote those of $\rho_D$, $\rho_{AC}$ and $\rho_{AD}$. The rationale behind this estimation is that the tripartite mutual information $I_{3(2)}$ (calculated in terms of 2-Renyi entropy) serves as an upper bound of the tripartite mutual information $I_3$ (calculated in terms of von Neumann entropy). In other words, the tripartite mutual information $I_{3(2)}$ satisfies the following inequality \cite{Ding2016Conditional}
\begin{equation}
\label{TMI-inequality}
-2 \log d_A \le I_3 \le I_{3(2)} \le 0.
\end{equation}
Therefore, if a unitary operator exhibits scrambling as indicated by the tripartite mutual information calculated in terms of the 2-Renyi entropy, then its actual scrambling power—measured via the von Neumann entropy—must be greater than or equal to that calculated in terms of the 2-Renyi entropy.

\section{Field expansion in collapsing spacetime}\label{field-expansion}
The non-uniqueness of the vacuum is a fundamental feature of quantum field theory in curved spacetime, setting it apart from its flat-spacetime counterpart \cite{Hawking1975Particle,Birrell1982Quantum,Wald1975particle,DEWITT1975Quantum}. In flat spacetime, the Poincar\'e group characterizes the fundamental symmetries of the manifold. This gives rise to a natural choice of coordinates—namely, the Cartesian coordinates $(t,x,y,z)$. Then the global timelike Killing vector, which is orthogonal to the Cauchy surface $t={\rm constant}$, allows for a unique and invariant definition of positive-frequency modes of a quantum field. 
In contrast, in curved spacetime, the manifold generally does not possess a global symmetry group such as the Poincar\'e group. 
Therefore, there is generally no preferred coordinates and no global timelike Killing vector, which makes it much harder to provide a unique and invariant definition of positive-frequency modes of a quantum field.
In some spacetimes, there are several asymptotically flat regions, where several approximate Killing vectors may exist. As a result, the definition of positive-frequency modes of a quantum field becomes ambiguous, and the mode expansion of the quantum field is non-unique.

Consider the scenario of a RN black hole formed through the gravitational collapse of spherically symmetric matter. The corresponding spacetime is commonly referred to as collapsing spacetime. As a curved spacetime, the collapsing spacetime does not yield a unique expansion of quantum fields. Due to the presence of two asymptotically flat regions, there are two distinct expansions of the field, which are inequivalent to each other. To illustrate this, consider a black hole with mass $M$ and charge $Q=Ze$, where $Z$ is an integer and $e$ is the elementary charge. In the distant past, before the matter becomes highly concentrated, the spacetime can be approximately described by flat Minkowski metric, with the line element
\begin{equation}
ds^2 = dt^2 - dx^2 - dy^2 -dz^2 = dt^2 - dr^2 - r^2(d\theta^2 + \sin^2\theta d\phi^2).
\end{equation}
After the collapse, in the distant future, the resulting black hole is described by the RN metric
\begin{align}
\label{RN-metric}
\nn
ds^2 =& \Big( 1-\frac{r_s}{r} + \frac{r_Q^2}{r^2} \Big) dt^2 - \Big( 1-\frac{r_s}{r} + \frac{r_Q^2}{r^2} \Big)^{-1} dr^2\\
&- r^2 ( d{\theta}^2 + \sin^2\theta d{\phi}^2 ),
\end{align}
where $r_s=2GM$ is the Schwarzschild radius, $r_Q^2=Q^2G$ is the charge-related length scale, $G=1/(8\pi M_{pl}^2)$ and $M_{pl}$ is the Planck mass.

The collapsing spacetime contains two asymptotically flat regions, denoted as $I^- \cup i^-$ (the union of past null infinity and past time-like infinity) and $I^+ \cup i^+$ (the union of future null infinity and future time-like infinity), as well as a future event horizon $H^+$. 
In each region, one can define a set of positive-frequency modes\footnote{For the collapsing spacetime, only the portion of the horizon outside the collapsing matter is stationary \cite{Wald1975particle}. As a result, there is no globally defined Killing vector on the horizon $H^+$, leading to an ambiguity in the definition of positive frequency modes on $H^+$. Nevertheless, one can always obtain a set of ``positive frequency'' modes by selecting a set of orthonormal solutions that vanish on $I^+ \cup i^+$. Although this selection is not unique, fortunately, the specific choice does not affect physical predictions on $I^+ \cup i^+$ \cite{Hawking1975Particle,Birrell1982Quantum}.}
as orthonormal solutions to the field equations.
The region $I^- \cup i^-$ forms a complete Cauchy surface for the entire spacetime, and the modes defined on it form a complete basis.
In contrast, the region $I^+ \cup i^+ \cup H^+$ is not a global Cauchy surface --- it only serves as a Cauchy surface for its own causal past. As a result, the modes defined on $I^+ \cup i^+ \cup H^+$ form a complete basis only for observables supported within this causal domain, but not for the entire spacetime. This reflects a key difference between the spacetimes describing the gravitational collapse to form RN and Schwarzschild black holes, as $I^+ \cup i^+ \cup H^+$ forms a complete Cauchy surface in the Schwarzschild case.

In the following, we consider the mode expansion of fermion and gauge fields in the collapsing spacetime, which provides the basis for analyzing the information scrambling within the frame work of quantum electrodynamics.

\subsection{Mode expansion of free field}
Denote the free fermion and gauge fields by $\psi$ and $A_{\mu}$. Let $\psi_{Q}^{1+}$ and $\psi_{Q}^{1-}$, where $Q$ labels the relevant quantum number $Q$ (including momentum, spin projection, etc.), denote the positive- and negative-frequency modes of the Dirac equation on $I^- \cup i^-$, with $\psi_{Q}^{1-} \ne (\psi_{Q}^{1+})^{\dagger}$. The set $\{\psi_{Q}^{1}\}=\{\psi_{Q}^{1+},\psi_{Q}^{1-}\}$ forms a complete orthonormal basis for the fermion field. Similarly, let $A_{\mu Q}^{1+}$ and $A_{\mu Q}^{1-}$ denote the positive- and negative-frequency modes of the gauge field equations on the same surfaces, with $A_{\mu Q}^{1-} = (A_{\mu Q}^{1+})^{\dagger}$. The corresponding set $\{A_{\mu Q}^{1}\}=\{A_{\mu Q}^{1-},\\ A_{\mu Q}^{1+}\}$ forms a complete basis for the gauge field. Then the fermion and gauge fields can be expanded as
\begin{align}
\label{psi_in}
& \psi = \sum_{Q} \Big\{\psi_{Q}^{1+}a_{Q}^{1} + \psi_{Q}^{1-}b_{Q}^{1\dagger}\Big\},\\[-1mm]
\label{A_in}
& A_{\mu} = \sum_{Q} \Big\{A_{\mu Q}^{1+}c_{Q}^{1} + A_{\mu Q}^{1-}c_{Q}^{1\dagger}\Big\},
\end{align}
where $\smash{a_{Q}^{1\dagger}}$ and $\smash{a_{Q}^{1}}$	are the creation and annihilation operators for fermions, $\smash{b_{Q}^{1\dagger}}$ and $\smash{b_{Q}^{1}}$ for anti-fermions, and $c_{Q}^{1\dagger}$ and $c_{Q}^{1}$ for photons. The superscript $1$ indicates that these operators are associated with the mode sets $\{\psi_{Q}^{1}\}$ and $\{A_{\mu Q}^{1}\}$ defined on $I^- \cup i^-$ (corresponding to distant past). The annihilation operators $a_{Q}^1$, $b_{Q}^1$ and $c_{Q}^1$ also define the vacuum in the distant past for all $Q$ via
\begin{equation}
a_{Q}^{1}\ket{0} = b_{Q}^{1}\ket{0} = c_{Q}^{1}\ket{0} = 0.
\end{equation}

Similarly, let $\{\psi_{Q}^{2}\}=\{\psi_{Q}^{2+},\psi_{Q}^{2-}\}$ denote the positive‑ and negative‑frequency mode set of the fermion field on $I^+ \cup i^+$, describing particles escaping to future infinity, and let $\{\psi_{Q}^{3}\}=\{\psi_{Q}^{3+},\psi_{Q}^{3-}\}$ denote the positive‑ and negative‑frequency mode set on the future horizon $H^+$, describing particles falling into the black hole. Likewise, let $\{A_{\mu Q}^{2}\}=\{A_{\mu Q}^{2-},A_{\mu Q}^{2+}\}$ and $\{A_{\mu Q}^{3}\}=\{A_{\mu Q}^{3-},A_{\mu Q}^{3+}\}$ denote the positive‑ and negative‑frequency mode sets of gauge field on $I^+ \cup i^+$ and $H^+$. Then the fermion and gauge fields can be expanded as
\begin{align}
\label{psi_out}
& \psi \!=\! \sum_{Q} \Big\{\psi_{Q}^{2+}a_{Q}^{2} + \psi_{Q}^{2-}b_{Q}^{2\dagger} + \psi_{Q}^{3+}a_{Q}^{3} + \psi_{Q}^{3-}b_{Q}^{3\dagger}\Big\},\\[-1mm]
\label{A_out}
& A_{\mu} \!=\! \sum_{Q} \Big\{A_{\mu Q}^{2+}c_{Q}^{2} + A_{\mu Q}^{2-}c_{Q}^{2\dagger} + A_{\mu Q}^{3+}c_{Q}^{3} + A_{\mu Q}^{3-}c_{Q}^{3\dagger}\Big\},
\end{align}
where $\smash{a_{Q}^{2\dagger}}$ and $\smash{a_{Q}^{2}}$, $\smash{b_{Q}^{2\dagger}}$ and $\smash{b_{Q}^{2}}$, $\smash{c_{Q}^{2\dagger}}$ and $\smash{c_{Q}^{2}}$, $\smash{a_{Q}^{3\dagger}}$ and $\smash{a_{Q}^{3}}$, $\smash{b_{Q}^{3\dagger}}$ and $\smash{b_{Q}^{3}}$, $\smash{c_{Q}^{3\dagger}}$ and $\smash{c_{Q}^{3}}$	are the corresponding creation and annihilation operators.
The superscript $2$ indicates that these operators are associated with the mode sets $\{\psi_{Q}^{2}\}$ and $\{A_{\mu Q}^{2}\}$ defined on $I^+ \cup i^+$ (corresponding to the asymptotically region of distant future), while the superscript $3$ indicates association with the mode sets $\{\psi_{Q}^{3}\}$ and $\{A_{\mu Q}^{3}\}$ defined on $H^+$ (corresponding to the event horizon of distant future). The annihilation operators $a_{Q}^2$, $b_{Q}^2$, $c_{Q}^2$, $a_{Q}^3$, $b_{Q}^3$ and $c_{Q}^3$ define another vacuum in the distant future for all $Q$ via
\begin{equation}
a_{Q}^{2}\ket{\bar{0}} \!=\! a_{Q}^{3}\ket{\bar{0}} \!=\! b_{Q}^{2}\ket{\bar{0}} \!=\! b_{Q}^{3}\ket{\bar{0}} \!=\! c_{Q}^{2}\ket{\bar{0}} \!=\! c_{Q}^{3}\ket{\bar{0}} \!=\! 0.
\end{equation}
Since the mode functions defined on $I^- \cup i^-$, and those on $I^+ \cup i^+ \cup H^+$, each form a complete basis in the causal past of $I^+ \cup i^+ \cup H^+$, they can be expanded in terms of one another. For example, the modes ${\psi_Q^2}$  and ${\psi_Q^3}$, ${A_{\mu Q}^2}$ and ${A_{\mu Q}^3}$ can be written as linear combinations of the modes ${\psi_Q^1}$ and ${A_{\mu Q}^1}$
\begin{align}
\label{dependence-past-future}
& \psi_{Q}^{2\pm} = \sum_{Q'} \Big\{ \alpha_{QQ'}^{2\pm}\psi_{Q'}^{1+} + \beta_{QQ'}^{2\pm}\psi_{Q'}^{1-} \Big\},\\[-1mm]
& \psi_{Q}^{3\pm} = \sum_{Q'} \Big\{ \alpha_{QQ'}^{3\pm}\psi_{Q'}^{1+} + \beta_{QQ'}^{3\pm}\psi_{Q'}^{1-} \Big\},\\[-1mm]
& A_{\mu Q}^{2+} = \sum_{Q'} \Big\{ \gamma_{QQ'}^{2+}A_{\mu Q'}^{1+} + \eta_{QQ'}^{2+}A_{\mu Q'}^{1-} \Big\},\\[-1mm]
& A_{\mu Q}^{3+} = \sum_{Q'} \Big\{ \gamma_{QQ'}^{3+}A_{\mu Q'}^{1+} + \eta_{QQ'}^{3+}A_{\mu Q'}^{1-} \Big\}.
\end{align}
These are Bogoliubov transformations with corresponding Bogoliubov coefficients $\alpha$, $\beta$, $\gamma$, and $\eta$. Substituting them into Eq.~\eqref{psi_out} and Eq.~\eqref{A_out} gives the Bogoliubov transformations between the two sets of creation and annihilation operators
\begin{align}
& a_{Q}^{1} = \sum_{Q'} \Big\{ a_{Q'}^2\alpha_{Q'Q}^{2+} + b_{Q'}^{2\dagger}\alpha_{Q'Q}^{2-} + a_{Q'}^3\alpha_{Q'Q}^{3+} + b_{Q'}^{3\dagger}\alpha_{Q'Q}^{3-} \Big\},\\[-1mm]
& b_{Q}^{1\dagger} = \sum_{Q'} \Big\{ a_{Q'}^2\beta_{Q'Q}^{2+} + b_{Q'}^{2\dagger}\beta_{Q'Q}^{2-} + a_{Q'}^3\beta_{Q'Q}^{3+} + b_{Q'}^{3\dagger}\beta_{Q'Q}^{3-} \Big\},\\[-1mm]
& c_{Q}^{1} = \sum_{Q'} \Big\{ c_{Q'}^2\gamma_{Q'Q}^{2+} + c_{Q'}^{2\dagger}\eta_{Q'Q}^{2-} + c_{Q'}^3\gamma_{Q'Q}^{3+} + c_{Q'}^{3\dagger}\eta_{Q'Q}^{3-} \Big\}.
\end{align}
These relations show that although the annihilation operators $a_Q^1$, $b_Q^1$, and $c_Q^1$ annihilate the vacuum state $\ket{0}$, they do not annihilate $\ket{\bar{0}}$.

Given the vacuum state and the associated creation and annihilation operators, one can construct the corresponding Fock states. The Fock states defined with respect to the vacuum in the distant past are given by
\begin{align}
\nn
& a_{Q}^{1\dagger}\ket{0} \!=\! \ket{1_{Q}^{e^{-}}}, \ b_{Q}^{1\dagger}\ket{0} \!=\! \ket{1_{Q}^{e^{+}}}, \ \ket{\{1_{Q}^{e^{\pm}}\}}\!=\!\prod_Q\!\ket{1_{Q}^{e^{\pm}}},\\[-1mm]
\nn
&c_{Q}^{1\dagger}\ket{0} \!=\! \ket{1_{Q}^{\gamma}}, \ \ket{n_{Q}^{\gamma}} \!=\! \frac{1}{\sqrt{n_{Q}^{\gamma}!}}\ket{1_{Q}^{\gamma},\cdots,1_{Q}^{\gamma}},\\[-1mm]
& \ket{\{n_{Q}^{\gamma}\}}\!=\!\prod_Q\!\ket{n_{Q}^{\gamma}}.
\end{align}
Here, $\ket{1_{Q}^{*}}$ ($*= e^{-}, e^{+}, \gamma$) denotes a single-particle state in the distant past labeled by quantum number $Q$, and $\ket{n_{Q}^{*}}$ denotes a multi-particle state with $n_{Q}^{*}$ particles in the same single-particle state $\ket{1_{Q}^{*}}$. 
By the Pauli exclusion principle, the fermion occupation numbers $\smash{n_{Q}^{e^{-}}}$ and $\smash{n_{Q}^{e^{+}}}$ can only take the values $0$ or $1$, but for notational convenience we continue to use them to indicate the electron and positron occupation numbers.
Therefore, the complete Fock basis on $I^- \cup i^-$ can be compactly denoted as $\{\ket{\{n_{Q}^{e^{-}}\}, \{n_{Q}^{e^{+}}\}, \{n_{Q}^{\gamma}}\}$. Similarly, the Fock states defined with respect to the vacuum in the distant future are given by
\begin{align}
\nn
& a_{Q}^{2\dagger}\ket{\bar{0}} \!=\! \ket{\bar{1}_{Q}^{e^{-}}}, \ b_{Q}^{2\dagger}\ket{\bar{0}} \!=\! \ket{\bar{1}_{Q}^{e^{+}}}, \ \ket{\{\bar{1}_{Q}^{e^{\pm}}\}}\!=\!\prod_Q\!\ket{\bar{1}_{Q}^{e^{\pm}}},\\[-1mm]
\nn
&c_{Q}^{2\dagger}\ket{\bar{0}} \!=\! \ket{\bar{1}_{Q}^{\gamma}}, \ \ket{\bar{n}_{Q}^{\gamma}} \!=\! \frac{1}{\sqrt{n_{Q}^{\gamma}!}}\ket{\bar{1}_{Q}^{\gamma},\cdots,\bar{1}_{Q}^{\gamma}},\\[-1mm]
& \ket{\{\bar{n}_{Q}^{\gamma}\}}\!=\!\prod_Q\!\ket{\bar{n}_{Q}^{\gamma}},
\end{align}
and
\begin{align}
\nn
& a_{Q}^{3\dagger}\ket{\bar{0}} \!=\! \ket{\tilde{1}_{Q}^{e^{-}}}, \ b_{Q}^{3\dagger}\ket{\bar{0}} \!=\! \ket{\tilde{1}_{Q}^{e^{+}}}, \ \ket{\{\tilde{1}_{Q}^{e^{\pm}}\}}\!=\!\prod_Q\!\ket{\tilde{1}_{Q}^{e^{\pm}}},\\[-1mm]
\nn
&c_{Q}^{3\dagger}\ket{\bar{0}} \!=\! \ket{\tilde{1}_{Q}^{\gamma}}, \ \ket{\tilde{n}_{Q}^{\gamma}} \!=\! \frac{1}{\sqrt{n_{Q}^{\gamma}!}}\ket{\tilde{1}_{Q}^{\gamma},\cdots,\tilde{1}_{Q}^{\gamma}},\\[-1mm]
& \ket{\{\bar{n}_{Q}^{\gamma}\}}\!=\!\prod_Q\!\ket{\bar{n}_{Q}^{\gamma}}.
\end{align}
Here, $\ket{\bar{1}_{Q}^{*}}$ and $\ket{\tilde{1}_{Q}^{*}}$ denote the single-particle states in the distant future, the former defined on $I^+ \cup i^+$ and the latter on $H^+$, while $\ket{\bar{n}_{Q}^{*}}$ and $\ket{\tilde{n}_{Q}^{*}}$ denote the corresponding multi-particle states.
Therefore, the complete Fock basis on $I^+ \cup i^+ \cup H^+$ can be compactly denoted as $\{\ket{\{\bar{n}_{\bar{Q}^{-}}^{e^{-}}\},\!\{\bar{n}_{\bar{Q}^{+}}^{e^{+}}\},\!\{\bar{n}_{\bar{Q}}^{\gamma}\},\!\{\tilde{n}_{\tilde{Q}^{-}}^{e^{-}}\},\!\{\tilde{n}_{\tilde{Q}^{+}}^{e^{+}}\},\!\{\tilde{n}_{\tilde{Q}}^{\gamma}\}}$.

\subsection{Mode expansion of interacting field}
Under the standard assumption that interactions vanish in the asymptotic limits $t\to\pm\infty$, interacting quantum fields can be treated as free fields in both the distant past and future \cite{Birrell1982Quantum,Birrell1980Analysis,Wald1975particle}. Accordingly, the fermion field $\psi(x)$ and the gauge field $A_{\mu}(x)$ asymptotically reduce to
\begin{align}
\label{incoming}
& \lim_{t\to-\infty} \psi(x) = \psi^{\rm in}(x), \quad \lim_{t\to+\infty} \psi(x) = \psi^{\rm out}(x),\\
\label{outgoing}
& \lim_{t\to-\infty} A_{\mu}(x) = A_{\mu}^{\rm in}(x), \quad \lim_{t\to+\infty} A_{\mu}(x) = A_{\mu}^{\rm out}(x),
\end{align}
where ``in'' and ``out'' denote the incoming and outgoing free fields, respectively.

As in the free-field case, each asymptotic field $\psi^{\rm in}$, $\psi^{\rm out}$, $A_{\mu}^{\rm in}$ and $A_{\mu}^{\rm out}$ admits two distinct mode expansions: one in terms of the mode sets $\{\psi_{Q}^{1}\}$, $\{A_{\mu Q}^{1}\}$, and another in terms of the mode sets $\{\psi_{Q}^{2}\}\cup\{\psi_{Q}^{3}\}$, $\{A_{\mu Q}^{2}\}\cup\{A_{\mu Q}^{3}\}$. Explicitly
\begin{align}
\nn
\psi^{\rm in/out} =& \sum_{Q} \Big\{\psi_{Q}^{1+}(a_{Q}^{1})_{\rm in/out} + \psi_{Q}^{1-}(b_{Q}^{1\dagger})_{\rm in/out}\Big\}\\[-1mm]
\nn
=& \sum_{Q} \Big\{\psi_{Q}^{2+}(a_{Q}^{2})_{\rm in/out} + \psi_{Q}^{2-}(b_{Q}^{2\dagger})_{\rm in/out}\\[-1mm]
& \qquad + \psi_{Q}^{3+}(a_{Q}^{3})_{\rm in/out} + \psi_{Q}^{3-}(b_{Q}^{3\dagger})_{\rm in/out}\Big\},\\[2mm]
\nn
A_{\mu}^{\rm in/out} =& \sum_{Q} \Big\{A_{\mu Q}^{1+}(c_{Q}^{1})_{\rm in/out} + A_{\mu Q}^{1-}(c_{Q}^{1\dagger})_{\rm in/out}\Big\}\\[-1mm]
\nn
=& \sum_{Q} \Big\{A_{\mu Q}^{2+}(c_{Q}^{2})_{\rm in/out} + A_{\mu Q}^{2-}(c_{Q}^{2\dagger})_{\rm in/out}\\[-1mm]
& \qquad + A_{\mu Q}^{3+}(c_{Q}^{3})_{\rm in/out} + A_{\mu Q}^{3-}(c_{Q}^{3\dagger})_{\rm in/out}\Big\}.
\end{align}
The corresponding annihilation operators define four vacuum states: $\ket{0}_{\rm in}$, $\ket{\bar{0}}_{\rm in}$, $\ket{0}_{\rm out}$ and $\ket{\bar{0}}_{\rm out}$. The states $\ket{0}_{\rm in/out }$ are annihilated by the operators associated with the mode sets $\{\psi_{Q}^{1}\}$, $\{A_{\mu Q}^{1}\}$, namely
\begin{align}
&(a_{Q}^{1})_{\rm in}\ket{0}_{\rm in} = (b_{Q}^{1})_{\rm in}\ket{0}_{\rm in} = (c_{Q}^{1})_{\rm in}\ket{0}_{\rm in} = 0,\\
&(a_{Q}^{1})_{\rm out}\ket{0}_{\rm out} = (b_{Q}^{1})_{\rm out}\ket{0}_{\rm out} = (c_{Q}^{1})_{\rm out}\ket{0}_{\rm out} = 0,
\end{align}
for all $Q$. Similarly, the states $\ket{\bar{0}}_{\rm in/out}$ are annihilated by the operators associated with the mode sets $\{\psi_{Q}^{2}\}\cup\{\psi_{Q}^{3}\}$, $\{A_{\mu Q}^{2}\}\cup\{A_{\mu Q}^{3}\}$, namely
\begin{align}
& (a_{Q}^{2})_{\rm in}\ket{\bar{0}}_{\rm in} = (b_{Q}^{2})_{\rm in}\ket{\bar{0}}_{\rm in} = (c_{Q}^{2})_{\rm in}\ket{\bar{0}}_{\rm in}=0, \\
& (a_{Q}^{3})_{\rm in}\ket{\bar{0}}_{\rm in} = (b_{Q}^{3})_{\rm in}\ket{\bar{0}}_{\rm in} = (c_{Q}^{3})_{\rm in}\ket{\bar{0}}_{\rm in} = 0,\\
& (a_{Q}^{2})_{\rm out}\ket{\bar{0}}_{\rm out} = (b_{Q}^{2})_{\rm out}\ket{\bar{0}}_{\rm out} = (c_{Q}^{2})_{\rm out}\ket{\bar{0}}_{\rm out}=0, \\
& (a_{Q}^{3})_{\rm out}\ket{\bar{0}}_{\rm out} = (b_{Q}^{3})_{\rm out}\ket{\bar{0}}_{\rm out} = (c_{Q}^{3})_{\rm out}\ket{\bar{0}}_{\rm out} = 0,
\end{align}
for all $Q$.

Although the fields $\psi^{\rm in}$, $\psi^{\rm out}$, $A_{\mu}^{\rm in}$ and $A_{\mu}^{\rm out}$ can, in principle, be expanded using different complete sets of modes, the expansion in terms of modes associated with the corresponding asymptotic region gives rise to physical states with clear particle interpretation. For instance, the physically meaningful expansion of $\psi^{\rm in}$ and $A_{\mu}^{\rm in}$ is in terms of the modes defined in the distant past, while that of $\psi^{\rm out}$ and $A_{\mu}^{\rm out}$ is in terms of the modes defined in the distant future. Expansions using the mode set associated with the opposite asymptotic region may still be useful in intermediate calculations, such as in constructing Bogoliubov transformations, but do not directly correspond to physical particle states.

Given the choice of asymptotic mode sets, one can construct physically meaningful Fock states for each asymptotic region by acting the corresponding creation operators on the associated vacuum states.
For the incoming region $I^- \cup i^-$, the physically meaningful Fock states are generated by acting the creation operators $(a_{Q}^{1\dagger})_{\rm in}$, $(b_{Q}^{1\dagger})_{\rm in}$ and $(c_{Q}^{1\dagger})_{\rm in}$ on the in-vacuum $\ket{0}_{\rm in}$, namely,
\begin{align}
\label{in-vacuum}
\nn
& (a_{Q}^{1\dagger})_{\rm in}\ket{0}_{\rm in} = \ket{1_{Q}^{e^{-}}}_{\rm in}, \ (b_{Q}^{1\dagger})_{\rm in}\ket{0}_{\rm in} = \ket{1_{Q}^{e^{+}}}_{\rm in},\\
& (c_{Q}^{1\dagger})_{\rm in}\ket{0}_{\rm in} = \ket{1_{Q}^{\gamma}}_{\rm in}.
\end{align}
Therefore, the complete and physically meaningful Fock basis on $I^- \cup i^-$ is denoted as $\{\ket{\{n_{Q^{-}}^{e^{-}}\},\!\{n_{Q^{+}}^{e^{+}}\},\!\{n_{Q}^{\gamma}\}}_{\rm in}\}$.
Likewise, the physically meaningful Fock states on the outgoing region $I^+ \cup i^+ \cup H^+$ are generated by acting the creation operators $(a_{Q}^{2\dagger})_{\rm out}$, $(b_{Q}^{2\dagger})_{\rm out}$, $(c_{Q}^{2\dagger})_{\rm out}$, $(a_{Q}^{3\dagger})_{\rm out}$, $(b_{Q}^{3\dagger})_{\rm out}$, and $(c_{Q}^{3\dagger})_{\rm out}$ on the out-vacuum $\ket{\bar{0}}_{\rm out}$, namely,
\begin{align}
\nn
& (a_{Q}^{2\dagger})_{\rm out}\ket{\bar{0}}_{\rm out} = \ket{\bar{1}_{Q}^{e^{-}}}_{\rm out}, \ (a_{Q}^{3\dagger})_{\rm out}\ket{\bar{0}}_{\rm out} = \ket{\tilde{1}_{Q}^{e^{-}}}_{\rm out},\\
\nn
& (b_{Q}^{2\dagger})_{\rm out}\ket{\bar{0}}_{\rm out} = \ket{\bar{1}_{Q}^{e^{+}}}_{\rm out}, \ (b_{Q}^{3\dagger})_{\rm out}\ket{\bar{0}}_{\rm out} = \ket{\tilde{1}_{Q}^{e^{+}}}_{\rm out},\\
&(c_{Q}^{2\dagger})_{\rm out}\ket{\bar{0}}_{\rm out} = \ket{\bar{1}_{Q}^{\gamma}}_{\rm out}, \ (c_{Q}^{3\dagger})_{\rm out}\ket{\bar{0}}_{\rm out} = \ket{\tilde{1}_{Q}^{\gamma}}_{\rm out},
\end{align}
with the bar and tilde indicating states defined on $I^+ \cup i^+$ and $H^+$, respectively.
Therefore, the complete and physically meaningful Fock basis on $I^+ \cup i^+ \cup H^+$ is denoted as $\{\ket{\{\bar{n}_{\bar{Q}^{-}}^{e^{-}}\},\!\{\bar{n}_{\bar{Q}^{+}}^{e^{+}}\},\!\{\bar{n}_{\bar{Q}}^{\gamma}\},\!\{\tilde{n}_{\tilde{Q}^{-}}^{e^{-}}\},\!\{\tilde{n}_{\tilde{Q}^{+}}^{e^{+}}\},\!\{\tilde{n}_{\tilde{Q}}^{\gamma}\}}_{\rm out}\}$.
For convenience, hereafter, we abbreviate the physically meaningful incoming Fock state $\ket{\{n_{Q^{-}}^{e^{-}}\},\!\{n_{Q^{+}}^{e^{+}}\},\{n_{Q}^{\gamma}\}}_{\rm in}$ as $\ket{\{n_{Q^{*}}^{*}\}}_{\rm in}$ and the physically meaningful outgoing Fock state $\ket{\{\bar{n}_{\bar{Q}^{-}}^{e^{-}}\},\!\{\bar{n}_{\bar{Q}^{+}}^{e^{+}}\},\!\{\bar{n}_{\bar{Q}}^{\gamma}\},\!\{\tilde{n}_{\tilde{Q}^{-}}^{e^{-}}\},\!\{\tilde{n}_{\tilde{Q}^{+}}^{e^{+}}\},\!\{\tilde{n}_{\tilde{Q}}^{\gamma}\}}_{\rm out}$ as $\ket{\{\bar{\tilde{n}}_{\bar{\tilde{Q}}^{*}}^{*}\}}_{\rm out}$. 

With the physically meaningful incoming Fock state $\ket{\{n_{Q^{*}}^{*}\}}_{\rm in}$ and the physically meaningful outgoing Fock state $\ket{\{\bar{\tilde{n}}_{\bar{\tilde{Q}}^{*}}^{*}\}}_{\rm out}$, the $S$-matrix element associated with them is then given by their inner product \cite{Birrell1982Quantum}
\begin{align}
\label{amplitude-Fock}
\nn
&{}_{\rm out}\braket{\{\bar{\tilde{n}}_{\bar{\tilde{Q}}^{*}}^{*}\}|\{n_{Q^{*}}^{*}\}}_{\rm in}\\[2mm]
\nn
&= \sum_{\{m_{q^{*}}\}} {}_{\rm out}\braket{\{\bar{\tilde{n}}_{\bar{\tilde{Q}}^{*}}^{*}\}|\{m_{q^{*}}^{*}\}}_{\rm out \ out}\braket{\{m_{q^{*}}^{*}\}|\{n_{Q^{*}}^{*}\}}_{\rm in}\\[-1mm]
&= \sum_{\{m_{q^{*}}\}} {}_{\rm out}\braket{\{\bar{\tilde{n}}_{\bar{\tilde{Q}}^{*}}^{*}\}|\{m_{q^{*}}^{*}\}}_{\rm out \ in}\braket{\{m_{q^{*}}^{*}\}|S|\{n_{Q^{*}}^{*}\}}_{\rm in}.
\end{align}
The first factor ${}_{\rm out}\braket{\{\bar{\tilde{n}}_{\bar{\tilde{Q}}^{*}}^{*}\}|\{m_{q^{*}}^{*}\}}_{\rm out}$ --- the inner product between outgoing states defined with respect to different mode sets --- is fully determined by the Bogoliubov coefficients and is independent of quantum interactions. In contrast, the second factor ${}_{\rm out}\braket{\{m_{q^{*}}^{*}\}|\{n_{Q^{*}}^{*}\}}_{\rm in}$ --- the inner product between incoming and outgoing states defined in terms of the same mode set --- encodes the effects of quantum interactions. Moreover, the factor ${}_{\rm out}\braket{\{m_{q^{*}}^{*}\}|\{n_{Q^{*}}^{*}\}}_{\rm in}$ is the curved-spacetime counterpart of the flat-spacetime $S$-matrix element in the Heisenberg picture, while the factor ${}_{\rm in}\braket{\{m_{q^{*}}^{*}\}|S|\{n_{Q^{*}}^{*}\}}_{\rm in}$ corresponds to that one in either the Schr\"odinger or the interaction pictures, where $S$ denotes the evolution operator of quantum states.
In practice, the calculations of the factor ${}_{\rm in}\braket{\{m_{q^{*}}^{*}\}|S|\{n_{Q^{*}}^{*}\}}_{\rm in}$ are typically performed in the interaction picture --- here the interaction can be separated from the free evolution, making perturbative expansions more straightforward.
In this case, the evolution operator $S$ for QED takes the form
\begin{equation}
\label{total-S}
S = P \exp \Big[ -i \int_{\Sigma_{\rm in}}^{\Sigma_{\rm out}} \sqrt{-g} \ e \bar{\psi}(x)\gamma^{\mu}A_{\mu}(x)\psi(x) d^4x \Big].
\end{equation}
The operator $P$ is the curved-spacetime counterpart of the flat-spacetime time-ordering operator \cite{Birrell1982Quantum}. The surface $\Sigma_{\rm in}$, corresponding to $I^- \cup i^-$, denotes the spacelike Cauchy surface in the distant past, while $\Sigma_{\rm out}$, corresponding to $I^+ \cup i^+ \cup H^+$, denotes the spacelike Cauchy surface in the distant future --- this latter serves as a Cauchy surface only for the causal past of $I^+ \cup i^+ \cup H^+$.

\section{Tripartite mutual information in collapsing spacetime}\label{TMI-collapsing}
As demonstrated in references \cite{Strominger2017Black,Carney2017Infrared,Carney2018Dressed}, the correlation between soft and hard degrees of freedom can make a pure state appear mixed, suggesting a possible route toward solving the black hole information paradox. This naturally leads to the question of how the presence of soft degrees of freedom induces the scrambling of information carried by hard degrees of freedom. Motivated by this, in our previous work we studied how the inclusion of soft photon degrees of freedom in the unitary evolution of QED in flat spacetime leads to information scrambling \cite{Su2024Scrambling}. Using the method outlined in Section \ref{information-scrambling}, we quantified the degree of scrambling via the tripartite mutual information, specifically, established an upper bound on the tripartite mutual information, thereby providing a lower bound on the amount of information that must be scrambled. In this paper, following the same approach, we extend the analysis to the collapsing spacetime, aiming to compare the scrambling behavior in the presence of soft photon degrees of freedom between flat and collapsing spacetimes.

There are two main differences between studying information scrambling in collapsing spacetime and in flat spacetime. 

First, as already shown in section \ref{field-expansion}, unlike in flat spacetime, the vacuum state of a quantum field in collapsing spacetime is not unique. This directly leads to the fact that the formal expression of the $S$-matrix element in collapsing spacetime contains some additional factors ${}_{\rm out}\braket{\{\bar{\tilde{n}}_{\bar{\tilde{Q}}^{*}}^{*}\}|\{m_{q^{*}}^{*}\}}_{\rm out}$ compared to the formal expression of the $S$-matrix element in flat spacetime --- see e.g. Eq.~\eqref{amplitude-Fock}. These additional factors, which are fully determined by the Bogoliubov coefficients, increase the complexity of the analysis. However, when the timescale of the gravitational collapse greatly exceeds that of the QED interaction, the structure of the $S$-matrix element --- as a sum of products of two types of factors --- together with the properties of Bogoliubov transformation and the perturbative nature of QED, makes it possible to analyze the contributions to the tripartite mutual information from the pure gravitational background, the pure QED interaction, and the gravitational corrections to QED separately. This greatly simplifies the evaluation of information scrambling.

\begin{figure}[h]
\centering
\includegraphics[width=8.5cm,height=3.25cm]{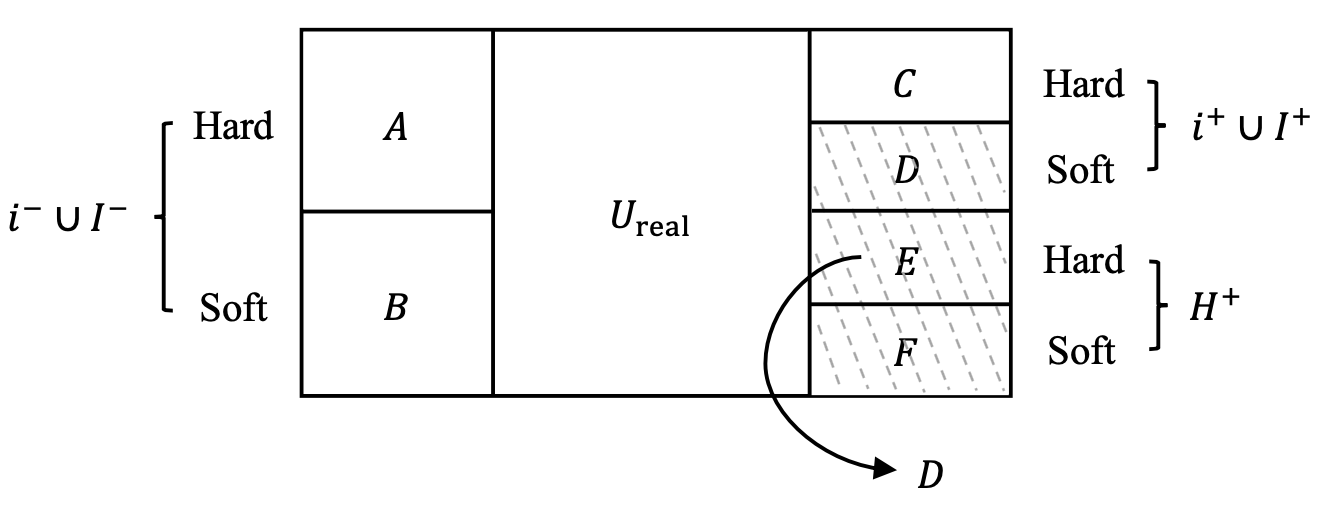}
\caption{The degrees of freedom in the shaded region, originally labeled $D$, $E$, and $F$, are inaccessible to an observer in the asymptotically flat region. To reflect this inaccessibility and to maintain consistency with the flat-spacetime case, we group them together and relabel the whole as $D$.}
\label{figure1}
\end{figure}

Second, the division into bipartite parts differs between flat and collapsing spacetimes. In flat spacetime, the division is straightforward: both the incoming and outgoing states are split into soft and hard degrees of freedom, labeled as $A,B$ and $C,D$, respectively. In collapsing spacetime, the outgoing states include not only particles that reach future infinity but also those that fall into the black hole. Each of these contains both soft and hard degrees of freedom, leading to four types of outgoing degrees of freedom, denoted by $C,D$ for escaping particles and $E,F$ for infalling ones. For an observer located in the asymptotically flat region after black hole formation, only the degrees of freedom labeled as $C$ are accessible, while $D$, $E$ and $F$ are effectively unobservable. 
It is thus natural to group $D$, $E$, and $F$ into a single effective part, which we relabel as $D$.
This grouping allows us to analyze the information scrambling in collapsing spacetime in a way similar to that in flat spacetime by evaluating the tripartite mutual information associated with a bipartite unitary operator $(U_{\rm real})_{AB \to CD}$. The subscript “real” in $U_{\rm real}$ indicates that this unitary operator includes the Bogoliubov transformation, in contrast to unitary operator $U$, which is associated solely with the QED evolution operator $S$. A schematic illustration of the division of degrees of freedom in collapsing spacetime is shown in Fig.~\ref{figure1}.

In the following, moving from our previous work in flat spacetime \cite{Su2024Scrambling}, and our understanding of the differences in studying information scrambling between flat and collapsing spacetimes, we extend the analysis from flat to collapsing spacetime.

\subsection{The structure of unitary operator}
Take the space spanned by the complete basis $\{\ket{\{n_{Q^{*}}^{*}\}}_{\rm in}\}$ as the input space, and the space spanned by the complete basis $\{\ket{\{\bar{\tilde{n}}_{\bar{\tilde{Q}}^{*}}^{*}\}}_{\rm out}\}$ as the output space. Due to the properties of complete orthonormal bases \cite{Sakurai2020Modern}, there always exists a unitary operator $U_{\rm real}$ that transforms one basis to the other. Denoting the states in the input space as $\ket{\Psi(\Sigma_{\rm in})}$, while the states in the output space as $\ket{\bar{\Psi}(\Sigma_{\rm out})}$, we find the relation
\begin{equation}
\label{incoming-outgoing-Fock}
\left\{
\begin{aligned}
    &\ket{\Psi(\Sigma_{\rm in})} = \ket{\{n_{Q^{*}}^{*}\}}_{\rm in},\\
    &\ket{\bar{\Psi}(\Sigma_{\rm out})} = U_{\rm real} \ket{\{n_{Q^{*}}^{*}\}}_{\rm in} \ \leftrightarrow \ \ket{\{\bar{\tilde{n}}_{\bar{\tilde{Q}}^{*}}^{*}\}}_{\rm out}.
\end{aligned}
\right.
\end{equation}
The bidirectional arrow ``$\leftrightarrow$'' indicates that the output state $U_{\rm real} \ket{\{n_{Q^{*}}^{*}\}}_{\rm in}$, obtained by applying the unitary operator $U_{\rm real}$ to the input state $\ket{\{n_{Q^{*}}^{*}\}}_{\rm in}$, can be expanded in terms of the complete basis associated with $\ket{\{\bar{\tilde{n}}_{\bar{\tilde{Q}}^{*}}^{*}\}}_{\rm out}$. If one compares the input and output states with the $S$-matrix elements in the Heisenberg picture, namely ${}_{\rm out}\braket{\{\bar{\tilde{n}}_{\bar{\tilde{Q}}^{*}}^{*}\}|\{n_{Q^{*}}^{*}\}}_{\rm in}$ --- see Eq. \eqref{amplitude-Fock} --- then one would find that the choice of input and output states can be understood by treating the state kets in the Heisenberg picture as input states and the basis kets as output states. With the definitions of the input and output states in place, the matrix representation of the unitary operator $U_{\rm real}$ can be recast as
\begin{align}
U_{\rm real} = \sum_{\{n_{Q^{*}}^{*}\}} U_{\rm real} \ket{\{n_{Q^{*}}^{*}\}}_{\rm in \ in}\bra{\{n_{Q^{*}}^{*}\}}.
\end{align}
By expanding $U_{\rm real}$ in the basis of the output states, one obtains $(U_{\rm real})_{\rm out}$
\begin{align}
\label{U-real-out-Fock}
\nn
(U_{\rm real})_{\rm out} =& \!\sum_{\{n_{Q^{*}}^{*}\}} \! \sum_{\{\bar{\tilde{n}}_{\bar{\tilde{Q}}^{*}_1}^{*}\}} \! \sum_{\{\bar{\tilde{n}}_{\bar{\tilde{Q}}^{*}_2}^{*}\}} {}_{\rm out} \bra{\{\bar{\tilde{n}}_{\bar{\tilde{Q}}^{*}_2}^{*}\}} U_{\rm real} \ket{\{n_{Q^{*}}^{*}\}}_{\rm in}\\
&{}_{\rm in} \braket{\{n_{Q^{*}}^{*}\}|\{\bar{\tilde{n}}_{\bar{\tilde{Q}}^{*}_1}^{*}\}}_{\rm out} \ \ket{\{\bar{\tilde{n}}_{\bar{\tilde{Q}}^{*}_2}^{*}\}}_{\rm out \ out} \bra{\{\bar{\tilde{n}}_{\bar{\tilde{Q}}^{*}_1}^{*}\}}.
\end{align}
The summation over $\{n_{Q^{*}}^{*}\}$ runs over all quantum numbers $Q^{*}$ and, for each $Q^{*}$, over all possible occupation numbers $n_{Q^{*}}^{*}$. The same applies to the summations over $\{\bar{n}_{\bar{Q}^{*}}^{*}\}$ and $\{\tilde{n}_{\tilde{Q}^{*}}^{*}\}$. The factor ${}_{\rm out} \bra{\{\bar{\tilde{n}}_{\bar{\tilde{Q}}^{*}}^{*}\}} U_{\rm real} \ket{\{n_{Q^{*}}^{*}\}}_{\rm in}$ represents the expansion coefficient of the outgoing state $U_{\rm real} \ket{\{n_{Q^{*}}^{*}\}}_{\rm in}$ in the complete basis $\ket{\{\bar{\tilde{n}}_{\bar{\tilde{Q}}^{*}}^{*}\}}_{\rm out}$. The factor ${}_{\rm in} \braket{\{n_{Q^{*}}^{*}\}|\{\bar{\tilde{n}}_{\bar{\tilde{Q}}^{*}}^{*}\}}_{\rm out}$ is the complex conjugate of the $S$-matrix element that appears in Eq.~\eqref{amplitude-Fock}. 

The $S$-matrix element ${}_{\rm out}\braket{\{\bar{\tilde{n}}_{\bar{\tilde{Q}}^{*}}^{*}\}|\{n_{Q^{*}}^{*}\}}_{\rm in}$ formally encodes all possible scattering processes. However, evaluating all such processes is impractical. In the weak-coupling regime, where perturbation theory applies, it suffices to consider only the leading-order contributions. In this case, the evolution operator $S$, appearing in the 
$S$-matrix element --- see Eq.~\eqref{amplitude-Fock} and Eq.~\eqref{total-S}) --- admits an expansion of the form
\begin{equation}
\label{leading-order-S}
S = \mathbb{I} + P \Big[ -i \int_{\Sigma_{\rm in}}^{\Sigma_{\rm out}} \sqrt{-g} \ e \bar{\psi}(x)\gamma^{\mu}A_{\mu}(x)\psi(x) d^4x \Big].
\end{equation}
The first term in the expansion is the identity operator, corresponding to the zeroth-order contribution. While the second term, involving only a single hard vertex, represents the first-order contribution. Here, only vertices associated with hard particles are counted, while those associated with soft particles are neglected, since soft degrees of freedom reflects asymptotic symmetries \cite{Bondi1962Gravitational,Sachs1962Gravitational,Weinberg1965Infrared}, which could be regarded as an intrinsic property. Therefore, the second term corresponds to the process that a single electron or positron, accompanied by a set of soft photons, scatters off an external electromagnetic field.

Since it is not feasible to account for all possible scattering processes, to provide a practical estimation of the tripartite mutual information, it is sufficient to focus on the leading-order contributions of the unitary operator. Therefore, in what follows, all discussions will be restricted to the leading order. Nonetheless, to preserve the unitarity of $U_{\rm real}$, maintain formal simplicity, and avoid ambiguity in the computations, we still use the expression introduced in Eq.~\eqref{U-real-out-Fock}, but keep in mind that it formally retains the full expression of $U_{\rm real}$ while representing only its leading-order contribution.

Before proceeding, it is necessary to consider the infrared (IR) divergence problem associated with the $S$-matrix elements. 
As mentioned in our previous work \cite{Su2024Scrambling}, without taking into account the existence of soft photons, the integration over the momenta of virtual photons in the $S$-matrix elements leads to divergences, known as IR divergences. Once the soft photons are taken into account, these IR divergences are canceled. 

Two commonly used approaches to deal with such divergences are the inclusive probability formalism \cite{Mandl1993QFT,Bloch1937Note,Yennie1961infrared,Weinberg1965Infrared} and the dressed state formalism \cite{Chung1965Infrared,Kibble1968Coherent,Kulish1970Asymptotic,Ware2013Construction,Furugori2021Dressed,Choi2018Soft}.
In the inclusive probability formalism, soft photon states are represented by orthogonal Fock states, the $S$-matrix elements involving soft photon Fock states are not IR finite, but the IR divergences cancels when consider the transition rates.
In the dressed state formalism, soft photon states are represented by dressed states, which can be considered as some special non-orthogonal coherent states, the $S$-matrix elements involving soft photon coherent states are IR finite.

For the study of information scrambling, the orthogonality of soft photon Fock states is essential for constructing Choi state. On the other hand, the IR-finite $S$-matrix elements constructed from soft photon coherent states greatly simplify the computation, as they are already IR finite by construction. In contrast, $S$-matrix elements built from soft photon Fock states always contain IR divergences that must be carefully canceled at a later stage. Although such cancellations are guaranteed in principle, they tend to make the calculation more cumbersome. 
Since the soft photon states in the two approaches differ only by a phase factor which vanishes at leading order \cite{Chung1965Infrared,Kulish1970Asymptotic}, and there is a correspondence between Fock states (particle number states) and coherent states \cite{Su2024Scrambling}, we are allowed, at leading order, to employ both approaches simultaneously.

Consequently, we could switch $U_{\rm real}$ from one formalism to the other, so that $(U_{\rm real})_{\rm out}$ in Eq.~\eqref{U-real-out-Fock}, the photon part of which is represented in Fock states, can equivalently be rewritten in terms of coherent states
\begin{align}
\label{U-real-out-coherent}
\nn
(U_{\rm real})_{\rm out} =& \sum_{\{n_{Q^{\pm}}^{e^{\pm}}\}} \! \sum_{\{\beta_{Q}\}} \! \sum_{\{\bar{\tilde{n}}_{\bar{\tilde{Q}}^{\pm}_1}^{\pm}\}} \! \sum_{\{\bar{\tilde{\beta}}_{\bar{\tilde{Q}}_1}\}} \! \sum_{\{\bar{\tilde{n}}_{\bar{\tilde{Q}}^{\pm}_2}^{\pm}\}} \! \sum_{\{\bar{\tilde{\beta}}_{\bar{\tilde{Q}}_2}\}}\\
\nn
&{}_{\rm out}\bra{\{\bar{\tilde{n}}_{\bar{\tilde{Q}}^{\pm}_2}^{e^{\pm}}\},\!\{\bar{\tilde{\beta}}_{\bar{\tilde{Q}}_2}\}} U_{\rm real} \ket{\{n_{Q^{\pm}}^{e^{\pm}}\},\!\{\beta_{Q}\}}_{\rm in}\\
\nn
&{}_{\rm in} \braket{\{n_{Q^{\pm}}^{e^{\pm}}\},\!\{\beta_{Q}\}|\{\bar{\tilde{n}}_{\bar{\tilde{Q}}^{\pm}_1}^{e^{\pm}}\},\!\{\bar{\tilde{\beta}}_{\bar{\tilde{Q}}_1}\}}_{\rm out}\\
&\ket{\{\bar{\tilde{n}}_{\bar{\tilde{Q}}^{\pm}_2}^{e^{\pm}}\},\!\{\bar{\tilde{\beta}}_{\bar{\tilde{Q}}_2}\}}_{\rm out \ out} \bra{\{\bar{\tilde{n}}_{\bar{\tilde{Q}}^{\pm}_1}^{e^{\pm}}\},\!\{\bar{\tilde{\beta}}_{\bar{\tilde{Q}}_1}\}}.
\end{align}
Here, shorthand notation is employed, where $\ket{\{n_{Q^{\pm}}^{e^{\pm}}\}}_{\rm in}\\\equiv\ket{\{n_{Q^{-}}^{e^{-}}\},\!\{n_{Q^{+}}^{e^{+}}\}}_{\rm in}$, $\ket{\{\bar{\tilde{n}}_{\bar{\tilde{Q}}^{\pm}}^{e^{\pm}}\}}_{\rm in}\equiv\ket{\{\bar{n}_{\bar{Q}^{\pm}}^{e^{\pm}}\},\{\tilde{n}_{\tilde{Q}^{\pm}}^{e^{\pm}}\}}_{\rm in}\equiv\ket{\{\bar{n}_{\bar{Q}^{-}}^{e^{-}}\},\!\{\bar{n}_{\bar{Q}^{+}}^{e^{+}}\},\{\tilde{n}_{\tilde{Q}^{-}}^{e^{-}}\},\!\{\tilde{n}_{\tilde{Q}^{+}}^{e^{+}}\}}_{\rm in}$. The notation $\ket{\{\beta_{Q}\}}_{\rm in}$ denotes the photon coherent state in the distant past, the explicit expression of which is given by
\begin{align}
\nn
&\ket{\{\beta_{Q}\}}=\prod_{Q}\ket{\beta_{Q}}=\exp \Big[\sum_{Q} \beta_{Q} (d_{Q}^{1{\dagger}})_{\rm in} - \beta_{Q}^{*} (d_{Q}^{1})_{\rm in}\Big] \ket{0}_{\rm in}\\
&=\exp \Big[\sum_{l,a} \beta_{a}^l (d_{l,a}^{1{\dagger}})_{\rm in} - \beta_{a}^{l*} (d_{l,a}^{1})_{\rm in}\Big] \ket{0}_{\rm in},
\end{align}
where, $\ket{0}_{\rm in}$ is the vacuum state in the distant past --- see Eq.~\eqref{in-vacuum} --- while $(d_{Q}^{1{\dagger}})_{\rm in}\equiv(d_{l,a}^{1{\dagger}})_{\rm in} = \int d^3k f_a(k) (c^{1{\dagger}}_{l,k})_{\rm in}$ is the creation operator for a photon cloud (coherent state) with momentum distribution $f_{a}(k)$ and polarization $l$, and $(c^{1{\dagger}}_{Q})_{\rm in}\equiv(c^{1{\dagger}}_{l,k})_{\rm in}$ is the creation operator for a single photon (Fock state) with momentum $k$ and polarization $l$ --- see Eq.\eqref{in-vacuum}. 
Similarly, the shorthand notation $\ket{\{\bar{\tilde{\beta}}_{\bar{\tilde{Q}}}\}}_{\rm out}\equiv\ket{\{\bar{\beta}_{\bar{Q}}\},\{\tilde{\beta}_{\tilde{Q}}\}}_{\rm out}$ is employed. The notations $\ket{\{\bar{\beta}_{\bar{Q}}\}}_{\rm out}$ and $\ket{\{\tilde{\beta}_{\tilde{Q}}\}}_{\rm out}$ denote the photon coherent states in the distant future, with the bar and tilde indicating states on the asymptotic region and the black hole horizon, respectively. The corresponding vacuum state and the associated creation operators acting on it are $\ket{\bar{0}}_{\rm out}$, $(d_{Q}^{2{\dagger}})_{\rm in}\equiv(d_{l,a}^{2{\dagger}})_{\rm in} = \int d^3k f_a(k) (c^{2{\dagger}}_{l,k})_{\rm in}$ and $(d_{Q}^{3{\dagger}})_{\rm in}\equiv(d_{l,a}^{3{\dagger}})_{\rm in} = \int d^3k f_a(k) (c^{3{\dagger}}_{l,k})_{\rm in}$, respectively. Moreover, the summation over $\{\beta_{Q}\}$ in Eq.~\eqref{U-real-out-coherent} runs over all quantum numbers $Q$ and, for each $Q$, over all possible coefficients $\beta$. The same applies to the summations over $\{\bar{\tilde{\beta}}_{\bar{\tilde{Q}}_1}\}$ and $\{\bar{\tilde{\beta}}_{\bar{\tilde{Q}}_2}\}$.

In the present context, the states in the input and output spaces, originally given in Eq.~\eqref{incoming-outgoing-Fock}, can be rewritten as
\begin{equation}
\label{incoming-outgoing-coherent}
\left\{\!
\begin{aligned}
    &\ket{\Psi(\Sigma_{\rm in})} \!=\! \ket{\{n_{Q^{\pm}}^{e^{\pm}}\},\!\{\beta_{Q}\}}_{\rm in},\\
    &\ket{\bar{\Psi}(\Sigma_{\rm out})} \!=\! U_{\rm real} \ket{\{n_{Q^{\pm}}^{e^{\pm}}\},\!\{\beta_{Q}\}}_{\rm in} \ \!\leftrightarrow\! \ \ket{\{\bar{\tilde{n}}_{\bar{\tilde{Q}}^{\pm}}^{e^{\pm}}\},\!\{\bar{\tilde{\beta}}_{\bar{\tilde{Q}}}\}}_{\rm out}.
\end{aligned}
\right.
\end{equation}
Moreover, ${}_{\rm out} \braket{\{\bar{\tilde{n}}_{\bar{\tilde{Q}}^{\pm}}^{e^{\pm}}\},\!\{\bar{\tilde{\beta}}_{\bar{\tilde{Q}}}\}|\{n_{Q^{\pm}}^{e^{\pm}}\},\!\{\beta_{Q}\}}_{\rm in}$, the corresponding $S$-matrix element, takes a form analogous to that of Eq.\eqref{amplitude-Fock}:
\begin{align}
\label{amplitude-coherent}
\nn
&{}_{\rm out}\braket{\{\bar{\tilde{n}}_{\bar{\tilde{Q}}^{\pm}}^{e^{\pm}}\},\!\{\bar{\tilde{\beta}}_{\bar{\tilde{Q}}}\}|\{n_{Q^{\pm}}^{e^{\pm}}\},\!\{\beta_{Q}\}}_{\rm in}\\[2mm]
\nn
&= \sum_{\{m_{q^{\pm}}^{e^{\pm}}\}} \sum_{\{\alpha_{q}\}} {}_{\rm out}\braket{\{\bar{\tilde{n}}_{\bar{\tilde{Q}}^{\pm}}^{e^{\pm}}\},\!\{\bar{\tilde{\beta}}_{\bar{\tilde{Q}}}\}|\{m_{q^{\pm}}^{e^{\pm}}\},\!\{\alpha_{q}\}}_{\rm out}\\[-2mm]
\nn
&{}_{\rm out}\braket{\{m_{q^{\pm}}^{e^{\pm}}\},\!\{\alpha_{q}\}|\{n_{Q^{\pm}}^{e^{\pm}}\},\!\{\beta_{Q}\}}_{\rm in}\\
\nn
&= \sum_{\{m_{q^{\pm}}^{e^{\pm}}\}} \sum_{\{\alpha_{q}\}} {}_{\rm out}\braket{\{\bar{\tilde{n}}_{\bar{\tilde{Q}}^{\pm}}^{e^{\pm}}\},\!\{\bar{\tilde{\beta}}_{\bar{\tilde{Q}}}\}|\{m_{q^{\pm}}^{e^{\pm}}\},\!\{\alpha_{q}\}}_{\rm out}\\[-2mm]
&{}_{\rm in}\braket{\{m_{q^{\pm}}^{e^{\pm}}\},\!\{\alpha_{q}\}|S|\{n_{Q^{\pm}}^{e^{\pm}}\},\!\{\beta_{Q}\}}_{\rm in}.
\end{align}

Comparing Eq.~\eqref{amplitude-Fock} and Eq.~\eqref{incoming-outgoing-Fock}, as well as Eq.~\eqref{amplitude-coherent} and Eq.~\eqref{incoming-outgoing-coherent}, one is naturally led to view the $S$ operator as connecting the input states to the intermediate ``out'' states --- those obtained by acting the creation operators $\{(a_{Q}^{1{\dagger}})_{\rm out}, (b_{Q}^{1{\dagger}})_{\rm out},(c_{Q}^{1{\dagger}})_{\rm out} \ {\rm or} \ (d_{Q}^{1{\dagger}})_{\rm out}\}$ on vacuum state $\ket{0}_{\rm out}$ --- namely, connecting $\ket{\{n_{Q^{*}}^{*}\}}_{\rm in}$ to $\ket{\{n_{Q^{*}}^{*}\}}_{\rm out}$, and likewise connecting $\ket{\{n_{Q^{\pm}}^{e^{\pm}}\},\!\{\beta_{Q}\}}_{\rm in}$ to $\ket{\{n_{Q^{\pm}}^{e^{\pm}}\},\!\{\beta_{Q}\}}_{\rm out}$.
To maintain consistency with the form of Eq.~\eqref{incoming-outgoing-Fock} and Eq.~\eqref{incoming-outgoing-coherent}, we define $U \equiv S^{\dagger}$ and denote $\ket{\Psi(\Sigma_{\rm out})}$ as the intermediate ``out'' states, so that the relation between the input states and the intermediate ``out'' states can be expressed as
\begin{equation}
\label{incoming-outgoing-Fock-U}
\left\{
\begin{aligned}
    &\ket{\Psi(\Sigma_{\rm in})} = \ket{\{n_{Q^{*}}^{*}\}}_{\rm in},\\
    &\ket{{\Psi}(\Sigma_{\rm out})} = U\ket{\{n_{Q^{*}}^{*}\}}_{\rm in}\equiv\ket{\{n_{Q^{*}}^{*}\}}_{\rm out},
\end{aligned}
\right.
\end{equation}
and
\begin{equation}
\label{incoming-outgoing-coherent-U}
\left\{\!
\begin{aligned}
    &\ket{\Psi(\Sigma_{\rm in})} \!=\! \ket{\{n_{Q^{\pm}}^{e^{\pm}}\},\!\{\beta_{Q}\}}_{\rm in},\\
    &\ket{{\Psi}(\Sigma_{\rm out})} \!=\! U \ket{\{n_{Q^{\pm}}^{e^{\pm}}\},\!\{\beta_{Q}\}}_{\rm in}\equiv\ket{\{n_{Q^{\pm}}^{e^{\pm}}\},\!\{\beta_{Q}\}}_{\rm out}.
\end{aligned}
\right.
\end{equation}
The intermediate ``out'' states $\ket{\Psi(\Sigma_{\rm out})}$ should not be confused with the output states $\ket{\bar{\Psi}(\Sigma_{\rm out})}$ --- the latter ones being obtained by the action of $\{(a_{Q}^{2{\dagger}})_{\rm out}, (b_{Q}^{2{\dagger}})_{\rm out}, \\(c_{Q}^{2{\dagger}})_{\rm out} \ {\rm or} \ (d_{Q}^{2{\dagger}})_{\rm out}\}$ and $\{(a_{Q}^{3{\dagger}})_{\rm out}, (b_{Q}^{3{\dagger}})_{\rm out},(c_{Q}^{3{\dagger}})_{\rm out} \ {\rm or} \\ (d_{Q}^{3{\dagger}})_{\rm out}\}$ on the vacuum state $\ket{\bar{0}}_{\rm out}$ --- appearing in Eq.~\eqref{incoming-outgoing-Fock} and Eq.~\eqref{incoming-outgoing-coherent}.
Clearly, the former are associated with the mode expansion of the free outgoing field in terms of the mode set defined on $I^- \cup i^-$, whereas the latter are associated with the mode expansion of the free outgoing field in terms of the mode set defined on $I^+ \cup i^+ \cup H^+$. 

Moreover, the intermediate ``in'' states $\ket{\bar{\Psi}(\Sigma_{\rm in})}$ also exist, and the relationship between these states and the input states $\ket{\Psi(\Sigma_{\rm in})}$ is analogous to that between the intermediate ``out'' states $\ket{\Psi(\Sigma_{\rm out})}$ and the output states $\ket{\bar{\Psi}(\Sigma_{\rm out})}$. Specifically, the states $\ket{\bar{\Psi}(\Sigma_{\rm in})}$ are obtained by action of the creation operators $\{(a_{Q}^{2{\dagger}})_{\rm in}, (b_{Q}^{2{\dagger}})_{\rm in}, \\ (c_{Q}^{2{\dagger}})_{\rm in} \ {\rm or} \ (d_{Q}^{2{\dagger}})_{\rm in}\}$ and $\{(a_{Q}^{3{\dagger}})_{\rm in}, (b_{Q}^{3{\dagger}})_{\rm in},(c_{Q}^{3{\dagger}})_{\rm in} \ {\rm or} \ (d_{Q}^{3{\dagger}})_{\rm in}\}$ on the vacuum state $\ket{\bar{0}}_{\rm in}$, and are associated with the mode expansion of the free incoming field in terms of the mode set defined on $I^+ \cup i^+ \cup H^+$. By contrast, the states $\ket{\Psi(\Sigma_{\rm in})}$ are obtained by acting the creation operators $\{(a_{Q}^{1{\dagger}})_{\rm in}, (b_{Q}^{1{\dagger}})_{\rm in}, (c_{Q}^{1{\dagger}})_{\rm in} \ {\rm or} \ (d_{Q}^{1{\dagger}})_{\rm in}\}$ on the vacuum state $\ket{0}_{\rm in}$, and are associated with the mode expansion of the free incoming field in terms of the mode set defined on $I^- \cup i^-$. 

Therefore, the $S$-matrix elements in Eq.~\eqref{amplitude-Fock} and Eq.~\eqref{amplitude-coherent} admit an alternative expansion, obtained by inserting intermediate ``in'' states instead of intermediate ``out'' states, which leads to
\begin{align}
\label{amplitude-Fock-alternative}
\nn
&{}_{\rm out}\braket{\{\bar{\tilde{n}}_{\bar{\tilde{Q}}^{*}}^{*}\}|\{n_{Q^{*}}^{*}\}}_{\rm in}\\[2mm]
\nn
&= \sum_{\{\bar{\tilde{m}}_{\bar{\tilde{q}}^{*}}^{*}\}} {}_{\rm out}\braket{\{\bar{\tilde{n}}_{\bar{\tilde{Q}}^{*}}^{*}\}|\{\bar{\tilde{m}}_{\bar{\tilde{q}}^{*}}^{*}\}}_{\rm in \ in}\braket{\{\bar{\tilde{m}}_{\bar{\tilde{q}}^{*}}^{*}\}|\{n_{Q^{*}}^{*}\}}_{\rm in}\\[-1mm]
&= \sum_{\{\bar{\tilde{m}}_{\bar{\tilde{q}}^{*}}^{*}\}} {}_{\rm in}\braket{\{\bar{\tilde{n}}_{\bar{\tilde{Q}}^{*}}^{*}\}|S|\{\bar{\tilde{m}}_{\bar{\tilde{q}}^{*}}^{*}\}}_{\rm in \ in}\braket{\{\bar{\tilde{m}}_{\bar{\tilde{q}}^{*}}^{*}\}|\{n_{Q^{*}}^{*}\}}_{\rm in},
\end{align}
and
\begin{align}
\label{amplitude-coherent-alternative}
\nn
&{}_{\rm out}\braket{\{\bar{\tilde{n}}_{\bar{\tilde{Q}}^{\pm}}^{e^{\pm}}\},\!\{\bar{\tilde{\beta}}_{\bar{\tilde{Q}}}\}|\{n_{Q^{\pm}}^{e^{\pm}}\},\!\{\beta_{Q}\}}_{\rm in}\\[2mm]
\nn
&= \sum_{\{\bar{\tilde{m}}_{\bar{\tilde{q}}^{\pm}}^{e^{\pm}}\}} \sum_{\{\bar{\tilde{\alpha}}_{\bar{\tilde{q}}}\}} {}_{\rm out}\braket{\{\bar{\tilde{n}}_{\bar{\tilde{Q}}^{\pm}}^{e^{\pm}}\},\!\{\bar{\tilde{\beta}}_{\bar{\tilde{Q}}}\}|\{\bar{\tilde{m}}_{\bar{\tilde{q}}^{\pm}}^{e^{\pm}}\},\!\{\bar{\tilde{\alpha}}_{\bar{\tilde{q}}}\}}_{\rm in}\\[-2mm]
\nn
&{}_{\rm in}\braket{\{\bar{\tilde{m}}_{\bar{\tilde{q}}^{\pm}}^{e^{\pm}}\},\!\{\bar{\tilde{\alpha}}_{\bar{\tilde{q}}}\}|\{n_{Q^{\pm}}^{e^{\pm}}\},\!\{\beta_{Q}\}}_{\rm in}\\
\nn
&= \sum_{\{\bar{\tilde{m}}_{\bar{\tilde{q}}^{\pm}}^{e^{\pm}}\}} \sum_{\{\bar{\tilde{\alpha}}_{\bar{\tilde{q}}}\}} {}_{\rm in}\braket{\{\bar{\tilde{n}}_{\bar{\tilde{Q}}^{\pm}}^{e^{\pm}}\},\!\{\bar{\tilde{\beta}}_{\bar{\tilde{Q}}}\}|S|\{\bar{\tilde{m}}_{\bar{\tilde{q}}^{\pm}}^{e^{\pm}}\},\!\{\bar{\tilde{\alpha}}_{\bar{\tilde{q}}}\}}_{\rm in}\\[-2mm]
&{}_{\rm in}\braket{\{\bar{\tilde{m}}_{\bar{\tilde{q}}^{\pm}}^{e^{\pm}}\},\!\{\bar{\tilde{\alpha}}_{\bar{\tilde{q}}}\}|\{n_{Q^{\pm}}^{e^{\pm}}\},\!\{\beta_{Q}\}}_{\rm in}.
\end{align}
The second equalities in Eq.~\eqref{amplitude-Fock-alternative} and Eq.~\eqref{amplitude-coherent-alternative} follow from the observation of the identities\footnote{Leveraging these identities and comparing Eq.~\eqref{amplitude-Fock-alternative} with Eq.~\eqref{amplitude-Fock}, and Eq.~\eqref{amplitude-coherent-alternative} with Eq.~\eqref{amplitude-coherent}, it becomes evident that the second equalities in Eq.~\eqref{amplitude-Fock-alternative} and Eq.~\eqref{amplitude-coherent-alternative} necessarily hold.}
\begin{align}
\label{identity-1}
&{}_{\rm in}\braket{\{\bar{\tilde{m}}_{\bar{\tilde{q}}^{*}}^{*}\}|\{n_{Q^{*}}^{*}\}}_{\rm in}={}_{\rm out}\braket{\{\bar{\tilde{m}}_{\bar{\tilde{q}}^{*}}^{*}\}|\{n_{Q^{*}}^{*}\}}_{\rm out},
\end{align}
and
\begin{align}
\label{identity-2}
\nn
&{}_{\rm in}\braket{\{\bar{\tilde{m}}_{\bar{\tilde{q}}^{\pm}}^{e^{\pm}}\},\!\{\bar{\tilde{\alpha}}_{\bar{\tilde{q}}}\}|\{n_{Q^{\pm}}^{e^{\pm}}\},\!\{\beta_{Q}\}}_{\rm in}\\
&={}_{\rm out}\braket{\{\bar{\tilde{m}}_{\bar{\tilde{q}}^{\pm}}^{e^{\pm}}\},\!\{\bar{\tilde{\alpha}}_{\bar{\tilde{q}}}\}|\{n_{Q^{\pm}}^{e^{\pm}}\},\!\{\beta_{Q}\}}_{\rm out}.
\end{align}
At the perturbative level, with $U=\mathbb{I} - i\mathcal{T}$, the zeroth-order terms of Eq~.\eqref{amplitude-Fock} and Eq.~\eqref{amplitude-coherent} yield precisely these identities.

Therefore, the operator $U=S^{\dagger}$ also connects the intermediate ``in'' states to the output states, namely,
\begin{equation}
\label{incoming-outgoing-Fock-U-bar}
\left\{
\begin{aligned}
    &\ket{\bar{\Psi}(\Sigma_{\rm in})} = \ket{\{\bar{\tilde{n}}_{\bar{\tilde{Q}}^{*}}^{*}\}}_{\rm in},\\
    &\ket{\bar{\Psi}(\Sigma_{\rm out})} = U\ket{\{\bar{\tilde{n}}_{\bar{\tilde{Q}}^{*}}^{*}\}}_{\rm in}\equiv\ket{\{\bar{\tilde{n}}_{\bar{\tilde{Q}}^{*}}^{*}\}}_{\rm out},
\end{aligned}
\right.
\end{equation}
and
\begin{equation}
\label{incoming-outgoing-coherent-U-bar}
\left\{\!
\begin{aligned}
    &\ket{\bar{\Psi}(\Sigma_{\rm in})} \!=\! \ket{\{\bar{\tilde{n}}_{\bar{\tilde{Q}}^{\pm}}^{e^{\pm}}\},\!\{\bar{\tilde{\beta}}_{\bar{\tilde{Q}}}\}}_{\rm in},\\
    &\ket{\bar{\Psi}(\Sigma_{\rm out})} \!=\! U \ket{\{\bar{\tilde{n}}_{\bar{\tilde{Q}}^{\pm}}^{e^{\pm}}\},\!\{\bar{\tilde{\beta}}_{\bar{\tilde{Q}}}\}}_{\rm in}\equiv\ket{\{\bar{\tilde{n}}_{\bar{\tilde{Q}}^{\pm}}^{e^{\pm}}\},\!\{\bar{\tilde{\beta}}_{\bar{\tilde{Q}}}\}}_{\rm out}.
\end{aligned}
\right.
\end{equation}
Moreover, the identities in Eq.~\eqref{identity-1} and Eq.~\eqref{identity-2} suggest that one could introduce an operator $U_0$ that connects input states to the intermediate ``in'' states, and the intermediate ``out'' states to output states, namely,
\begin{equation}
\label{incoming-Fock-U0}
\left\{
\begin{aligned}
    &\ket{\Psi(\Sigma_{\rm in})} = \ket{\{n_{Q^{*}}^{*}\}}_{\rm in},\\
    &\ket{\bar{\Psi}(\Sigma_{\rm in})} = U_{0} \ket{\{n_{Q^{*}}^{*}\}}_{\rm in} \ \leftrightarrow \ \ket{\{\bar{\tilde{n}}_{\bar{\tilde{Q}}^{*}}^{*}\}}_{\rm in},
\end{aligned}
\right.
\end{equation}
and
\begin{equation}
\label{outgoing-Fock-U0}
\left\{
\begin{aligned}
    &\ket{\Psi(\Sigma_{\rm out})} = \ket{\{n_{Q^{*}}^{*}\}}_{\rm out},\\
    &\ket{\bar{\Psi}(\Sigma_{\rm out})} = U_{0} \ket{\{n_{Q^{*}}^{*}\}}_{\rm out} \ \leftrightarrow \ \ket{\{\bar{\tilde{n}}_{\bar{\tilde{Q}}^{*}}^{*}\}}_{\rm out},
\end{aligned}
\right.
\end{equation}
and
\begin{equation}
\label{incoming-coherent-U0}
\left\{\!
\begin{aligned}
    &\ket{\Psi(\Sigma_{\rm in})} \!=\! \ket{\{n_{Q^{\pm}}^{e^{\pm}}\},\!\{\beta_{Q}\}}_{\rm in},\\
    &\ket{\bar{\Psi}(\Sigma_{\rm in})} \!=\! U_{0} \ket{\{n_{Q^{\pm}}^{e^{\pm}}\},\!\{\beta_{Q}\}}_{\rm in} \ \!\leftrightarrow\! \ \ket{\{\bar{\tilde{n}}_{\bar{\tilde{Q}}^{\pm}}^{e^{\pm}}\},\!\{\bar{\tilde{\beta}}_{\bar{\tilde{Q}}}\}}_{\rm in},
\end{aligned}
\right.
\end{equation}
and
\begin{equation}
\label{outgoing-coherent-U0}
\left\{\!
\begin{aligned}
    &\ket{\Psi(\Sigma_{\rm out})} \!=\! \ket{\{n_{Q^{\pm}}^{e^{\pm}}\},\!\{\beta_{Q}\}}_{\rm out},\\
    &\ket{\bar{\Psi}(\Sigma_{\rm out})} \!=\! U_{0} \ket{\{n_{Q^{\pm}}^{e^{\pm}}\},\!\{\beta_{Q}\}}_{\rm out} \ \!\leftrightarrow\! \ \ket{\{\bar{\tilde{n}}_{\bar{\tilde{Q}}^{\pm}}^{e^{\pm}}\},\!\{\bar{\tilde{\beta}}_{\bar{\tilde{Q}}}\}}_{\rm out}.
\end{aligned}
\right.
\end{equation}
These analyses further imply that $U_{\rm real}$, $U$ (encoding the QED interaction), and $U_0$ (encoding the Bogoliubov transformation) satisfy the following relation
\begin{equation}
\label{factor}
U_{\rm real} = U_0 U = U U_0.
\end{equation}
Here we assume both $U$ and $U_0$ to be unitary.

Furthermore, the Bogolubov coefficients in the Friedmann–Lemaitre–Robertson–Walker spacetime can be effectively obtained by treating the interaction between the inflaton --- whose dynamics drives cosmic inflation --- and the Standard Model particles as an effective potential experienced by the latter \cite{Kofman1997Towards,Birrell1982Quantum}. It is then natural to expect that the Bogolubov coefficients in the collapsing spacetime can be obtained in a similar manner, by modeling the interaction between certain degrees of freedom --- whose dynamics describes the gravitational collapse --- and the Standard Model particles as an effective potential experienced by the latter ones. 
Therefore, we assume $U_0=\mathbb{I}-iG_0$, where $-iG_0$ represents the interaction between Standard Model particles and the degrees of freedom driving the gravitational collapse. 
When $-iG_0$ is treated as an effective potential experienced by the Standard Model particles, the overall effect of $\mathbb{I}-iG_0$ is expected to reproduce the Bogoliubov transformation\footnote{Although it is well known that the Bogoliubov transformation is not unitary, this does not contradict the assumption that $U_0$ is unitary, as the non-unitarity of the Bogoliubov transformation is most likely a consequence of replacing the interaction in $U_0$ with an effective potential.}.
Therefore, by combining the assumption $U_0=\mathbb{I}-iG_0$ with the perturbative expansion of $U$, namely, $U=\mathbb{I}-i\mathcal{T}$, Eq.~\eqref{factor} can be rewritten as
\begin{equation}
\label{factor-sum}
U_{\rm real} = \mathbb{I} - i\mathcal{T} - iG_0 - G_0\mathcal{T} = \mathbb{I} - iG_0 - i\mathcal{T} - \mathcal{T}G_0.
\end{equation}

\subsection{The expression of Choi state and tripartite mutual information}
Building on the analysis of the unitary operator $U_{\rm real}$ in the previous subsection, we now proceed to estimate its scrambling power through the tripartite mutual information.
Based on the definition of the Choi state in Sec.~\ref{information-scrambling} --- see Eq.~\eqref{Choi-state-definition} --- and the expressions of $(U_{\rm real})_{\rm out}$ in the inclusive probability --- see Eq.~\eqref{U-real-out-Fock} --- and dressed state --- see Eq.~\eqref{U-real-out-coherent} --- formalisms, we obtain the Choi state corresponding to the unitary operator $U_{\rm real}$, which is given by
\begin{align}
\nn
\ket{U_{\rm real}} &= (\mathbb{I}_{\rm in} \otimes (U_{\rm real})_{\rm out}) \ket{I}\\
\nn
&= \frac{1}{\sqrt{D}} \sum_{\rm q.n.} {}_{\rm in} \braket{\{n_{Q^{\pm}_4}^{e^{\pm}}\},\!\{\beta_{Q_4}\}|\{n_{Q^{*}}^{*}\}}_{\rm in}\\[-1mm]
\nn
&{}_{\rm out} \braket{\{\bar{\tilde{n}}_{\bar{\tilde{Q}}^{\pm}_1}^{e^{\pm}}\},\!\{\bar{\tilde{\beta}}_{\bar{\tilde{Q}}_1}\}|U_{\rm real}|\{n_{Q^{\pm}_3}^{e^{\pm}}\},\!\{\beta_{Q_3}\}}_{\rm in}\\
\nn
& {}_{\rm in} \braket{\{n_{Q^{\pm}_3}^{e^{\pm}}\},\!\{\beta_{Q_3}\}|\{\bar{\tilde{n}}_{\bar{\tilde{Q}}^{\pm}_2}^{e^{\pm}}\},\!\{\bar{\tilde{\beta}}_{\bar{\tilde{Q}}_2}\}}_{\rm out}\\[-1mm]
\nn
& {}_{\rm out} \braket{\{\bar{\tilde{n}}_{\bar{\tilde{Q}}^{\pm}_2}^{e^{\pm}}\},\!\{\bar{\tilde{\beta}}_{\bar{\tilde{Q}}_2}\}|U_{\rm real}|\{n_{Q^{*}}^{*}\}}_{\rm in}\\
& \ket{\{n_{Q^{\pm}_4}^{e^{\pm}}\},\!\{\beta_{Q_4}\}}_{\rm in} \otimes \ket{\{\bar{\tilde{n}}_{\bar{\tilde{Q}}^{\pm}_1}^{e^{\pm}}\},\!\{\bar{\tilde{\beta}}_{\bar{\tilde{Q}}_1}\}}_{\rm out},
\end{align}
with
\begin{equation}
\ket{I} = \sum_{\rm q.n.} \ket{\{n_{Q^{*}}^{*}\}}_{\rm in} \otimes U_{\rm real} \ket{\{n_{Q^{*}}^{*}\}}_{\rm in},
\end{equation}
the entangled state constructed from the input and output states in the inclusive probability formalism --- see Eq.~\eqref{incoming-outgoing-Fock}. 
For convenience, we use the simplified notation $\sum_{\text{q.n.}}$, which runs over all repeated sets of quantum numbers, including $\{n_{Q^{*}}^{*}\}$, $\{\bar{\tilde{n}}_{\bar{\tilde{Q}}^{\pm}_1}^{e^{\pm}}\}$, and so on. Then the corresponding density matrix $\rho$ follows immediately
\begin{align}
\nn
& \rho_{\rm real} = \ket{U_{\rm real}} \bra{U_{\rm real}}\\
\nn
& = \frac{1}{D} \sum_{\text{q.n.}} \braket{\{{Q^{\pm}_4}\},\!\{{Q_4}\}|\{{Q^{*}}\}} \braket{\{{q^{*}}\}|\{{q^{\pm}_4}\},\!\{{q_4}\}} \\[-2mm]
\nn
& \braket{{\bar{\tilde{Q}}^{\pm}_1}\},\!\{{\bar{\tilde{Q}}_1}\}|U_{\rm real}|\{{Q^{\pm}_3}\},\!\{{Q_3}\}} \braket{\{{Q^{\pm}_3}\},\!\{{Q_3}\}|\{{\bar{\tilde{Q}}^{\pm}_2}\},\!\{{\bar{\tilde{Q}}_2}\}}\\
\nn
& \braket{{\bar{\tilde{Q}}^{\pm}_2}\},\!\{{\bar{\tilde{Q}}_2}\}|U_{\rm real}|\{{Q^{*}}\}} \braket{\{{q^{*}}\}|U_{\rm real}^{\dagger}|\{{\bar{\tilde{q}}^{\pm}_2}\},\!\{{\bar{\tilde{q}}_2}\}}\\
\nn
& \braket{\{{\bar{\tilde{q}}^{\pm}_2}\},\!\{{\bar{\tilde{q}}_2}\}|\{{q^{\pm}_3}\},\!\{{q_3}\}} \braket{\{{q^{\pm}_3}\},\!\{{q_3}\}|U_{\rm real}^{\dagger}|{\bar{\tilde{q}}^{\pm}_1}\},\!\{{\bar{\tilde{q}}_1}\}}\\
& \ket{\{{Q^{\pm}_4}\},\!\{{Q_4}\}} \bra{\{{q^{\pm}_4}\},\!\{{q_4}\}} \otimes \ket{\{{\bar{\tilde{Q}}^{\pm}_1}\},\!\{{\bar{\tilde{Q}}_1}\}} \bra{\{{\bar{\tilde{q}}^{\pm}_1}\},\!\{{\bar{\tilde{q}}_1}\}}.
\end{align}
For simplicity, the labels $n$ and $\beta$ have been omitted in the formula, since it is clear from the preceding discussion that the superscripts $*$, $\pm$, and the absence of a superscript denote different types of states. In addition, the subscripts ``in'' and ``out'' have also been omitted, as all states appearing here are physical input and output states rather than intermediate ones, and the presence of bars and tildes already suffices to distinguish output states from input states. Whenever intermediate states are needed, we will indicate them explicitly to avoid any ambiguity.

According to the definition of the tripartite mutual information in Sec.~\ref{information-scrambling} and the bipartite partition illustrated in Fig.~\ref{figure1} at the beginning of Sec.~\ref{TMI-collapsing}, we trace over the corresponding parts of the density matrix $\rho_{\rm real}$ to obtain the reduced matrices $\rho_{AC}$, $\rho_{AD}$, $\rho_C$ and $\rho_D$. Since these calculations are purely formal and straightforward, we present the results directly as follows:
\begin{align}
\label{rhoAC}
\nn
& \rho_{AC} = {\rm tr}_{BD}(\rho_{\rm real})\\
\nn
& = \frac{1}{D} \sum_{\text{q.n.}} \braket{\{{Q^{\pm}_4}\},\!\{{Q_4}\}|\{{Q^{*}}\}} \braket{\{{q^{*}}\}|\{{q^{\pm}_4}\},\!\{{q_4}\}} \braket{\{{q_4}\}|\{{Q_4}\}}\\[-2mm]
\nn
& \braket{\{{\bar{\tilde{Q}}^{\pm}_1}\},\!\{{\bar{\tilde{Q}}_1}\}|U_{\rm real}|\{{Q^{\pm}_3}\},\!\{{Q_3}\}} \braket{\{{Q^{\pm}_3}\},\!\{{Q_3}\}|\{{\bar{\tilde{Q}}^{\pm}_2}\},\!\{{\bar{\tilde{Q}}_2}\}}\\
\nn
& \braket{\{{\bar{\tilde{Q}}^{\pm}_2}\},\!\{{\bar{\tilde{Q}}_2}\}|U_{\rm real}|\{{Q^{*}}\}} \braket{\{{q^{*}}\}|U_{\rm real}^{\dagger}|\{{\bar{\tilde{q}}^{\pm}_2}\},\!\{{\bar{\tilde{q}}_2}\}}\\
\nn
& \braket{\{{\bar{\tilde{q}}^{\pm}_2}\},\!\{{\bar{\tilde{q}}_2}\}|\{{q^{\pm}_3}\},\!\{{q_3}\}} \braket{\{{q^{\pm}_3}\},\!\{{q_3}\}|U_{\rm real}^{\dagger}|\{{{\tilde{q}}^{\pm}_1}\},\!\{{\bar{\tilde{q}}_1}\}}\\
& \braket{\{{{\tilde{q}}^{\pm}_1}\},\!\{{\bar{\tilde{q}}_1}\}|\{{{\tilde{Q}}^{\pm}_1}\},\!\{{\bar{\tilde{Q}}_1}\}} \ \ket{\{{Q^{\pm}_4}\}} \bra{\{{q^{\pm}_4}} \otimes \ket{\{{\bar{{Q}}^{\pm}_1}\}} \bra{\{{\bar{{q}}^{\pm}_1}\}},
\end{align}
and
\begin{align}
\label{rhoAD}
\nn
& \rho_{AD} = {\rm tr}_{BC}(\rho_{\rm real})\\
\nn
& = \frac{1}{D} \sum_{\text{q.n.}} \braket{\{{Q^{\pm}_4}\},\!\{{Q_4}\}|\{{Q^{*}}\}} \braket{\{{q^{*}}\}|\{{q^{\pm}_4}\},\!\{{q_4}\}} \braket{\{{q_4}\}|\{{Q_4}\}}\\[-2mm]
\nn
& \braket{\{{\bar{\tilde{Q}}^{\pm}_1}\},\!\{{\bar{\tilde{Q}}_1}\}|U_{\rm real}|\{{Q^{\pm}_3}\},\!\{{Q_3}\}} \braket{\{{Q^{\pm}_3}\},\!\{{Q_3}\}|\{{\bar{\tilde{Q}}^{\pm}_2}\},\!\{{\bar{\tilde{Q}}_2}\}}\\
\nn
& \braket{\{{\bar{\tilde{Q}}^{\pm}_2}\},\!\{{\bar{\tilde{Q}}_2}\}|U_{\rm real}|\{{Q^{*}}\}} \braket{\{{q^{*}}\}|U_{\rm real}^{\dagger}|\{{\bar{\tilde{q}}^{\pm}_2}\},\!\{{\bar{\tilde{q}}_2}\}}\\
\nn
& \braket{\{{\bar{\tilde{q}}^{\pm}_2}\},\!\{{\bar{\tilde{q}}_2}\}|\{{q^{\pm}_3}\},\!\{{q_3}\}} \braket{\{{q^{\pm}_3}\},\!\{{q_3}\}|U_{\rm real}^{\dagger}|\{{\bar{\tilde{q}}^{\pm}_1}\},\!\{{\bar{\tilde{q}}_1}\}}\\
& \braket{\{{\bar{{q}}^{\pm}_1}\}|\{{\bar{{Q}}^{\pm}_1}\}} \ \ket{\{{Q^{\pm}_4}\}}\bra{\{{q^{\pm}_4}\}} \otimes \ket{\{{{\tilde{Q}}^{\pm}_1}\},\!\{{\bar{\tilde{Q}}_1}\}}\bra{\{{{\tilde{q}}^{\pm}_1}\},\!\{{\bar{\tilde{q}}_1}\}},
\end{align}
and
\begin{equation}
\rho_{C} = {\rm tr}_{A}(\rho_{AC}) = \frac{d_D}{D} \sum_{\text{q.n.}} \ket{\{{\bar{{Q}}^{\pm}_1}\}} \bra{\{{\bar{{Q}}^{\pm}_1}\}},
\end{equation}
and
\begin{align}
\rho_{D} = {\rm tr}_{A}(\rho_{AD}) = \frac{d_C}{D} \sum_{\text{q.n.}} \ket{\{{{\tilde{Q}}^{\pm}_1}\},\!\{{\bar{\tilde{Q}}_1}\}} \bra{\{{{\tilde{Q}}^{\pm}_1}\},\!\{{\bar{\tilde{Q}}_1}\}}.
\end{align}
Here $d_C$ and $d_D$ denote the dimensions of parts $C$ and $D$ in Fig.~\ref{figure1}.
Since $\ket{\{{\bar{\tilde{Q}}^{\pm}}\},\!\{{\bar{\tilde{Q}}}\}}\equiv\ket{\{{\bar{{Q}}^{\pm}}\},\!\{{\bar{{Q}}}\},\!\{{{\tilde{Q}}^{\pm}}\},\!\{{{\tilde{Q}}}\}}$, tracing out the $D$ part of $\ket{\{{\bar{\tilde{Q}}^{\pm}}\},\!\{{\bar{\tilde{Q}}}\}}$ leaves $\ket{\{{\bar{{Q}}^{\pm}}\}}$, while tracing out the $C$ part of $\ket{\{{\bar{\tilde{Q}}^{\pm}}\},\!\{{\bar{\tilde{Q}}}\}}$ leaves $\ket{\{{\bar{{Q}}}\},\\\!\{{{\tilde{Q}}^{\pm}}\},\!\{{{\tilde{Q}}}\}} \equiv\ket{\{{{\tilde{Q}}^{\pm}}\},\!\{{\bar{\tilde{Q}}}\}}$.

Subsequently, using the definition of the 2-Renyi entropy in Sec.~\ref{information-scrambling}, we obtain the entropies $S_2(AC)$, $S_2(AD)$, $S_2(C)$, and $S_2(D)$ corresponding to the above reduced density matrices. Similar to the above calculations, these calculations also are purely formal and straightforward, so we present the results directly, as shown below:
\begin{align}
\label{S2(AC)-form}
\nn
& S_2(AC) = -\log{\rm tr}\rho_{AC}^2\\
\nn
& = -\log \Big[ \frac{1}{D^2} \sum_{\text{q.n.}}\\[-2mm]
\nn
& \braket{\{{\bar{\tilde{Q}}^{\pm}_1}\},\!\{{\bar{\tilde{Q}}_1}\}|U_{\rm real}|\{{Q^{\pm}_3}\},\!\{{Q_3}\}} \braket{\{{Q^{\pm}_3}\},\!\{{Q_3}\}|\{{\bar{\tilde{Q}}^{\pm}_2}\},\!\{{\bar{\tilde{Q}}_2}\}}\\
\nn
& \braket{\{{\bar{\tilde{Q}}^{\pm}_2}\},\!\{{\bar{\tilde{Q}}_2}\}|U_{\rm real}|\{{Q^{\pm}}\},\!\{{Q}\}} \braket{\{{q^{\pm}}\},\!\{{Q}\}|U_{\rm real}^{\dagger}|\{{\bar{\tilde{q}}^{\pm}_2}\},\!\{{\bar{\tilde{q}}_2}\}}\\
\nn
& \braket{\{{\bar{\tilde{q}}^{\pm}_2}\},\!\{{\bar{\tilde{q}}_2}\}|\{{q^{\pm}_3}\},\!\{{q_3}\}} \braket{\{{q^{\pm}_3}\},\!\{{q_3}\}|U_{\rm real}^{\dagger}|\{{\bar{{G}}^{\pm}_1}\};\!\{{{\tilde{Q}}^{\pm}_1}\},\!\{{\bar{\tilde{Q}}_1}\}}\\
\nn
& \braket{\{{\bar{\tilde{G}}^{\pm}_1}\},\!\{{\bar{\tilde{G}}_1}\}|U_{\rm real}|\{{G^{\pm}_3}\},\!\{{G_3}\}} \braket{\{{G^{\pm}_3}\},\!\{{G_3}\}|\{{\bar{\tilde{G}}^{\pm}_2}\},\!\{{\bar{\tilde{G}}_2}\}}\\
\nn
& \braket{\{{\bar{\tilde{G}}^{\pm}_2}\},\!\{{\bar{\tilde{G}}_2}\}|U_{\rm real}|\{{q^{\pm}}\},\!\{{q}\}} \braket{\{{Q^{\pm}}\},\!\{{q}\}|U_{\rm real}^{\dagger}|\{{\bar{\tilde{g}}^{\pm}_2}\},\!\{{\bar{\tilde{g}}_2}\}}\\
& \braket{\{{\bar{\tilde{g}}^{\pm}_2}\},\!\{{\bar{\tilde{g}}_2}\}|\{{g^{\pm}_3}\},\!\{{g_3}\}} \braket{\{{g^{\pm}_3}\},\!\{{g_3}\}|U_{\rm real}^{\dagger}|\{{\bar{{Q}}^{\pm}_1}\};\!\{{{\tilde{G}}^{\pm}_1}\},\!\{{\bar{\tilde{G}}_1}\}} \Big],
\end{align}
and
\begin{align}
\label{S2(AD)-form}
\nn
& S_2(AD) = -\log{\rm tr} \rho_{AD}^2\\
\nn
& = -\log \Big[ \frac{1}{D^2} \sum_{\text{q.n.}}\\[-2mm]
\nn
& \braket{\{{\bar{\tilde{Q}}^{\pm}_1}\},\!\{{\bar{\tilde{Q}}_1}\}|U_{\rm real}|\{{Q^{\pm}_3}\},\!\{{Q_3}\}} \braket{\{{Q^{\pm}_3}\},\!\{{Q_3}\}|\{{\bar{\tilde{Q}}^{\pm}_2}\},\!\{{\bar{\tilde{Q}}_2}\}}\\
\nn
& \braket{\{{\bar{\tilde{Q}}^{\pm}_2}\},\!\{{\bar{\tilde{Q}}_2}\}|U_{\rm real}|\{{Q^{\pm}}\},\!\{{Q}\}} \braket{\{{q^{\pm}}\},\!\{{Q}\}|U_{\rm real}^{\dagger}|\{{\bar{\tilde{q}}^{\pm}_2}\},\!\{{\bar{\tilde{q}}_2}\}}\\
\nn
& \braket{\{{\bar{\tilde{q}}^{\pm}_2}\},\!\{{\bar{\tilde{q}}_2}\}|\{{q^{\pm}_3}\},\!\{{q_3}\}} \braket{\{{q^{\pm}_3}\},\!\{{q_3}\}|U_{\rm real}^{\dagger}|\{{\bar{{Q}}^{\pm}_1}\};\!\{{{\tilde{G}}^{\pm}_1}\},\!\{{\bar{\tilde{G}}_1}\}}\\
\nn
& \braket{\{{\bar{\tilde{G}}^{\pm}_1}\},\!\{{\bar{\tilde{G}}_1}\}|U_{\rm real}|\{{G^{\pm}_3}\},\!\{{G_3}\}} \braket{\{{G^{\pm}_3}\},\!\{{G_3}\}|\{{\bar{\tilde{G}}^{\pm}_2}\},\!\{{\bar{\tilde{G}}_2}\}}\\
\nn
& \braket{\{{\bar{\tilde{G}}^{\pm}_2}\},\!\{{\bar{\tilde{G}}_2}\}|U_{\rm real}|\{{q^{\pm}}\},\!\{{q}\}} \braket{\{{Q^{\pm}}\},\!\{{q}\}|U_{\rm real}^{\dagger}|\{{\bar{\tilde{g}}^{\pm}_2}\},\!\{{\bar{\tilde{g}}_2}\}}\\
& \braket{\{{\bar{\tilde{g}}^{\pm}_2}\},\!\{{\bar{\tilde{g}}_2}\}|\{{g^{\pm}_3}\},\!\{{g_3}\}} \braket{\{{g^{\pm}_3}\},\!\{{g_3}\}|U_{\rm real}^{\dagger}|\{{\bar{{G}}^{\pm}_1}\};\!\{{{\tilde{Q}}^{\pm}_1}\},\!\{{\bar{\tilde{Q}}_1}\}} \Big],
\end{align}
and
\begin{equation}
S_2(C) = -\log{\rm tr} \rho_{C}^2 = -\log \Big[ \frac{d_C d_D^2}{D^2} \Big] = \log d_C,
\end{equation}
and
\begin{equation}
S_2(D) = -\log{\rm tr} \rho_{D}^2 = -\log \Big[ \frac{d_C^2 d_D}{D^2} \Big] = \log d_D,
\end{equation}
where
\begin{align}
\nn
&\ket{\{{\bar{{G}}^{\pm}}\};\!\{{{\tilde{Q}}^{\pm}}\},\!\{{\bar{\tilde{Q}}}\}} \equiv \ket{\{{\bar{{G}}^{\pm}}\},\!\{{\bar{{Q}}}\},\!\{{{\tilde{Q}}^{\pm}}\},\!{\{\tilde{Q}}\}}\\
\nn
&=\ket{\{{\bar{{G}}^{\pm}}\}}\otimes\ket{\{{{\tilde{Q}}^{\pm}}\},\!\{{\bar{\tilde{Q}}}\}}\\
\nn
&=\sum_{\{{\bar{\tilde{K}}}^{\pm}\}} \sum_{\{{\bar{\tilde{K}}}\}} \braket{\{{\bar{{K}}^{\pm}}\}|\{{\bar{{G}}^{\pm}}\}} \braket{\{{{\tilde{K}}^{\pm}}\},\!\{{\bar{\tilde{K}}}\}|\{{{\tilde{Q}}^{\pm}}\},\!\{{\bar{\tilde{Q}}}\}}\\
& \ket{\{{\bar{{K}}^{\pm}}\}}\otimes\ket{\{{{\tilde{K}}^{\pm}}\},\!\{{\bar{\tilde{K}}}\}}.
\end{align}

Since $U_{\rm real}$ can be written as the sum of four terms, $\mathbb{I}$, $-iG_0$, $- i\mathcal{T}$, and $-G_0\mathcal{T}$, substituting $U_{\rm real} = \mathbb{I} - iG_0 - i\mathcal{T} - G_0\mathcal{T}$ into the preceding expressions for $S_2(AC)$ and $S_2(AD)$, and keeping only their leading order, $S_2(AC)$ and $S_2(AD)$ can be expressed as the sum of four contributions, namely
\begin{align}
\nn
S_2(AC) &= -\log\Big[ F(AC;\mathbb{I})+F(AC;G_0)+F(AC;\mathcal{T})\\
&+F(AC;G_0\mathcal{T})+\cdots\Big],\\
\nn
S_2(AD) &= -\log \Big[ F(AD;\mathbb{I})+F(AD;G_0)+F(AD;\mathcal{T})\\
&+F(AD;G_0\mathcal{T})+\cdots\Big].
\end{align}

The $F(AC;\mathbb{I})$ and $F(AD;\mathbb{I})$ represent the contributions from the identity $\mathbb{I}$ in $U_{\rm real}$. 
They can be obtained by selecting, from the expressions of $S_2(AC)$ and $S_2(AD)$, the term that contain eight $\mathbb{I}$, and then removing some completeness relations, such as
\begin{align}
& \sum_{\rm q.n.}\ket{\{{Q^{\pm}}\}}\bra{\{{Q^{\pm}}\},}=\mathbb{I}_A,\\
& \sum_{\rm q.n.}\ket{\{{Q}\}}\bra{\{{Q}\}}=\mathbb{I}_B,\\
& \sum_{\rm q.n.}\ket{\{{\bar{{Q}}^{\pm}}\}}\bra{\{{\bar{{Q}}^{\pm}}\}}=\mathbb{I}_C,\\
& \sum_{\rm q.n.}\ket{\{{{\tilde{Q}}^{\pm}}\},\!\{{\bar{\tilde{Q}}}\}}\bra{\{{{\tilde{Q}}^{\pm}}\},\!\{{\bar{\tilde{Q}}}\}}=\mathbb{I}_D,
\end{align}
with the additional observation that when $U_{\rm real}$ reduces to $\mathbb{I}$, the sets $\{\psi_{Q}^{2}\}+\{\psi_{Q}^{3}\}$ and $\{A_{\mu Q}^{2}\}+\{A_{\mu Q}^{3}\}$ reduce to $\{\psi_{Q}^{1}\}$ and $\{A_{\mu Q}^{1}\}$. This gives
\begin{align}
F(AC;\mathbb{I}) = \frac{d_{C_0}^2 d_{E_0} d_{D_0}^2 d_{F_0}^2}{D^2},
\end{align}
and 
\begin{align}
F(AD;\mathbb{I}) = \frac{d_{C_0} d_{E_0}^2 d_{D_0} d_{F_0}}{D^2}.
\end{align}
Here $d_{C_0}$, $d_{D_0}$, $d_{E_0}$, and $d_{F_0}$ denote the dimensions of the original parts $C$, $D$, $E$, and $F$ in Fig.~\ref{figure1}. Their relation to the dimensions of the merged parts $C$ and $D$ is given by $d_C=d_{C_0}$ and $d_D=d_{D_0}d_{E_0}d_{F_0}$.

The $F(AC;G_0)$ and $F(AD;G_0)$ represent the contributions from $G_0$ in $U_{\rm real}$. They can be obtained by collecting, from the expressions of $S_2(AC)$ and $S_2(AD)$, all the terms that contain one $G_0$ and seven $\mathbb{I}$, and then removing some completeness relations. This gives
\begin{align}
\nn
&F(AC;G_0)\\
\nn
&= 2 \frac{d_{C_0}^2 d_{E_0} d_{D_0}^2 d_{F_0}^2}{D^3} \sum_{\text{q.n.}} \Big[ -i \braket{\{{Q^{\pm}}\},\!\{{Q}\}|G_0|\{{Q^{\pm}}\},\!\{{Q}\}}\\[-2mm]
&+ i \braket{\{{Q^{\pm}}\},\!\{{Q}\}|G_0^{\dagger}|\{{Q^{\pm}}\},\!\{{Q}\}} \Big],
\end{align}
and
\begin{align}
\nn
& F(AD;G_0)\\
\nn
&= 2 \frac{d_{C_0} d_{E_0}^2 d_{D_0} d_{F_0}}{D^3} \sum_{\text{q.n.}} \Big[ -i \braket{\{{Q^{\pm}}\},\!\{{Q}\}|G_0|\{{Q^{\pm}}\},\!\{{Q}\}}\\[-2mm]
&+ i \braket{\{{Q^{\pm}}\},\!\{{Q}\}|G_0^{\dagger}|\{{Q^{\pm}}\},\!\{{Q}\}} \Big].
\end{align}
Using the unitarity of $U_0(=\mathbb{I}-iG_0)$,
\begin{align}
U_0U_0^{\dagger}=\mathbb{I}=\mathbb{I}-iG_0+iG_0^{\dagger}+G_0G_0^{\dagger},
\end{align}
the terms inside the square brackets can be simplified to
\begin{align}
\nn
& -i \braket{\{{Q^{\pm}}\},\!\{{Q}\}|G_0|\{{Q^{\pm}}\},\!\{{Q}\}}\\
\nn
& + i \braket{\{{Q^{\pm}}\},\!\{{Q}\}|G_0^{\dagger}|\{{Q^{\pm}}\},\!\{{Q}\}}\\
& = - \sum_{\{{G^{\pm}}\}}\sum_{\{{G}\}} {}_{\rm in} \big|\braket{\{{G^{\pm}}\},\!\{{G}\}|G_0|\{{Q^{\pm}}\},\!\{{Q}\}}\big|^2.
\end{align}
Therefore, $F(AC;G_0)$ and $F(AD;G_0)$ can be rewritten as
\begin{align}
\nn
&F(AC;G_0)\\
&= - 2 \frac{d_{C_0}^2 d_{E_0} d_{D_0}^2 d_{F_0}^2}{D^3} \sum_{\text{q.n.}} \big|\braket{\{{G^{\pm}}\},\!\{{G}\}|G_0|\{{Q^{\pm}}\},\!\{{Q}\}}\big|^2,\\[-2mm]
\nn
& F(AD;G_0)\\
&= - 2 \frac{d_{C_0} d_{E_0}^2 d_{D_0} d_{F_0}}{D^3} \sum_{\text{q.n.}} \big|\braket{\{{G^{\pm}}\},\!\{{G}\}|G_0|\{{Q^{\pm}}\},\!\{{Q}\}}\big|^2.
\end{align}

Similarly, the contributions from $\mathcal{T}$ and $G_0\mathcal{T}$ can be obtained by collecting the terms that contain one $\mathcal{T}$ or one $G_0\mathcal{T}$ together with seven $\mathbb{I}$, namely,
\begin{align}
\nn
&F(AC;\mathcal{T})\\
&= - 2 \frac{d_{C_0}^2 d_{E_0} d_{D_0}^2 d_{F_0}^2}{D^3} \sum_{\text{q.n.}} \big|\braket{\{{G^{\pm}}\},\!\{{G}\}|\mathcal{T}|\{{Q^{\pm}}\},\!\{{Q}\}}\big|^2,\\[-2mm]
\nn
& F(AD;\mathcal{T})\\
&= - 2 \frac{d_{C_0} d_{E_0}^2 d_{D_0} d_{F_0}}{D^3} \sum_{\text{q.n.}} \big|\braket{\{{G^{\pm}}\},\!\{{G}\}|\mathcal{T}|\{{Q^{\pm}}\},\!\{{Q}\}}\big|^2,\\[-2mm]
\nn
&F(AC;G_0\mathcal{T})\\
&= - 2 \frac{d_{C_0}^2 d_{E_0} d_{D_0}^2 d_{F_0}^2}{D^3} \sum_{\text{q.n.}} \big|\braket{\{{G^{\pm}}\},\!\{{G}\}|G_0\mathcal{T}|\{{Q^{\pm}}\},\!\{{Q}\}}\big|^2,\\[-2mm]
\nn
& F(AD;G_0\mathcal{T})\\
&= - 2 \frac{d_{C_0} d_{E_0}^2 d_{D_0} d_{F_0}}{D^3} \sum_{\text{q.n.}} \big|\braket{\{{G^{\pm}}\},\!\{{G}\}|G_0\mathcal{T}|\{{Q^{\pm}}\},\!\{{Q}\}}\big|^2.
\end{align}

Gathering the expressions obtained above, we can represent $S_2(AC)$ and $S_2(AD)$ as
\begin{align}
\nn
S_2(AC) &= -\log\Big[ \frac{d_{C_0}^2 d_{E_0} d_{D_0}^2 d_{F_0}^2}{D^2} \Big(1\\[-1mm]
\nn
& - \frac{2}{D} \sum_{\text{q.n.}} \big|\braket{\{{G^{\pm}}\},\!\{{G}\}|G_0|\{{Q^{\pm}}\},\!\{{Q}\}}\big|^2\\[-2mm]
\nn
& - \frac{2}{D} \sum_{\text{q.n.}} \big|\braket{\{{G^{\pm}}\},\!\{{G}\}|\mathcal{T}|\{{Q^{\pm}}\},\!\{{Q}\}}\big|^2\\[-2mm]
& - \frac{2}{D} \sum_{\text{q.n.}} \big|\braket{\{{G^{\pm}}\},\!\{{G}\}|G_0\mathcal{T}|\{{Q^{\pm}}\},\!\{{Q}\}}\big|^2 +\cdots \Big),
\end{align}
\begin{align}
\nn
S_2(AD) &= -\log\Big[ \frac{d_{C_0} d_{E_0}^2 d_{D_0} d_{F_0}}{D^2} \Big(1\\[-1mm]
\nn
& - \frac{2}{D} \sum_{\text{q.n.}} \big|\braket{\{{G^{\pm}}\},\!\{{G}\}|G_0|\{{Q^{\pm}}\},\!\{{Q}\}}\big|^2\\[-2mm]
\nn
& - \frac{2}{D} \sum_{\text{q.n.}} \big|\braket{\{{G^{\pm}}\},\!\{{G}\}|\mathcal{T}|\{{Q^{\pm}}\},\!\{{Q}\}}\big|^2\\[-2mm]
& - \frac{2}{D} \sum_{\text{q.n.}} \big|\braket{\{{G^{\pm}}\},\!\{{G}\}|G_0\mathcal{T}|\{{Q^{\pm}}\},\!\{{Q}\}}\big|^2 +\cdots \Big).
\end{align}

Consequently, the tripartite mutual information measured by the 2-Renyi entropy --- defined in Eq.~\eqref{TMI-2-Renyi} --- can be written as
\begin{align}
\nn
& I_{3(2)} = \log d_C + \log d_D\\
\nn
&+ \log \Big[ \frac{d_{C_0}^2 d_{E_0} d_{D_0}^2 d_{F_0}^2}{D^2} \Big] + \log \Big[ \frac{d_{C_0} d_{E_0}^2 d_{D_0} d_{F_0}}{D^2} \Big]\\
\nn
& +2\log \Big[ 1 - \frac{2}{D} \sum_{\text{q.n.}} \big|\braket{\{{G^{\pm}}\},\!\{{G}\}|G_0|\{{Q^{\pm}}\},\!\{{Q}\}}\big|^2\\[-1mm]
\nn
& - \frac{2}{D} \sum_{\text{q.n.}} \big|\braket{\{{G^{\pm}}\},\!\{{G}\}|\mathcal{T}|\{{Q^{\pm}}\},\!\{{Q}\}}\big|^2 \Big]\\[-1mm]
\nn
& - \frac{2}{D} \sum_{\text{q.n.}} \big|\braket{\{{G^{\pm}}\},\!\{{G}\}||G_0\mathcal{T}|\{{Q^{\pm}}\},\!\{{Q}\}}\big|^2 +\cdots \Big]\\[-1mm]
\nn
&= - \frac{2}{D} \sum_{\text{q.n.}} \big|\braket{\{{G^{\pm}}\},\!\{{G}\}|G_0|\{{Q^{\pm}}\},\!\{{Q}\}}\big|^2\\[-1mm]
\nn
& - \frac{2}{D} \sum_{\text{q.n.}} \big|\braket{\{{G^{\pm}}\},\!\{{G}\}|\mathcal{T}|\{{Q^{\pm}}\},\!\{{Q}\}}\big|^2\\[-1mm]
& - \frac{2}{D} \sum_{\text{q.n.}} \big|\braket{\{{G^{\pm}}\},\!\{{G}\}|G_0\mathcal{T}|\{{Q^{\pm}}\},\!\{{Q}\}}\big|^2 +\cdots.
\end{align}
Here we have employed the properties of the logarithmic function, together with the series expansion of $\log(1\\+x)$. We denote the first term on the right-hand side of the second equality by 
$I_{3(2)}^0$ (the contribution from the purely gravitational background), the second term by 
$I_{3(2)}^1$ (the purely QED contribution), and the third term by $I_{3(2)}^{01}$ (the correction to QED induced by the gravitational background). One then naturally observes that in flat spacetime we have 
$I_{3(2)}^0=I_{3(2)}^{01}=0$, so that the tripartite mutual information reduces to $I_{3(2)}=I_{3(2)}^1$. In contrast, in collapsing spacetime both $I_{3(2)}^0\ne0$ and $I_{3(2)}^{01}\ne0$, leading to 
$I_{3(2)}=I_{3(2)}^0+I_{3(2)}^1+I_{3(2)}^{01}<I_{3(2)}^1$.

\subsection{The estimation of tripartite mutual information}
We now turn to estimating the magnitudes of $I_{3(2)}^0$, $I_{3(2)}^1$, and $I_{3(2)}^{01}$. 
This requires, as a central step, evaluating the matrix elements $\braket{\{{G^{\pm}}\},\!\{{G}\}|G_0|\{{Q^{\pm}}\},\!\{{Q}\}}$, 
$\braket{\{{G^{\pm}}\},\!\{{G}\}|\mathcal{T}|\{{Q^{\pm}}\},\!\{{Q}\}}$, and $\braket{\{{G^{\pm}}\},\!\{{G}\}|G_0\mathcal{T}|\{{Q^{\pm}}\},\\\!\{{Q}\}}$. 

For $I_{3(2)}^1$, since the identity operator in the expansion of $U_{\rm real}$	does not contribute to the tripartite mutual information, evaluating $I_{3(2)}^1$ is equivalent to estimating the tripartite mutual information in flat spacetime. In this case, the leading order contribution of $\braket{\{{G^{\pm}}\},\!\{{G}\}|\\\mathcal{T}|\{{Q^{\pm}}\},\!\{{Q}\}}$ corresponds to the scattering of a single electron or positron, with soft-photon dressing, by an external electromagnetic field. This is exactly the same as in our previous work --- see Ref.~\cite{Su2024Scrambling}. Therefore, we simply quote the result here, namely, 
\begin{equation}
\label{I_{3(2)}^1}
I_{3(2)}^{1} \simeq - 2\pi^2 (Ze^2)^2 A_1 (2\pi)^3 \frac{T}{m^2 V}\simeq - (10^1)\frac{T}{Vm^2}.
\end{equation}
Here, $m$ denotes the mass of the electron or positron, $V$ denotes the ``space box'' and $T$ the ``time box'', both of which are indispensable in field-theoretic calculations, as they fix the quantization of the momenta and eliminate the divergences arising from the product of two delta functions.

For $I_{3(2)}^0$ and $I_{3(2)}^{01}$, $G_0$ can be viewed as an effective gravitational potential that encodes collapsing dynamics. In this description, the leading order contribution of $\braket{\{{G^{\pm}}\},\!\{{G}\}|G_0|\{{Q^{\pm}}\},\!\{{Q}\}}$ describes the scattering of a single electron or positron, dressed with soft photons, by this effective gravitational potential, and the leading order contribution of $\braket{\{{G^{\pm}}\},\!\{{G}\}|G_0\mathcal{T}|\{{Q^{\pm}}\},\!\{{Q}\}}$ captures the extra contribution of the external electromagnetic field caused by the same effective gravitational potential.
Since the output states $U_{\rm real}\ket{\{{Q^{\pm}}\},\!\{{Q}\}}$, which can be expressed as
\begin{equation}
\label{output}
U_{\rm real}\ket{\{{Q^{\pm}}\},\!\{{Q}\}}=(\mathbb{I} - iG_0 - i\mathcal{T} - G_0\mathcal{T})\ket{\{{Q^{\pm}}\},\!\{{Q}\}},
\end{equation}
evolve from the input states $\ket{\{{Q^{\pm}}\},\!\{{Q}\}}$, i.e., the solutions to the equation of motion (EOM) for fermions and gauge fields in the distant future evolve from those in the distant past, one could estimate the above two contributions by analyzing this evolution.

Note that each of these two contributions consists of two parts: one from particles that traverse the collapsing matter and eventually escape to future infinity, and the other from particles that fail to escape and are ultimately captured by the black hole. For particles that escape to future infinity, their EOM can be directly obtained. 
In contrast, for those captured by the black hole, their EOM cannot be easily determined, since the physics inside the event horizon is still not well understood.
As a result, our present analysis must be restricted to the parts from particles that escape to future infinity.
Nevertheless, we will show later that, even under this restriction, the tripartite mutual information can still be estimated.

Moreover, when evaluating the matrix elements, the role of soft photons is to cancel IR divergences and introduce a small correction, which can be neglected at leading order\footnote{Although soft photons can be neglected in the evaluation of matrix elements, they cannot be omitted when discussing the expression of the tripartite mutual information, where the bipartite partition of the input and output states is crucial.}. Hence, in the following calculations we focus on the fermions while ignoring the gauge fields. 

For the fermions that traverse the collapsing matter and eventually escape to future infinity, in the distant past, the EOM can be approximated by the Dirac equation in flat spacetime, which can be cast into a Hamiltonian form, namely \cite{Rose1961Relativistic},
\begin{align}
\label{Dirac-equation-distant-past}
\nn
i\partial_{t}\psi &= \Big[ - i\gamma^0\gamma^1 \frac{1}{r} \partial_{r} r - i\gamma^0\gamma^2 \frac{1}{r\sin^{1/2}\theta} \partial_{\theta}\sin^{1/2}\theta\\
\nn
&- i\gamma^0\gamma^3 \frac{1}{r\sin\theta} \partial_{\phi} + \gamma^0 m \Big] \psi\\
&= H_0\psi.
\end{align}
In the distant future, outside the event horizon, the EOM is given by the Dirac equation in the RN spacetime, which can likewise be cast into a Hamiltonian form, namely \cite{Boulware1975Spin,Noble2016Dirac},
\begin{align}
\label{Dirac-equation-distant-future}
\nn
i \partial_{t}\psi &= \varphi\Big[ - i\gamma^0\gamma^1 \frac{1}{r}
\varphi^{1/2} \partial_{r} \varphi^{1/2} r - i\gamma^0\gamma^2 \frac{1}{r\sin^{1/2}\theta} \partial_{\theta}\sin^{1/2}\theta\\
\nn
&- i\gamma^0\gamma^3 \frac{1}{r\sin\theta} \partial_{\phi} + \gamma^0 m - \frac{Qe}{r} \Big] \psi\\
&=H\psi,
\end{align}
where $\varphi=\big(1-\frac{r_s}{r}+\frac{r_Q^2}{r^2}\big)^{1/2}$, with $r_s$ and $r_Q^2$ defined exactly as in Eq.~\eqref{RN-metric}. 
Moreover, since the above EOM is valid outside the event horizon, namely, $r>r_s$, both ${r_s}/{r}$ and ${r_Q^2}/{r^2}$ in $\varphi$ are small --- we must have that $r_Q^2 \le r_s$, as $r_Q^2 > r_s$ would result in a forbidden naked singularity. Therefore, to the leading order, the above equation can be recast as
\begin{align}
\label{Dirac-equation-distant-future-perturb}
i \partial_{t}\psi &= (H_0+H_1)\psi,
\end{align}
where $H_0$ is the same one in Eq.~\eqref{Dirac-equation-distant-past}, and 
\begin{align}
\label{H-1}
\nn
H_1 &= \Big( - \frac{r_s}{2r} + \frac{r_Q^2}{2r^2} \Big) H_0 - i \gamma^0\gamma^1 \Big[ \Big( - \frac{r_s}{2r} + \frac{r_Q^2}{2r^2} \Big) \partial_r - \frac{1}{4} \frac{r_s}{r^2} \Big]\\
&- \Big(1 - \frac{r_s}{2r} + \frac{r_Q^2}{2r^2} \Big)\frac{Qe}{r}.
\end{align}

We expect that, during the collapse, the fermion state gradually evolves from a solution of Eq.~\eqref{Dirac-equation-distant-past} to a solution of Eq.~\eqref{Dirac-equation-distant-future-perturb}. Moreover, it seems that the solution of Eq.~\eqref{Dirac-equation-distant-future-perturb} could be obtained from the solution of Eq.~\eqref{Dirac-equation-distant-past} via perturbation theory. Therefore, if we denote the solution of Eq.~\eqref{Dirac-equation-distant-past} by $\ket{\psi_i^{(0)}}$, then the solution of Eq.~\eqref{Dirac-equation-distant-future-perturb}, denoted by $\ket{\psi_j}$, could take the form
\begin{align}
\nn
\label{psi}
\ket{\psi_j} &= \ket{\psi_j^{(0)}} + \sum_{i(i \ne j)} \frac{\braket{\psi_i^{(0)}|H_1|\psi_j^{(0)}}}{\omega_j^{(0)}-\omega_i^{(0)}} \ket{\psi_i^{(0)}}\\
&= \Bigg( \mathbb{I} + \sum_{i(i \ne j)}\frac{\ket{\psi_i^{(0)}}\bra{\psi_i^{(0)}}H_1}{\omega_j^{(0)}-\omega_i^{(0)}} \Bigg) \ket{\psi_j^{(0)}},\\
\omega_j &= \omega_j^{(0)}+\omega_j^{(1)}=\omega_j^{(0)}+\braket{\psi_j^{(0)}|H_1|\psi_j^{(0)}},
\end{align}
according to perturbation theory.

However, there is a subtlety here: the $r$–integration in $\braket{\psi_i^{(0)}|H_1|\psi_j^{(0)}}$ is defined over the whole domain of $r$, while $H_1$ is only valid in the region $r>r_s$, so how to deal with the integral in the region $r<r_s$ becomes a problem. 

To address this, let us recall the process of gravitational collapse. In the distant past, the distribution of collapsing matter is approximately homogeneous, so a traversing particle can be assumed near the center of the collapsing matter, i.e., around $r=0$. As the collapse proceeds, the particle propagates outward toward $r\to\infty$. 
Suppose that when the particle reaches $r=R$ (with $R>r_s$), a RN black hole forms.
Then, in the region $r\in [R,\infty)$, the particle’s EOM is indeed given by Eq.~\eqref{Dirac-equation-distant-future-perturb}, meaning that the particle experiences the RN geometry characterized by the mass $M$ and charge $Q$. 
What about the region $r\in [0,R)$? During the collapse, the matter distribution becomes increasingly concentrated, so the particle effectively experiences a RN geometry with mass $M(r)$ and charge $Q(r)$ that grow with $r$. 
We thus assume that these effective quantities are proportional to the volume enclosed within radius $r$, namely, $M(r)=\rho_M 4\pi r^3/3$ and $Q(r)= e \rho_q 4\pi r^3/3$, where $\rho_M$ and $\rho_q$ are determined by requiring $M(r\!=\!R)=M$ and $Q(r\!=\!R)=Q$. Furthermore, to ensure the validity of the perturbative treatment, we assume that the particle’s radial position $r$ always remains larger than the corresponding effective Schwarzschild radius $r_s(r)(=2GM(r))$.

Therefore, under the above assumption, we can extend the definition of $H_1$ into the region $r<r_s$ , thereby completing the integral in $\braket{\psi_i^{(0)}|H_1|\psi_j^{(0)}}$.
Then, the new $H_1$ is obtained by replacing $r_s$, $r_Q^2$, $M$ and $Q$ in the original $H_1$ --- shown in Eq.~\eqref{H-1} --- with $r_s(r)$, $r_Q^2(r)$, $M(r)$ and $Q(r)$, where $r_s(r)=r_s$ for $r\in[R,\infty)$ and $r_s(r)={8\pi G\rho_M r^3}/{3}$ for $r\in[0,R)$, and a similar piecewise definition applies to $r_Q^2(r)$, $M(r)$ and $Q(r)$.
In this way, the integration over $r$ in $\braket{\psi_i^{(0)}|H_1|\psi_j^{(0)}}$ symbolically represents the process of the particle traversing through the collapsing matter.

By comparing Eq.~\eqref{psi} with Eq.~\eqref{output}, one naturally expects that some terms in the inner product
\begin{align}
\label{inner-product}
\nn
&\int dt \ \braket{\psi_i^{(0)}|\psi_j} \ {\rm e}^{-i(\omega_j-\omega_i^{(0)})t}\\
&= \Big(\delta_{ij}+ \frac{\braket{\psi_i^{(0)}|H_1|\psi_j^{(0)}}}{(\omega_j^{(0)}-\omega_i^{(0)})}\Big)\delta(\omega_j^{(0)}+\omega_j^{(1)}-\omega_i^{(0)}),
\end{align}
yields the contributions of matrix elements $\braket{\{{G^{\pm}}\},\!\{{G}\}|\\G_0|\{{Q^{\pm}}\},\!\{{Q}\}}$ and $\braket{\{{G^{\pm}}\},\!\{{G}\}|G_0\mathcal{T}|\{{Q^{\pm}}\},\!\{{Q}\}}$ from particles that escape to future infinity with the small corrections arising from soft photons omitted.

The nonperturbative solution $\ket{\psi_i^{(0)}}$ in the above expression, namely the solution of the Dirac equation in flat spacetime, is well known \cite{Rose1961Relativistic,Boulware1975Spin,Greiner2000Relativistic,Strange1998Relativistic,Kordt2012Single} and takes the form
\begin{align}
\psi(r,t)=
\begin{pmatrix}
f(r,\omega,\kappa)\chi_{\kappa}^{\mu}(\theta,\phi)\\
ig(r,\omega,\kappa)\chi_{-\kappa}^{\mu}(\theta,\phi)
\end{pmatrix}
{\rm e}^{-i{\omega}t},
\end{align}
where $f(r,\omega,\kappa)$ and $g(r,\omega,\kappa)$ are the radial functions, and $\chi_{\kappa}^{\mu}(\theta,\phi)$ and $\chi_{-\kappa}^{\mu}(\theta,\phi)$ are the spin-angular functions. The explicit forms of these functions are given in \ref{A-A}.

Using the properties of the spin–angular functions and those of the gamma matrices, the matrix element $\braket{\psi_i^{(0)}|H_1|\psi_j^{(0)}}$ --- in this expression, and in the one that follows, the dependence of $r_s(r)$, $r_Q^2(r)$, $M(r)$ and $Q(r)$ on  $r$ is not written explicitly for notational simplicity --- can be simplified to
\begin{align}
\nn
&\braket{\psi_i^{(0)}|H_1|\psi_j^{(0)}}\\
\nn
&= \int r^2 dr d\Omega \Big[ \Big( - \frac{r_s}{2r} + \frac{r_Q^2}{2r^2} \Big) \omega_j^{(0)} - \Big(1 - \frac{r_s}{2r} + \frac{r_Q^2}{2r^2} \Big)\frac{Qe}{r}\Big]\\
& \Big[ f(r,\omega_i^{(0)},\kappa_i)f(r,\omega_j^{(0)},\kappa_j) + g(r,\omega_i^{(0)},\kappa_i)g(r,\omega_j^{(0)},\kappa_j) \Big].
\end{align}
It is then straightforward to observe that the second term of $H_1$ in Eq.~\eqref{H-1} does not contribute to the matrix element $\braket{\psi_i^{(0)}|H_1|\psi_j^{(0)}}$. We therefore denote the first term of $H_1$ by $H_1^1$ and the third term by $H_1^2$, namely,
\begin{align}
H_1^1 = \Big( - \frac{r_s}{2r} + \frac{r_Q^2}{2r^2} \Big) H_0, \quad H_1^2 = - \Big(1 - \frac{r_s}{2r} + \frac{r_Q^2}{2r^2} \Big)\frac{Qe}{r}.
\end{align}
Since the contribution of $H_1^1$ arises purely from the gravitational background, whereas that of $H_1^2$ accounts for the influence of the gravitational background on the external electromagnetic field, it follows that the matrix element $\braket{\psi_i^{(0)}|H_1^1|\psi_j^{(0)}}$ corresponds to the contribution of the matrix element $\braket{\{{G^{\pm}}\},\!\{{G}\}|G_0|\{{Q^{\pm}}\},\!\{{Q}\}}$ from particles that escape to future infinity once the subleading corrections from soft photons are neglected, while the matrix element $\braket{\psi_i^{(0)}|H_1^2|\psi_j^{(0)}}$ corresponds to that of the matrix element $\braket{\{{G^{\pm}}\},\!\{{G}\}|G_0\mathcal{T}|\{{Q^{\pm}}\},\!\{{Q}\}}$ from particles that escape to future infinity with the same omission.

By performing a direct calculation and assuming $2 n GM|\boldsymbol{p}|_{\rm max}<1$ --- where $|\boldsymbol{p}|_{\rm max} \sim 10^{-8} M_{pl}$ is the maximum momentum, and $n=R/r_s \sim 10^{1}$ is the ratio of $R$ to the Schwarzschild radius $r_s$ after black hole has formed --- we obtain the contributions of $I_{3(2)}^0$ and $I_{3(2)}^{01}$ from particles that escape to future infinity (the accessible contributions) --- denoted as ${(I_{3(2)}^0)}^{\rm acc.}$ and ${(I_{3(2)}^{01})}^{\rm acc.}$ --- as follows:
\begin{align}
\label{I_{3(2)}^0-acc}
\nn
&{(I_{3(2)}^0)}^{\rm acc.}\\
\nn
&\simeq - \frac{4T}{2\pi d_A} \int d|p_i| d|p_j| \ |\braket{\psi_i^{(0)}\big|H_1^1|\psi_j^{(0)}}\big|^2/(\omega_j^{(0)}-\omega_i^{(0)})^2\\
\nn
&\delta(\omega_j^{(0)}+\omega_j^{(1)}-\omega_i^{(0)})\\
\nn
&\simeq - 8\pi G^2 M^2 m^2 \frac{T}{Vm^2} +\frac{16 G Z^4 e^4}{9 \pi^2 R}\frac{T}{V^{4/3}}\\
&\simeq - (10^{-2})\frac{T}{Vm^2},
\end{align}
and
\begin{align}
\label{I_{3(2)}^{01}-acc}
\nn
&{(I_{3(2)}^{01})}^{\rm acc.}\\
\nn
&\simeq - \frac{4T}{2\pi d_A} \int d|p_i| d|p_j| \ |\braket{\psi_i^{(0)}\big|H_1^2|\psi_j^{(0)}}\big|^2/(\omega_j^{(0)}-\omega_i^{(0)})^2\\
\nn
&\delta(\omega_j^{(0)}+\omega_j^{(1)}-\omega_i^{(0)})\\
\nn
&\simeq - \frac{16 Z e^2}{\pi} G^2 M^2 m^2 \frac{T}{Vm^2} - \frac{128 Z^3 e^4 G^3 M^2 }{3 \pi^3 R}\frac{T}{V^{4/3}}\\
&\simeq - (10^{-3})\frac{T}{Vm^2}.
\end{align}
In the final dimensional estimate, we invoke the condition $2GMm<1$, which follows from $2 n GM|p|_{\rm max}<1$. Since $M$ is macroscopic, it must satisfy $M > M_{pl} > m$, implying that the corresponding Compton wavelength obeys $L_{M} < L_{pl} < L_m$. Consequently, the Schwarzschild radius $r_s$ satisfies $r_s = 2L_{pl}^2/L_{M} > L_{pl}$. At the same time, the requirement $2GMm < 1$ further enforces $r_s < L_m$, which in turn yields the inequality $2{L_{pl}^2}/(8\pi)<{L_{M}L_m}$. Therefore, these constraints require that the mass $M$ must lie within the range $M_{pl} \ (\sim 10^{-5}g) < M < 8\pi M_{pl}^2/m \ (\sim 10^{19}g)$.

For the fermions captured by the black hole, their states can likewise be described by the solutions to the EOM inside the event horizon in the distant future. Denoting the solution for captured particles as $\ket{\psi'_j}$, and recalling the solution $\ket{\psi_j}$ for particles escaping to future infinity, the complete output state takes the form $\ket{\psi_j}+\ket{\psi'_j}$. If one believes that the amplitudes in the strong-gravity regime can be treated perturbatively in some manner (for instance, via the AdS/CFT correspondence \cite{Maldacena1997Large}), then by writing the perturbation as $H'={H'}_1^1+{H'}_1^2$, one obtains
\begin{align}
\nn
&I_{3(2)}^0\simeq - \frac{4T}{2\pi d_A} \int d|p_i| d|p_j| \ \delta(\omega_j^{(0)}+\omega_j^{(1)}-\omega_i^{(0)})\\
&\Big(|\braket{\psi_i^{(0)}\big|H_1^1|\psi_j^{(0)}}\big|^2+|\braket{\psi_i^{(0)}\big|{H'}_1^1|\psi_j^{(0)}}\big|^2\Big)/(\omega_j^{(0)}-\omega_i^{(0)})^2,
\end{align}
and
\begin{align}
\nn
&I_{3(2)}^{01}\simeq - \frac{4T}{2\pi d_A} \int d|p_i| d|p_j| \ \delta(\omega_j^{(0)}+\omega_j^{(1)}-\omega_i^{(0)})\\
&\Big(|\braket{\psi_i^{(0)}\big|H_1^2|\psi_j^{(0)}}\big|^2+|\braket{\psi_i^{(0)}\big|{H'}_1^2|\psi_j^{(0)}}\big|^2\Big)/(\omega_j^{(0)}-\omega_i^{(0)})^2,
\end{align}
where the orthogonality condition $\braket{\psi_j|\psi'_j}=0$ has been used. As the contributions from $H_1^1$ and $H_1^2$ have been denoted as ${(I_{3(2)}^0)}^{\rm acc.}$ and ${(I_{3(2)}^{01})}^{\rm acc.}$, respectively, we now denote the contributions from ${H'}_1^1$ and ${H'}_1^2$ as ${(I_{3(2)}^0)}^{\rm inacc.}$ and ${(I_{3(2)}^{01})}^{\rm inacc.}$, respectively. Consequently, since both ${(I_{3(2)}^0)}^{\rm inacc.}$ and ${(I_{3(2)}^{01})}^{\rm inacc.}$ are negative, it follows that
\begin{align}
{(I_{3(2)})}^{\rm acc.} ={(I_{3(2)}^0)}^{\rm acc.}+{(I_{3(2)}^1)}^{\rm acc.}+{(I_{3(2)}^{01})}^{\rm acc.} \ge I_{3(2)},
\end{align}
where $I_{3(2)}^1$ is relabeled as ${(I_{3(2)}^1)}^{\rm acc.}$ for notational consistency. Therefore, even considering only 
${(I_{3(2)})}^{\rm acc.}$, it is sufficient to estimate the tripartite mutual information, namely, to provide an upper bound for it.

\section{Conclusion}\label{conclusion}
Before an appropriate regularization procedure is applied, the dimension $d_A$ of the part $A$ in Fig. \ref{figure1} is infinite, which sets the lower bound of the tripartite mutual information in Eq. \eqref{TMI-inequality} to negative infinity. By contrast, the ``space box'' volume $V$ and ``time box'' volume $T$ in Eq. \eqref{I_{3(2)}^1}, Eq. \eqref{I_{3(2)}^0-acc} and Eq. \eqref{I_{3(2)}^{01}-acc} are finite. 
Therefore, the tripartite mutual information we obtained is negative, finite, and substantially larger than the lower bound associated with maximal scrambling in Eq. \eqref{TMI-inequality}. 
This aligns with our expectation that either in flat or collapsing spacetime, when the scattering process involves soft degrees of freedom, the corresponding unitary operator scrambles the information carried by the incoming hard degrees of freedom.

Compared with our previous work in flat spacetime \cite{Su2024Scrambling}, the crucial distinction is that in flat spacetime scrambling disappears once QED interactions are switched off, whereas in collapsing spacetime it persists even in their absence. This persistence arises from the non-uniqueness of the vacuum in a collapsing spacetime: modes defined with respect to different vacua are related by Bogoliubov transformations, which mix and delocalize the information carried by incoming hard degrees of freedom with that of soft degrees of freedom during evolution. Consequently, whenever the evolution operator contains elements such as interactions or nontrivial transformations, the presence of soft degrees of freedom enables the scrambling of information initially encoded in the hard degrees of freedom.

In our calculation of the gravitational contribution to the tripartite mutual information, we have not included contributions from the region inside the event horizon. This is due to two reasons: first, no standard description of the physics inside the horizon is currently available; second, any such contribution would necessarily be non-positive. As a result, our calculation provides an upper bound on the tripartite mutual information in terms of the 2-Renyi entropy. Since the tripartite mutual information in terms of the 2-Renyi entropy upper-bounds the corresponding quantity in terms of von Neumann entropy, the actual tripartite mutual information should be less than or equal to our estimate. Consequently, the degree of information scrambling, measured by the absolute value of the tripartite mutual information, is at least as large as our estimate.

Since the correlations between soft and hard degrees of freedom explain why a pure state may appear thermal to an observer \cite{Strominger2017Black,Carney2017Infrared,Carney2018Dressed,Carney2018need}, our analysis is expected to provide new insights into the black hole information paradox. Within our framework, it is natural to imagine that the degree of information scrambling continuously changes as the gravitational background evolves. If we regard each moment during the formation and evaporation of a RN black hole as corresponding to a RN black hole with a time-dependent effective mass and charge, we would expect the degree of scrambling to first increase and then decrease. Consequently, once the black hole formed from the collapse has fully evaporated, the degree of information scrambling should return to that of flat spacetime. Of course, what exactly happens when the black hole mass decreases to the Planck mass remains unclear, so that our speculation cannot be entirely accurate. Nevertheless, this perspective provides new insights how information scrambling occurs during black hole formation, and offers new conjectures how it might be eventually restored.

While our explicit discussion has focused on soft photons, the framework can naturally be extended to soft gravitons, where a similar scrambling mechanism is expected, though the degree of scrambling may differ quantitatively. Such an extension may help clarify which soft hairs are most relevant to the soft-hair resolution. More broadly, our analysis can be generalized to scenarios in which the information paradox is addressed through correlations among other types of degrees of freedom, suggesting that our approach possesses a certain degree of generality.

In summary, following the lines of references \cite{Hawking2016Soft,Strominger2017Black,Carney2017Infrared,Carney2018Dressed,Carney2018need}, through a series of studies—both our previous work in flat spacetime and the present work in collapsing spacetime—we have investigated the similarities and differences of information scrambling in flat and collapsing backgrounds. Our results confirm the relevance of soft degrees of freedom for the black hole information paradox and indicate that the source of scrambling originates from both interactions and the nontrivial structure of spacetime. This provides a foundation for further understanding the black hole information paradox, specifically how information scrambling occurs and evolves.

\section*{Acknowledgments}
\noindent
A.M. wishes to thank Chris Fields and Matteo Lulli for several discussions on information scrambling and the black hole information paradox. A.M. and X.-L. S. wish to thank Ugo Moschella for discussions and suggestions on estimating the tripartite mutual information. A.H. acknowledges support from NSF award no. 2014000.

\appendix

\section{The solution of Dirac equation in flat spacetime}
\label{A-A}
The Dirac equation in flat spacetime, namely Eq. \eqref{Dirac-equation-distant-past}, can be recast in the form \cite{Rose1961Relativistic}
\begin{align}
\label{Dirac-equation-distant-past-another-form}
\nn
&i\partial_{t}\psi = H_0\psi = (\boldsymbol{\alpha}\cdot\boldsymbol{p} + \beta m) \psi\\
\nn
&= \Big[ -i\alpha_{r} \Big(\partial_{r}+\frac{1}{r}\Big) + i\alpha_{r}\frac{1}{r}(1+\boldsymbol{\sigma}\cdot\boldsymbol{L}) + \beta m \Big] \psi\\
&= \Big[-i\gamma^5\sigma_r\Big(\partial_r+\frac{1}{r}-\frac{1}{r}(1+\boldsymbol{\sigma}\cdot\boldsymbol{L})\Big)+\beta m\Big]\psi,
\end{align}
where $\beta=\gamma^0$ and $\boldsymbol{\alpha}=\gamma^5\boldsymbol{\sigma}$, with $\boldsymbol{\sigma}=\{\sigma_i\}$ ($i=x,y,z$) denoting the standard Pauli matrices, and $\gamma^5$ and $\gamma^0$ being the gamma matrices, which in the Dirac representation take the following form
\begin{equation} 
\gamma^5=
\begin{pmatrix}
0 & 1 \\
1 & 0
\end{pmatrix}, \quad
\gamma^0=
\begin{pmatrix}
1 & 0 \\
0 & -1
\end{pmatrix}.
\end{equation}
Moreover, $\alpha_r$ and $\sigma_r$, the component of $\boldsymbol{\alpha}$ and $\boldsymbol{\sigma}$ along the radial direction, are given by $\alpha_r=\boldsymbol{\alpha}\cdot\hat{\boldsymbol{r}}$ and $\sigma_r=\boldsymbol{\sigma}\cdot\hat{\boldsymbol{r}}$, and the operator $(1+\boldsymbol{\sigma}\cdot\boldsymbol{L})$ is introduced to distinguish the two classes of spinors, with $\boldsymbol{L}$ denoting the orbital angular momentum.

The solution of Eq.~\eqref{Dirac-equation-distant-past-another-form} takes the form \cite{Kordt2012Single}
\begin{align}
\label{solution}
\psi(r,t)=
\begin{pmatrix}
f(r,\omega,\kappa)\chi_{\kappa}^{\mu}(\theta,\phi)\\
ig(r,\omega,\kappa)\chi_{-\kappa}^{\mu}(\theta,\phi)
\end{pmatrix}
{\rm e}^{-i{\omega}t}.
\end{align}
Here, $\chi_{\kappa}^{\mu}(\theta,\phi)$ and $\chi_{-\kappa}^{\mu}(\theta,\phi)$ are the spin-angular functions, which are eigenfunctions of the operator $(1+\boldsymbol{\sigma}\cdot\boldsymbol{L})$, namely,
\begin{align}
&(1+\boldsymbol{\sigma}\cdot\boldsymbol{L}) \ \chi_{\kappa}^{\mu}(\theta,\phi) = -\kappa \  \chi_{\kappa}^{\mu}(\theta,\phi),\\
&(1+\boldsymbol{\sigma}\cdot\boldsymbol{L}) \ \chi_{-\kappa}^{\mu}(\theta,\phi) = \kappa \  \chi_{-\kappa}^{\mu}(\theta,\phi),
\end{align}
their explicit expressions being given by
\begin{align}
&\chi_{\kappa}^{\mu}(\theta,\phi) = \sum_{m} C(l,\frac{1}{2},j;\mu-m,m,\mu) Y_{l}^{\mu-m}(\theta,\phi) \chi^{m},\\
&\chi_{-\kappa}^{\mu}(\theta,\phi) = \sum_{m} C(\bar{l},\frac{1}{2},j;\mu-m,m,\mu) Y_{\bar{l}}^{\mu-m}(\theta,\phi) \chi^{m},
\end{align}
with
\begin{align}
l=\left\{\begin{aligned}
&\kappa, &\kappa>0,\\
&-\kappa-1, &\kappa<0,
\end{aligned}\right. \qquad
\bar{l}=\left\{\begin{aligned}
&l-1=\kappa-1, &\kappa>0,\\
&l+1=-\kappa, &\kappa<0.
\end{aligned}\right.
\end{align}
Here, $C(l/\bar{l},\frac{1}{2},j;\mu-m,m,\mu)$ denote the Clebsch-Gordan coefficients, with $j$, $l/\bar{l}$ and $\frac{1}{2}$ are quantum numbers associated with $\boldsymbol{J}$ (the total angular momentum operator), $\boldsymbol{L}$ (the orbital angular momentum operator) and $\boldsymbol{S}=\boldsymbol{\sigma}/2$ (the spin operator), respectively, while $\mu$, $\mu-m$ and $m$ are the eigenvalues of $\boldsymbol{J}_z$ ($z$ component of $\boldsymbol{J}$), $\boldsymbol{L}_z$ ($z$ component of $\boldsymbol{L}$) and $\boldsymbol{S}_z$ ($z$ component of $\boldsymbol{S}$), respectively. 
Additionally, $Y_{l/\bar{l}}^{\mu-m}(\theta,\phi)$ are the spherical harmonic functions, and $\chi^{m}$ are the eigenfunctions of $\boldsymbol{S}^2$ and $S_z$, namely
\begin{align}
\chi^{m=1/2} = \begin{pmatrix}
1 \\
0
\end{pmatrix},
\quad
\chi^{m=-1/2} = \begin{pmatrix}
0 \\
1
\end{pmatrix}.
\end{align}
While $f(r)$ and $g(r)$ in Eq.\eqref{solution} are the radial functions, their explicit expressions are given by
\begin{align}
f(r,\omega,\kappa) = A j_l(|\boldsymbol{p}|r), \quad g(r,\omega,\kappa) = A \frac{\kappa}{|\kappa|}\frac{|\boldsymbol{p}|}{\omega+m}j_{\bar{l}}(|\boldsymbol{p}|r), 
\end{align}
where $j_{l/\bar{l}}(|\boldsymbol{p}|r)$ are spherical Bessel functions, $\boldsymbol{p}$ and $\omega$ are the particle's 3-momentum and energy, related by $\boldsymbol{p}^2=\omega^2-m^2$.

\section{The integral involving spherical Bessel functions}
\label{A-B}
In the calculation of ${(I_{3(2)}^0)}^{\rm acc.}$ and ${(I_{3(2)}^{01})}^{\rm acc.}$ --- see Eqs.~\eqref{I_{3(2)}^0-acc}-\eqref{I_{3(2)}^{01}-acc} --- one encounters integrals of the type
\begin{align}
\label{integral}
\int r^2 dr \ r^a j_n(|\boldsymbol{p}_i| r) j_n(|\boldsymbol{p}_j| r).
\end{align}
When evaluating $\braket{\psi_j^{(0)}|H_1|\psi_j^{(0)}}$, it is clear that $|\boldsymbol{p}_i|=|\boldsymbol{p}_j|$. In this case,  the evaluation is straightforward, and analytic results can generally be obtained.
By contrast, when evaluating $\braket{\psi_i^{(0)}|H_1|\psi_j^{(0)}}$, one instead encounters the case $|\boldsymbol{p}_i|\neq|\boldsymbol{p}_j|$. In this case, the simple analytic solutions are typically unavailable.
Fortunately, at the perturbative level, namely, $|\boldsymbol{p}_j|=|\boldsymbol{p}_i|+\delta|\boldsymbol{p}_i|$, the spherical Bessel function $j_n(|\boldsymbol{p}_j|r)$ can be expanded in a Taylor series around $|\boldsymbol{p}_i|$, yielding\cite{Boas1966Mathematical}
\begin{align}
\nn
&j_n(|\boldsymbol{p}_j|r) = j_n(|\boldsymbol{p}_i|r)+\delta|\boldsymbol{p}_i|r \ j'_n(|\boldsymbol{p}_i|r)\\
&=j_n(|\boldsymbol{p}_i|r) + n\delta|\boldsymbol{p}_i|/|\boldsymbol{p}_1|j_n(|\boldsymbol{p}_i|r) - \delta|\boldsymbol{p}_i|r j_{n+1
}(|\boldsymbol{p}_i|r).
\end{align}
Under these circumstances, the integral appearing in Eq.~\eqref{integral} can be approximated as
\begin{align}
\nn
&\int r^2 dr \ r^a j_n(|\boldsymbol{p}_i|r)\\
&\times[j_n(|\boldsymbol{p}_i|r) + n\delta|\boldsymbol{p}_i|/|\boldsymbol{p}_i|j_n(|\boldsymbol{p}_i|r)- \delta|\boldsymbol{p}_i|r j_{n+1
}(|\boldsymbol{p}_i|r)],
\end{align}
which can then be evaluated using the standard analytical techniques.

\bibliography{main}

\begin{thebibliography}{62}%
\makeatletter
\providecommand \@ifxundefined [1]{%
 \@ifx{#1\undefined}
}%
\providecommand \@ifnum [1]{%
 \ifnum #1\expandafter \@firstoftwo
 \else \expandafter \@secondoftwo
 \fi
}%
\providecommand \@ifx [1]{%
 \ifx #1\expandafter \@firstoftwo
 \else \expandafter \@secondoftwo
 \fi
}%
\providecommand \natexlab [1]{#1}%
\providecommand \enquote  [1]{``#1''}%
\providecommand \bibnamefont  [1]{#1}%
\providecommand \bibfnamefont [1]{#1}%
\providecommand \citenamefont [1]{#1}%
\providecommand \href@noop [0]{\@secondoftwo}%
\providecommand \href [0]{\begingroup \@sanitize@url \@href}%
\providecommand \@href[1]{\@@startlink{#1}\@@href}%
\providecommand \@@href[1]{\endgroup#1\@@endlink}%
\providecommand \@sanitize@url [0]{\catcode `\\12\catcode `\$12\catcode
  `\&12\catcode `\#12\catcode `\^12\catcode `\_12\catcode `\%12\relax}%
\providecommand \@@startlink[1]{}%
\providecommand \@@endlink[0]{}%
\providecommand \url  [0]{\begingroup\@sanitize@url \@url }%
\providecommand \@url [1]{\endgroup\@href {#1}{\urlprefix }}%
\providecommand \urlprefix  [0]{URL }%
\providecommand \Eprint [0]{\href }%
\providecommand \doibase [0]{http://dx.doi.org/}%
\providecommand \selectlanguage [0]{\@gobble}%
\providecommand \bibinfo  [0]{\@secondoftwo}%
\providecommand \bibfield  [0]{\@secondoftwo}%
\providecommand \translation [1]{[#1]}%
\providecommand \BibitemOpen [0]{}%
\providecommand \bibitemStop [0]{}%
\providecommand \bibitemNoStop [0]{.\EOS\space}%
\providecommand \EOS [0]{\spacefactor3000\relax}%
\providecommand \BibitemShut  [1]{\csname bibitem#1\endcsname}%
\let\auto@bib@innerbib\@empty
\bibitem [{\citenamefont {Harlow}(2016)}]{Harlow2016Jerusalem}%
  \BibitemOpen
  \bibfield  {author} {\bibinfo {author} {\bibfnamefont {D.}~\bibnamefont
  {Harlow}},\ }\href@noop {} {\bibfield  {journal} {\bibinfo  {journal}
  {Reviews of Modern Physics}\ }\textbf {\bibinfo {volume} {88}},\ \bibinfo
  {pages} {015002} (\bibinfo {year} {2016})}\BibitemShut {NoStop}%
\bibitem [{\citenamefont {Marolf}(2017)}]{Marolf2017black}%
  \BibitemOpen
  \bibfield  {author} {\bibinfo {author} {\bibfnamefont {D.}~\bibnamefont
  {Marolf}},\ }\href@noop {} {\bibfield  {journal} {\bibinfo  {journal}
  {Reports on Progress in Physics}\ }\textbf {\bibinfo {volume} {80}},\
  \bibinfo {pages} {092001} (\bibinfo {year} {2017})}\BibitemShut {NoStop}%
\bibitem [{\citenamefont {Polchinski}(2017)}]{Polchinski2016Black}%
  \BibitemOpen
  \bibfield  {author} {\bibinfo {author} {\bibfnamefont {J.}~\bibnamefont
  {Polchinski}},\ }in\ \href@noop {} {\emph {\bibinfo {booktitle} {{Theoretical
  Advanced Study Institute in Elementary Particle Physics}: {New Frontiers in
  Fields and Strings}}}}\ (\bibinfo {year} {2017})\ \Eprint
  {http://arxiv.org/abs/1609.04036} {arXiv:1609.04036} \BibitemShut {NoStop}%
\bibitem [{\citenamefont {Chen}\ \emph {et~al.}(2015)\citenamefont {Chen},
  \citenamefont {Ong},\ and\ \citenamefont {Yeom}}]{Chen2015Black}%
  \BibitemOpen
  \bibfield  {author} {\bibinfo {author} {\bibfnamefont {P.}~\bibnamefont
  {Chen}}, \bibinfo {author} {\bibfnamefont {Y.}~\bibnamefont {Ong}}, \ and\
  \bibinfo {author} {\bibfnamefont {D.-h.}\ \bibnamefont {Yeom}},\ }\href@noop
  {} {\bibfield  {journal} {\bibinfo  {journal} {Physics Reports}\ }\textbf
  {\bibinfo {volume} {603}},\ \bibinfo {pages} {1–45} (\bibinfo {year}
  {2015})}\BibitemShut {NoStop}%
\bibitem [{\citenamefont {Balasubramanian}\ and\ \citenamefont
  {Czech}(2011)}]{Balasubramanian2011Quantitativ}%
  \BibitemOpen
  \bibfield  {author} {\bibinfo {author} {\bibfnamefont {V.}~\bibnamefont
  {Balasubramanian}}\ and\ \bibinfo {author} {\bibfnamefont {B.}~\bibnamefont
  {Czech}},\ }\href@noop {} {\bibfield  {journal} {\bibinfo  {journal}
  {Classical and Quantum Gravity}\ }\textbf {\bibinfo {volume} {28}},\ \bibinfo
  {pages} {163001} (\bibinfo {year} {2011})}\BibitemShut {NoStop}%
\bibitem [{\citenamefont {Hawking}(1975)}]{Hawking1975Particle}%
  \BibitemOpen
  \bibfield  {author} {\bibinfo {author} {\bibfnamefont {S.~W.}\ \bibnamefont
  {Hawking}},\ }\href@noop {} {\bibfield  {journal} {\bibinfo  {journal}
  {Communications in Mathematical Physics}\ }\textbf {\bibinfo {volume} {43}},\
  \bibinfo {pages} {199} (\bibinfo {year} {1975})}\BibitemShut {NoStop}%
\bibitem [{\citenamefont {Hawking}(1976)}]{Hawking1976Breakdown}%
  \BibitemOpen
  \bibfield  {author} {\bibinfo {author} {\bibfnamefont {S.~W.}\ \bibnamefont
  {Hawking}},\ }\href@noop {} {\bibfield  {journal} {\bibinfo  {journal}
  {Physical Review D}\ }\textbf {\bibinfo {volume} {14}},\ \bibinfo {pages}
  {2460} (\bibinfo {year} {1976})}\BibitemShut {NoStop}%
\bibitem [{\citenamefont {Page}(1993{\natexlab{a}})}]{Page1993Information}%
  \BibitemOpen
  \bibfield  {author} {\bibinfo {author} {\bibfnamefont {D.~N.}\ \bibnamefont
  {Page}},\ }\href@noop {} {\bibfield  {journal} {\bibinfo  {journal} {Physical
  Review Letters}\ }\textbf {\bibinfo {volume} {71}},\ \bibinfo {pages}
  {3743–3746} (\bibinfo {year} {1993}{\natexlab{a}})}\BibitemShut {NoStop}%
\bibitem [{\citenamefont {Page}(1993{\natexlab{b}})}]{Page1993Average}%
  \BibitemOpen
  \bibfield  {author} {\bibinfo {author} {\bibfnamefont {D.~N.}\ \bibnamefont
  {Page}},\ }\href@noop {} {\bibfield  {journal} {\bibinfo  {journal} {Physical
  Review Letters}\ }\textbf {\bibinfo {volume} {71}},\ \bibinfo {pages} {1291}
  (\bibinfo {year} {1993}{\natexlab{b}})}\BibitemShut {NoStop}%
\bibitem [{\citenamefont {Almheiri}\ \emph {et~al.}(2021)\citenamefont
  {Almheiri}, \citenamefont {Hartman}, \citenamefont {Maldacena}, \citenamefont
  {Shaghoulian},\ and\ \citenamefont {Tajdini}}]{Almheiri2020entropy}%
  \BibitemOpen
  \bibfield  {author} {\bibinfo {author} {\bibfnamefont {A.}~\bibnamefont
  {Almheiri}}, \bibinfo {author} {\bibfnamefont {T.}~\bibnamefont {Hartman}},
  \bibinfo {author} {\bibfnamefont {J.}~\bibnamefont {Maldacena}}, \bibinfo
  {author} {\bibfnamefont {E.}~\bibnamefont {Shaghoulian}}, \ and\ \bibinfo
  {author} {\bibfnamefont {A.}~\bibnamefont {Tajdini}},\ }\href@noop {}
  {\bibfield  {journal} {\bibinfo  {journal} {Reviews of Modern Physics}\
  }\textbf {\bibinfo {volume} {93}},\ \bibinfo {pages} {035002} (\bibinfo
  {year} {2021})}\BibitemShut {NoStop}%
\bibitem [{\citenamefont {Susskind}\ \emph {et~al.}(1993)\citenamefont
  {Susskind}, \citenamefont {Thorlacius},\ and\ \citenamefont
  {Uglum}}]{Susskind1993stretched}%
  \BibitemOpen
  \bibfield  {author} {\bibinfo {author} {\bibfnamefont {L.}~\bibnamefont
  {Susskind}}, \bibinfo {author} {\bibfnamefont {L.}~\bibnamefont
  {Thorlacius}}, \ and\ \bibinfo {author} {\bibfnamefont {J.}~\bibnamefont
  {Uglum}},\ }\href@noop {} {\bibfield  {journal} {\bibinfo  {journal}
  {Physical Review D}\ }\textbf {\bibinfo {volume} {48}},\ \bibinfo {pages}
  {3743–3761} (\bibinfo {year} {1993})}\BibitemShut {NoStop}%
\bibitem [{\citenamefont {Almheiri}\ \emph
  {et~al.}(2013{\natexlab{a}})\citenamefont {Almheiri}, \citenamefont {Marolf},
  \citenamefont {Polchinski},\ and\ \citenamefont {Sully}}]{Almheiri2013Black}%
  \BibitemOpen
  \bibfield  {author} {\bibinfo {author} {\bibfnamefont {A.}~\bibnamefont
  {Almheiri}}, \bibinfo {author} {\bibfnamefont {D.}~\bibnamefont {Marolf}},
  \bibinfo {author} {\bibfnamefont {J.}~\bibnamefont {Polchinski}}, \ and\
  \bibinfo {author} {\bibfnamefont {J.}~\bibnamefont {Sully}},\ }\href@noop {}
  {\bibfield  {journal} {\bibinfo  {journal} {Journal of High Energy Physics}\
  ,\ \bibinfo {pages} {62}} (\bibinfo {year} {2013}{\natexlab{a}})}\BibitemShut
  {NoStop}%
\bibitem [{\citenamefont {Almheiri}\ \emph
  {et~al.}(2013{\natexlab{b}})\citenamefont {Almheiri}, \citenamefont {Marolf},
  \citenamefont {Polchinski}, \citenamefont {Stanford},\ and\ \citenamefont
  {Sully}}]{Almheiri2013apologia}%
  \BibitemOpen
  \bibfield  {author} {\bibinfo {author} {\bibfnamefont {A.}~\bibnamefont
  {Almheiri}}, \bibinfo {author} {\bibfnamefont {D.}~\bibnamefont {Marolf}},
  \bibinfo {author} {\bibfnamefont {J.}~\bibnamefont {Polchinski}}, \bibinfo
  {author} {\bibfnamefont {D.}~\bibnamefont {Stanford}}, \ and\ \bibinfo
  {author} {\bibfnamefont {J.}~\bibnamefont {Sully}},\ }\href@noop {}
  {\bibfield  {journal} {\bibinfo  {journal} {Journal of High Energy Physics}\
  }\textbf {\bibinfo {volume} {2013}},\ \bibinfo {pages} {18} (\bibinfo {year}
  {2013}{\natexlab{b}})}\BibitemShut {NoStop}%
\bibitem [{\citenamefont {Maldacena}(1998)}]{Maldacena1997Large}%
  \BibitemOpen
  \bibfield  {author} {\bibinfo {author} {\bibfnamefont {J.~M.}\ \bibnamefont
  {Maldacena}},\ }\href@noop {} {\bibfield  {journal} {\bibinfo  {journal}
  {Adv. Theor. Math. Phys.}\ }\textbf {\bibinfo {volume} {2}},\ \bibinfo
  {pages} {231} (\bibinfo {year} {1998})}\BibitemShut {NoStop}%
\bibitem [{\citenamefont {Van~Raamsdonk}(2010)}]{VanRaamsdonk2010Building}%
  \BibitemOpen
  \bibfield  {author} {\bibinfo {author} {\bibfnamefont {M.}~\bibnamefont
  {Van~Raamsdonk}},\ }\href@noop {} {\bibfield  {journal} {\bibinfo  {journal}
  {International Journal of Modern Physics D}\ }\textbf {\bibinfo {volume}
  {19}},\ \bibinfo {pages} {2429} (\bibinfo {year} {2010})}\BibitemShut
  {NoStop}%
\bibitem [{\citenamefont {Maldacena}\ and\ \citenamefont
  {Susskind}(2013)}]{Maldacena2013Cool}%
  \BibitemOpen
  \bibfield  {author} {\bibinfo {author} {\bibfnamefont {J.}~\bibnamefont
  {Maldacena}}\ and\ \bibinfo {author} {\bibfnamefont {L.}~\bibnamefont
  {Susskind}},\ }\href@noop {} {\bibfield  {journal} {\bibinfo  {journal}
  {Fortschritte der Physik}\ }\textbf {\bibinfo {volume} {61}},\ \bibinfo
  {pages} {781–811} (\bibinfo {year} {2013})}\BibitemShut {NoStop}%
\bibitem [{\citenamefont {Verlinde}(2020)}]{Herman2020ER=EPR}%
  \BibitemOpen
  \bibfield  {author} {\bibinfo {author} {\bibfnamefont {H.}~\bibnamefont
  {Verlinde}},\ }\href@noop {} {\enquote {\bibinfo {title} {{ER = EPR}
  revisited: On the entropy of an einstein-rosen bridge},}\ } (\bibinfo {year}
  {2020}),\ \Eprint {http://arxiv.org/abs/2003.13117} {arXiv:2003.13117}
  \BibitemShut {NoStop}%
\bibitem [{\citenamefont {Almheiri}\ \emph {et~al.}(2020)\citenamefont
  {Almheiri}, \citenamefont {Mahajan}, \citenamefont {Maldacena},\ and\
  \citenamefont {Zhao}}]{Almheiri2020Page}%
  \BibitemOpen
  \bibfield  {author} {\bibinfo {author} {\bibfnamefont {A.}~\bibnamefont
  {Almheiri}}, \bibinfo {author} {\bibfnamefont {R.}~\bibnamefont {Mahajan}},
  \bibinfo {author} {\bibfnamefont {J.}~\bibnamefont {Maldacena}}, \ and\
  \bibinfo {author} {\bibfnamefont {Y.}~\bibnamefont {Zhao}},\ }\href@noop {}
  {\bibfield  {journal} {\bibinfo  {journal} {Journal of High Energy Physics}\
  }\textbf {\bibinfo {volume} {2020}},\ \bibinfo {pages} {149} (\bibinfo {year}
  {2020})}\BibitemShut {NoStop}%
\bibitem [{\citenamefont {Almheiri}\ \emph {et~al.}(2019)\citenamefont
  {Almheiri}, \citenamefont {Engelhardt}, \citenamefont {Marolf},\ and\
  \citenamefont {Maxfield}}]{Almheiri2019entropy}%
  \BibitemOpen
  \bibfield  {author} {\bibinfo {author} {\bibfnamefont {A.}~\bibnamefont
  {Almheiri}}, \bibinfo {author} {\bibfnamefont {N.}~\bibnamefont
  {Engelhardt}}, \bibinfo {author} {\bibfnamefont {D.}~\bibnamefont {Marolf}},
  \ and\ \bibinfo {author} {\bibfnamefont {H.}~\bibnamefont {Maxfield}},\
  }\href@noop {} {\bibfield  {journal} {\bibinfo  {journal} {Journal of High
  Energy Physics}\ }\textbf {\bibinfo {volume} {2019}},\ \bibinfo {pages} {63}
  (\bibinfo {year} {2019})}\BibitemShut {NoStop}%
\bibitem [{\citenamefont {Hawking}\ \emph {et~al.}(2016)\citenamefont
  {Hawking}, \citenamefont {Perry},\ and\ \citenamefont
  {Strominger}}]{Hawking2016Soft}%
  \BibitemOpen
  \bibfield  {author} {\bibinfo {author} {\bibfnamefont {S.~W.}\ \bibnamefont
  {Hawking}}, \bibinfo {author} {\bibfnamefont {M.~J.}\ \bibnamefont {Perry}},
  \ and\ \bibinfo {author} {\bibfnamefont {A.}~\bibnamefont {Strominger}},\
  }\href@noop {} {\bibfield  {journal} {\bibinfo  {journal} {Physical Review
  Letters}\ }\textbf {\bibinfo {volume} {116}},\ \bibinfo {pages} {231301}
  (\bibinfo {year} {2016})}\BibitemShut {NoStop}%
\bibitem [{\citenamefont {Strominger}(2017)}]{Strominger2017Black}%
  \BibitemOpen
  \bibfield  {author} {\bibinfo {author} {\bibfnamefont {A.}~\bibnamefont
  {Strominger}},\ }\href@noop {} {\enquote {\bibinfo {title} {Black hole
  information revisited},}\ } (\bibinfo {year} {2017}),\ \Eprint
  {http://arxiv.org/abs/1706.07143} {arXiv:1706.07143} \BibitemShut {NoStop}%
\bibitem [{\citenamefont {Carney}\ \emph {et~al.}(2017)\citenamefont {Carney},
  \citenamefont {Chaurette}, \citenamefont {Neuenfeld},\ and\ \citenamefont
  {Semenoff}}]{Carney2017Infrared}%
  \BibitemOpen
  \bibfield  {author} {\bibinfo {author} {\bibfnamefont {D.}~\bibnamefont
  {Carney}}, \bibinfo {author} {\bibfnamefont {L.}~\bibnamefont {Chaurette}},
  \bibinfo {author} {\bibfnamefont {D.}~\bibnamefont {Neuenfeld}}, \ and\
  \bibinfo {author} {\bibfnamefont {G.~W.}\ \bibnamefont {Semenoff}},\
  }\href@noop {} {\bibfield  {journal} {\bibinfo  {journal} {Physical Review
  Letters}\ }\textbf {\bibinfo {volume} {119}},\ \bibinfo {pages} {180502}
  (\bibinfo {year} {2017})}\BibitemShut {NoStop}%
\bibitem [{\citenamefont {Carney}\ \emph
  {et~al.}(2018{\natexlab{a}})\citenamefont {Carney}, \citenamefont
  {Chaurette}, \citenamefont {Neuenfeld},\ and\ \citenamefont
  {Semenoff}}]{Carney2018Dressed}%
  \BibitemOpen
  \bibfield  {author} {\bibinfo {author} {\bibfnamefont {D.}~\bibnamefont
  {Carney}}, \bibinfo {author} {\bibfnamefont {L.}~\bibnamefont {Chaurette}},
  \bibinfo {author} {\bibfnamefont {D.}~\bibnamefont {Neuenfeld}}, \ and\
  \bibinfo {author} {\bibfnamefont {G.~W.}\ \bibnamefont {Semenoff}},\
  }\href@noop {} {\bibfield  {journal} {\bibinfo  {journal} {Physical Review
  D}\ }\textbf {\bibinfo {volume} {97}},\ \bibinfo {pages} {025007} (\bibinfo
  {year} {2018}{\natexlab{a}})}\BibitemShut {NoStop}%
\bibitem [{\citenamefont {Su}\ \emph {et~al.}(2024)\citenamefont {Su},
  \citenamefont {Hamma},\ and\ \citenamefont {Marcian\`o}}]{Su2024Scrambling}%
  \BibitemOpen
  \bibfield  {author} {\bibinfo {author} {\bibfnamefont {X.-L.}\ \bibnamefont
  {Su}}, \bibinfo {author} {\bibfnamefont {A.}~\bibnamefont {Hamma}}, \ and\
  \bibinfo {author} {\bibfnamefont {A.}~\bibnamefont {Marcian\`o}},\
  }\href@noop {} {\bibfield  {journal} {\bibinfo  {journal} {The European
  Physical Journal C}\ }\textbf {\bibinfo {volume} {84}},\ \bibinfo {pages}
  {1002} (\bibinfo {year} {2024})}\BibitemShut {NoStop}%
\bibitem [{\citenamefont {Hosur}\ \emph {et~al.}(2016)\citenamefont {Hosur},
  \citenamefont {Qi}, \citenamefont {Roberts},\ and\ \citenamefont
  {Yoshida}}]{Hosur2016Chaos}%
  \BibitemOpen
  \bibfield  {author} {\bibinfo {author} {\bibfnamefont {P.}~\bibnamefont
  {Hosur}}, \bibinfo {author} {\bibfnamefont {X.-L.}\ \bibnamefont {Qi}},
  \bibinfo {author} {\bibfnamefont {D.~A.}\ \bibnamefont {Roberts}}, \ and\
  \bibinfo {author} {\bibfnamefont {B.}~\bibnamefont {Yoshida}},\ }\href@noop
  {} {\bibfield  {journal} {\bibinfo  {journal} {Journal of High Energy
  Physics}\ }\textbf {\bibinfo {volume} {2016}},\ \bibinfo {pages} {4}
  (\bibinfo {year} {2016})}\BibitemShut {NoStop}%
\bibitem [{\citenamefont {Harrow}\ \emph {et~al.}(2021)\citenamefont {Harrow},
  \citenamefont {Kong}, \citenamefont {Liu}, \citenamefont {Mehraban},\ and\
  \citenamefont {Shor}}]{Harrow2021Separation}%
  \BibitemOpen
  \bibfield  {author} {\bibinfo {author} {\bibfnamefont {A.~W.}\ \bibnamefont
  {Harrow}}, \bibinfo {author} {\bibfnamefont {L.}~\bibnamefont {Kong}},
  \bibinfo {author} {\bibfnamefont {Z.-W.}\ \bibnamefont {Liu}}, \bibinfo
  {author} {\bibfnamefont {S.}~\bibnamefont {Mehraban}}, \ and\ \bibinfo
  {author} {\bibfnamefont {P.~W.}\ \bibnamefont {Shor}},\ }\href@noop {}
  {\bibfield  {journal} {\bibinfo  {journal} {Physical Review X Quantum}\
  }\textbf {\bibinfo {volume} {2}},\ \bibinfo {pages} {020339} (\bibinfo {year}
  {2021})}\BibitemShut {NoStop}%
\bibitem [{\citenamefont {Leone}\ \emph
  {et~al.}(2021{\natexlab{a}})\citenamefont {Leone}, \citenamefont {Oliviero},\
  and\ \citenamefont {Hamma}}]{Leone2021Isospectral}%
  \BibitemOpen
  \bibfield  {author} {\bibinfo {author} {\bibfnamefont {L.}~\bibnamefont
  {Leone}}, \bibinfo {author} {\bibfnamefont {S.~F.~E.}\ \bibnamefont
  {Oliviero}}, \ and\ \bibinfo {author} {\bibfnamefont {A.}~\bibnamefont
  {Hamma}},\ }\href@noop {} {\bibfield  {journal} {\bibinfo  {journal}
  {Entropy}\ }\textbf {\bibinfo {volume} {23}},\ \bibinfo {pages} {1073}
  (\bibinfo {year} {2021}{\natexlab{a}})}\BibitemShut {NoStop}%
\bibitem [{\citenamefont {Gharibyan}\ \emph {et~al.}(2018)\citenamefont
  {Gharibyan}, \citenamefont {Hanada}, \citenamefont {Shenker},\ and\
  \citenamefont {Tezuka}}]{Gharibyan2018Onset}%
  \BibitemOpen
  \bibfield  {author} {\bibinfo {author} {\bibfnamefont {H.}~\bibnamefont
  {Gharibyan}}, \bibinfo {author} {\bibfnamefont {M.}~\bibnamefont {Hanada}},
  \bibinfo {author} {\bibfnamefont {S.~H.}\ \bibnamefont {Shenker}}, \ and\
  \bibinfo {author} {\bibfnamefont {M.}~\bibnamefont {Tezuka}},\ }\href@noop {}
  {\bibfield  {journal} {\bibinfo  {journal} {Journal of High Energy Physics}\
  }\textbf {\bibinfo {volume} {2018}},\ \bibinfo {pages} {124} (\bibinfo {year}
  {2018})}\BibitemShut {NoStop}%
\bibitem [{\citenamefont {Lashkari}\ \emph {et~al.}(2013)\citenamefont
  {Lashkari}, \citenamefont {Stanford}, \citenamefont {Hastings}, \citenamefont
  {Osborne},\ and\ \citenamefont {Hayden}}]{Lashkari2013Towards}%
  \BibitemOpen
  \bibfield  {author} {\bibinfo {author} {\bibfnamefont {N.}~\bibnamefont
  {Lashkari}}, \bibinfo {author} {\bibfnamefont {D.}~\bibnamefont {Stanford}},
  \bibinfo {author} {\bibfnamefont {M.}~\bibnamefont {Hastings}}, \bibinfo
  {author} {\bibfnamefont {T.}~\bibnamefont {Osborne}}, \ and\ \bibinfo
  {author} {\bibfnamefont {P.}~\bibnamefont {Hayden}},\ }\href@noop {}
  {\bibfield  {journal} {\bibinfo  {journal} {Journal of High Energy Physics}\
  }\textbf {\bibinfo {volume} {2013}},\ \bibinfo {pages} {22} (\bibinfo {year}
  {2013})}\BibitemShut {NoStop}%
\bibitem [{\citenamefont {Maldacena}\ \emph {et~al.}(2016)\citenamefont
  {Maldacena}, \citenamefont {Shenker},\ and\ \citenamefont
  {Stanford}}]{Maldacena2016bound}%
  \BibitemOpen
  \bibfield  {author} {\bibinfo {author} {\bibfnamefont {J.}~\bibnamefont
  {Maldacena}}, \bibinfo {author} {\bibfnamefont {S.~H.}\ \bibnamefont
  {Shenker}}, \ and\ \bibinfo {author} {\bibfnamefont {D.}~\bibnamefont
  {Stanford}},\ }\href@noop {} {\bibfield  {journal} {\bibinfo  {journal}
  {Journal of High Energy Physics}\ }\textbf {\bibinfo {volume} {2016}},\
  \bibinfo {pages} {106} (\bibinfo {year} {2016})}\BibitemShut {NoStop}%
\bibitem [{\citenamefont {Leone}\ \emph
  {et~al.}(2021{\natexlab{b}})\citenamefont {Leone}, \citenamefont {Oliviero},
  \citenamefont {Zhou},\ and\ \citenamefont {Hamma}}]{Leone2021Quantum}%
  \BibitemOpen
  \bibfield  {author} {\bibinfo {author} {\bibfnamefont {L.}~\bibnamefont
  {Leone}}, \bibinfo {author} {\bibfnamefont {S.~F.~E.}\ \bibnamefont
  {Oliviero}}, \bibinfo {author} {\bibfnamefont {Y.}~\bibnamefont {Zhou}}, \
  and\ \bibinfo {author} {\bibfnamefont {A.}~\bibnamefont {Hamma}},\
  }\href@noop {} {\bibfield  {journal} {\bibinfo  {journal} {Quantum}\ }\textbf
  {\bibinfo {volume} {5}},\ \bibinfo {pages} {453} (\bibinfo {year}
  {2021}{\natexlab{b}})}\BibitemShut {NoStop}%
\bibitem [{\citenamefont {Hayden}\ and\ \citenamefont
  {Preskill}(2007)}]{Hayden2007Black}%
  \BibitemOpen
  \bibfield  {author} {\bibinfo {author} {\bibfnamefont {P.}~\bibnamefont
  {Hayden}}\ and\ \bibinfo {author} {\bibfnamefont {J.}~\bibnamefont
  {Preskill}},\ }\href@noop {} {\bibfield  {journal} {\bibinfo  {journal}
  {Journal of High Energy Physics}\ }\textbf {\bibinfo {volume} {2007}},\
  \bibinfo {pages} {120} (\bibinfo {year} {2007})}\BibitemShut {NoStop}%
\bibitem [{\citenamefont {Leone}\ \emph {et~al.}(2022)\citenamefont {Leone},
  \citenamefont {Oliviero}, \citenamefont {Piemontese}, \citenamefont {True},\
  and\ \citenamefont {Hamma}}]{Leone2022Retrieving}%
  \BibitemOpen
  \bibfield  {author} {\bibinfo {author} {\bibfnamefont {L.}~\bibnamefont
  {Leone}}, \bibinfo {author} {\bibfnamefont {S.~F.~E.}\ \bibnamefont
  {Oliviero}}, \bibinfo {author} {\bibfnamefont {S.}~\bibnamefont
  {Piemontese}}, \bibinfo {author} {\bibfnamefont {S.}~\bibnamefont {True}}, \
  and\ \bibinfo {author} {\bibfnamefont {A.}~\bibnamefont {Hamma}},\
  }\href@noop {} {\bibfield  {journal} {\bibinfo  {journal} {Physical Review
  A}\ }\textbf {\bibinfo {volume} {106}},\ \bibinfo {pages} {062434} (\bibinfo
  {year} {2022})}\BibitemShut {NoStop}%
\bibitem [{\citenamefont {Roberts}\ and\ \citenamefont
  {Yoshida}(2017)}]{Roberts2017Chaos}%
  \BibitemOpen
  \bibfield  {author} {\bibinfo {author} {\bibfnamefont {D.~A.}\ \bibnamefont
  {Roberts}}\ and\ \bibinfo {author} {\bibfnamefont {B.}~\bibnamefont
  {Yoshida}},\ }\href@noop {} {\bibfield  {journal} {\bibinfo  {journal}
  {Journal of High Energy Physics}\ }\textbf {\bibinfo {volume} {2017}},\
  \bibinfo {pages} {121} (\bibinfo {year} {2017})}\BibitemShut {NoStop}%
\bibitem [{\citenamefont {Oliviero}\ \emph {et~al.}(2021)\citenamefont
  {Oliviero}, \citenamefont {Leone}, \citenamefont {Caravelli},\ and\
  \citenamefont {Hamma}}]{Oliviero2021Random}%
  \BibitemOpen
  \bibfield  {author} {\bibinfo {author} {\bibfnamefont {S.~F.~E.}\
  \bibnamefont {Oliviero}}, \bibinfo {author} {\bibfnamefont {L.}~\bibnamefont
  {Leone}}, \bibinfo {author} {\bibfnamefont {F.}~\bibnamefont {Caravelli}}, \
  and\ \bibinfo {author} {\bibfnamefont {A.}~\bibnamefont {Hamma}},\
  }\href@noop {} {\bibfield  {journal} {\bibinfo  {journal} {SciPost Physics}\
  }\textbf {\bibinfo {volume} {10}},\ \bibinfo {pages} {076} (\bibinfo {year}
  {2021})}\BibitemShut {NoStop}%
\bibitem [{\citenamefont {Ding}\ \emph {et~al.}(2016)\citenamefont {Ding},
  \citenamefont {Hayden},\ and\ \citenamefont {Walter}}]{Ding2016Conditional}%
  \BibitemOpen
  \bibfield  {author} {\bibinfo {author} {\bibfnamefont {D.}~\bibnamefont
  {Ding}}, \bibinfo {author} {\bibfnamefont {P.}~\bibnamefont {Hayden}}, \ and\
  \bibinfo {author} {\bibfnamefont {M.}~\bibnamefont {Walter}},\ }\href@noop {}
  {\bibfield  {journal} {\bibinfo  {journal} {Journal of High Energy Physics}\
  }\textbf {\bibinfo {volume} {2016}},\ \bibinfo {pages} {145} (\bibinfo {year}
  {2016})}\BibitemShut {NoStop}%
\bibitem [{\citenamefont {Birrell}\ and\ \citenamefont
  {Davies}(1982)}]{Birrell1982Quantum}%
  \BibitemOpen
  \bibfield  {author} {\bibinfo {author} {\bibfnamefont {N.~D.}\ \bibnamefont
  {Birrell}}\ and\ \bibinfo {author} {\bibfnamefont {P.~C.~W.}\ \bibnamefont
  {Davies}},\ }\href@noop {} {\emph {\bibinfo {title} {Quantum Fields in Curved
  Space}}}\ (\bibinfo  {publisher} {Cambridge University Press},\ \bibinfo
  {address} {Cambridge},\ \bibinfo {year} {1982})\BibitemShut {NoStop}%
\bibitem [{\citenamefont {Wald}(1975)}]{Wald1975particle}%
  \BibitemOpen
  \bibfield  {author} {\bibinfo {author} {\bibfnamefont {R.~M.}\ \bibnamefont
  {Wald}},\ }\href@noop {} {\bibfield  {journal} {\bibinfo  {journal}
  {Communications in Mathematical Physics}\ }\textbf {\bibinfo {volume} {45}},\
  \bibinfo {pages} {9} (\bibinfo {year} {1975})}\BibitemShut {NoStop}%
\bibitem [{\citenamefont {DeWitt}(1975)}]{DEWITT1975Quantum}%
  \BibitemOpen
  \bibfield  {author} {\bibinfo {author} {\bibfnamefont {B.~S.}\ \bibnamefont
  {DeWitt}},\ }\href@noop {} {\bibfield  {journal} {\bibinfo  {journal}
  {Physics Reports}\ }\textbf {\bibinfo {volume} {19}},\ \bibinfo {pages} {295}
  (\bibinfo {year} {1975})}\BibitemShut {NoStop}%
\bibitem [{\citenamefont {Birrell}\ and\ \citenamefont
  {Taylor}(1980)}]{Birrell1980Analysis}%
  \BibitemOpen
  \bibfield  {author} {\bibinfo {author} {\bibfnamefont {N.~D.}\ \bibnamefont
  {Birrell}}\ and\ \bibinfo {author} {\bibfnamefont {J.~G.}\ \bibnamefont
  {Taylor}},\ }\href@noop {} {\bibfield  {journal} {\bibinfo  {journal}
  {Journal of Mathematical Physics}\ }\textbf {\bibinfo {volume} {21}},\
  \bibinfo {pages} {1740} (\bibinfo {year} {1980})}\BibitemShut {NoStop}%
\bibitem [{\citenamefont {Sakurai}\ and\ \citenamefont
  {Napolitano}(2020)}]{Sakurai2020Modern}%
  \BibitemOpen
  \bibfield  {author} {\bibinfo {author} {\bibfnamefont {J.~J.}\ \bibnamefont
  {Sakurai}}\ and\ \bibinfo {author} {\bibfnamefont {J.}~\bibnamefont
  {Napolitano}},\ }\href@noop {} {\emph {\bibinfo {title} {Modern Quantum
  Mechanics}}}\ (\bibinfo  {publisher} {Cambridge University Press},\ \bibinfo
  {address} {Cambridge},\ \bibinfo {year} {2020})\BibitemShut {NoStop}%
\bibitem [{\citenamefont {Bondi}\ \emph {et~al.}(1962)\citenamefont {Bondi},
  \citenamefont {van~der Burg},\ and\ \citenamefont
  {Metzner}}]{Bondi1962Gravitational}%
  \BibitemOpen
  \bibfield  {author} {\bibinfo {author} {\bibfnamefont {H.}~\bibnamefont
  {Bondi}}, \bibinfo {author} {\bibfnamefont {M.~G.~J.}\ \bibnamefont {van~der
  Burg}}, \ and\ \bibinfo {author} {\bibfnamefont {A.~W.~K.}\ \bibnamefont
  {Metzner}},\ }\href@noop {} {\bibfield  {journal} {\bibinfo  {journal}
  {Proceedings of the Royal Society of London. Series A}\ }\textbf {\bibinfo
  {volume} {269}},\ \bibinfo {pages} {21} (\bibinfo {year} {1962})}\BibitemShut
  {NoStop}%
\bibitem [{\citenamefont {Sachs}(1962)}]{Sachs1962Gravitational}%
  \BibitemOpen
  \bibfield  {author} {\bibinfo {author} {\bibfnamefont {R.~K.}\ \bibnamefont
  {Sachs}},\ }\href@noop {} {\bibfield  {journal} {\bibinfo  {journal}
  {Proceedings of the Royal Society of London. Series A}\ }\textbf {\bibinfo
  {volume} {270}},\ \bibinfo {pages} {103} (\bibinfo {year}
  {1962})}\BibitemShut {NoStop}%
\bibitem [{\citenamefont {Weinberg}(1965)}]{Weinberg1965Infrared}%
  \BibitemOpen
  \bibfield  {author} {\bibinfo {author} {\bibfnamefont {S.}~\bibnamefont
  {Weinberg}},\ }\href@noop {} {\bibfield  {journal} {\bibinfo  {journal}
  {Physical Review}\ }\textbf {\bibinfo {volume} {140}},\ \bibinfo {pages}
  {B516} (\bibinfo {year} {1965})}\BibitemShut {NoStop}%
\bibitem [{\citenamefont {Mandl}\ and\ \citenamefont
  {Shaw}(1993)}]{Mandl1993QFT}%
  \BibitemOpen
  \bibfield  {author} {\bibinfo {author} {\bibfnamefont {F.}~\bibnamefont
  {Mandl}}\ and\ \bibinfo {author} {\bibfnamefont {G.}~\bibnamefont {Shaw}},\
  }\href@noop {} {\emph {\bibinfo {title} {Quantum Field Theory}}}\ (\bibinfo
  {publisher} {John Wiley and Sons},\ \bibinfo {address} {Chichester},\
  \bibinfo {year} {1993})\BibitemShut {NoStop}%
\bibitem [{\citenamefont {Bloch}\ and\ \citenamefont
  {Nordsieck}(1937)}]{Bloch1937Note}%
  \BibitemOpen
  \bibfield  {author} {\bibinfo {author} {\bibfnamefont {F.}~\bibnamefont
  {Bloch}}\ and\ \bibinfo {author} {\bibfnamefont {A.}~\bibnamefont
  {Nordsieck}},\ }\href@noop {} {\bibfield  {journal} {\bibinfo  {journal}
  {Physical Review}\ }\textbf {\bibinfo {volume} {52}},\ \bibinfo {pages} {54}
  (\bibinfo {year} {1937})}\BibitemShut {NoStop}%
\bibitem [{\citenamefont {Yennie}\ \emph {et~al.}(1961)\citenamefont {Yennie},
  \citenamefont {Frautschi},\ and\ \citenamefont {Suura}}]{Yennie1961infrared}%
  \BibitemOpen
  \bibfield  {author} {\bibinfo {author} {\bibfnamefont {D.~R.}\ \bibnamefont
  {Yennie}}, \bibinfo {author} {\bibfnamefont {S.~C.}\ \bibnamefont
  {Frautschi}}, \ and\ \bibinfo {author} {\bibfnamefont {H.}~\bibnamefont
  {Suura}},\ }\href@noop {} {\bibfield  {journal} {\bibinfo  {journal} {Annals
  of Physics}\ }\textbf {\bibinfo {volume} {13}},\ \bibinfo {pages} {379}
  (\bibinfo {year} {1961})}\BibitemShut {NoStop}%
\bibitem [{\citenamefont {Chung}(1965)}]{Chung1965Infrared}%
  \BibitemOpen
  \bibfield  {author} {\bibinfo {author} {\bibfnamefont {V.}~\bibnamefont
  {Chung}},\ }\href@noop {} {\bibfield  {journal} {\bibinfo  {journal}
  {Physical Review}\ }\textbf {\bibinfo {volume} {140}},\ \bibinfo {pages}
  {B1110} (\bibinfo {year} {1965})}\BibitemShut {NoStop}%
\bibitem [{\citenamefont {Kibble}(1968)}]{Kibble1968Coherent}%
  \BibitemOpen
  \bibfield  {author} {\bibinfo {author} {\bibfnamefont {T.~W.~B.}\
  \bibnamefont {Kibble}},\ }\href@noop {} {\bibfield  {journal} {\bibinfo
  {journal} {Journal of Mathematical Physics}\ }\textbf {\bibinfo {volume}
  {9}},\ \bibinfo {pages} {315} (\bibinfo {year} {1968})}\BibitemShut {NoStop}%
\bibitem [{\citenamefont {Kulish}\ and\ \citenamefont
  {Faddeev}(1970)}]{Kulish1970Asymptotic}%
  \BibitemOpen
  \bibfield  {author} {\bibinfo {author} {\bibfnamefont {P.~P.}\ \bibnamefont
  {Kulish}}\ and\ \bibinfo {author} {\bibfnamefont {L.~D.}\ \bibnamefont
  {Faddeev}},\ }\href@noop {} {\bibfield  {journal} {\bibinfo  {journal}
  {Theoretical and Mathematical Physics}\ }\textbf {\bibinfo {volume} {4}},\
  \bibinfo {pages} {745} (\bibinfo {year} {1970})}\BibitemShut {NoStop}%
\bibitem [{\citenamefont {Ware}\ \emph {et~al.}(2013)\citenamefont {Ware},
  \citenamefont {Saotome},\ and\ \citenamefont
  {Akhoury}}]{Ware2013Construction}%
  \BibitemOpen
  \bibfield  {author} {\bibinfo {author} {\bibfnamefont {J.}~\bibnamefont
  {Ware}}, \bibinfo {author} {\bibfnamefont {R.}~\bibnamefont {Saotome}}, \
  and\ \bibinfo {author} {\bibfnamefont {R.}~\bibnamefont {Akhoury}},\
  }\href@noop {} {\bibfield  {journal} {\bibinfo  {journal} {Journal of High
  Energy Physics}\ }\textbf {\bibinfo {volume} {2013}},\ \bibinfo {pages} {159}
  (\bibinfo {year} {2013})}\BibitemShut {NoStop}%
\bibitem [{\citenamefont {Furugori}\ and\ \citenamefont
  {Nojiri}(2021)}]{Furugori2021Dressed}%
  \BibitemOpen
  \bibfield  {author} {\bibinfo {author} {\bibfnamefont {H.}~\bibnamefont
  {Furugori}}\ and\ \bibinfo {author} {\bibfnamefont {S.}~\bibnamefont
  {Nojiri}},\ }\href@noop {} {\bibfield  {journal} {\bibinfo  {journal}
  {Physical Review D}\ }\textbf {\bibinfo {volume} {104}},\ \bibinfo {pages}
  {125004} (\bibinfo {year} {2021})}\BibitemShut {NoStop}%
\bibitem [{\citenamefont {Choi}\ and\ \citenamefont
  {Akhoury}(2018)}]{Choi2018Soft}%
  \BibitemOpen
  \bibfield  {author} {\bibinfo {author} {\bibfnamefont {S.}~\bibnamefont
  {Choi}}\ and\ \bibinfo {author} {\bibfnamefont {R.}~\bibnamefont {Akhoury}},\
  }\href@noop {} {\bibfield  {journal} {\bibinfo  {journal} {Journal of High
  Energy Physics}\ }\textbf {\bibinfo {volume} {2018}},\ \bibinfo {pages} {74}
  (\bibinfo {year} {2018})}\BibitemShut {NoStop}%
\bibitem [{\citenamefont {Kofman}\ \emph {et~al.}(1997)\citenamefont {Kofman},
  \citenamefont {Linde},\ and\ \citenamefont
  {Starobinsky}}]{Kofman1997Towards}%
  \BibitemOpen
  \bibfield  {author} {\bibinfo {author} {\bibfnamefont {L.}~\bibnamefont
  {Kofman}}, \bibinfo {author} {\bibfnamefont {A.}~\bibnamefont {Linde}}, \
  and\ \bibinfo {author} {\bibfnamefont {A.~A.}\ \bibnamefont {Starobinsky}},\
  }\href@noop {} {\bibfield  {journal} {\bibinfo  {journal} {Physical Review
  D}\ }\textbf {\bibinfo {volume} {56}},\ \bibinfo {pages} {3258–3295}
  (\bibinfo {year} {1997})}\BibitemShut {NoStop}%
\bibitem [{\citenamefont {Rose}(1961)}]{Rose1961Relativistic}%
  \BibitemOpen
  \bibfield  {author} {\bibinfo {author} {\bibfnamefont {M.~E.}\ \bibnamefont
  {Rose}},\ }\href@noop {} {\emph {\bibinfo {title} {Relativistic Electron
  Theory}}}\ (\bibinfo  {publisher} {John Wiley and Sons},\ \bibinfo {address}
  {Hoboken},\ \bibinfo {year} {1961})\BibitemShut {NoStop}%
\bibitem [{\citenamefont {Boulware}(1975)}]{Boulware1975Spin}%
  \BibitemOpen
  \bibfield  {author} {\bibinfo {author} {\bibfnamefont {D.~G.}\ \bibnamefont
  {Boulware}},\ }\href@noop {} {\bibfield  {journal} {\bibinfo  {journal}
  {Physical Review D}\ }\textbf {\bibinfo {volume} {12}},\ \bibinfo {pages}
  {350} (\bibinfo {year} {1975})}\BibitemShut {NoStop}%
\bibitem [{\citenamefont {Noble}\ and\ \citenamefont
  {Jentschura}(2016)}]{Noble2016Dirac}%
  \BibitemOpen
  \bibfield  {author} {\bibinfo {author} {\bibfnamefont {J.~H.}\ \bibnamefont
  {Noble}}\ and\ \bibinfo {author} {\bibfnamefont {U.~D.}\ \bibnamefont
  {Jentschura}},\ }\href@noop {} {\bibfield  {journal} {\bibinfo  {journal}
  {Phys. Rev. A}\ }\textbf {\bibinfo {volume} {93}},\ \bibinfo {pages} {032108}
  (\bibinfo {year} {2016})}\BibitemShut {NoStop}%
\bibitem [{\citenamefont {Greiner}(2000)}]{Greiner2000Relativistic}%
  \BibitemOpen
  \bibfield  {author} {\bibinfo {author} {\bibfnamefont {W.}~\bibnamefont
  {Greiner}},\ }\href@noop {} {\emph {\bibinfo {title} {{Relativistic quantum
  mechanics}}}}\ (\bibinfo  {publisher} {Springer-Verlag},\ \bibinfo {address}
  {Berlin},\ \bibinfo {year} {2000})\BibitemShut {NoStop}%
\bibitem [{\citenamefont {Strange}(1998)}]{Strange1998Relativistic}%
  \BibitemOpen
  \bibfield  {author} {\bibinfo {author} {\bibfnamefont {P.}~\bibnamefont
  {Strange}},\ }\href@noop {} {\emph {\bibinfo {title} {Relativistic quantum
  mechanics}}}\ (\bibinfo  {publisher} {Cambridge University Press},\ \bibinfo
  {address} {Cambridge},\ \bibinfo {year} {1998})\BibitemShut {NoStop}%
\bibitem [{\citenamefont {Kordt}(2012)}]{Kordt2012Single}%
  \BibitemOpen
  \bibfield  {author} {\bibinfo {author} {\bibfnamefont {P.}~\bibnamefont
  {Kordt}},\ }in\ \href@noop {} {\emph {\bibinfo {booktitle} {Single-site Green
  function of the Dirac equation for full-potential electron scattering}}},\
  Vol.~\bibinfo {volume} {34}\ (\bibinfo {year} {2012})\BibitemShut {NoStop}%
\bibitem [{\citenamefont {Carney}\ \emph
  {et~al.}(2018{\natexlab{b}})\citenamefont {Carney}, \citenamefont
  {Chaurette}, \citenamefont {Neuenfeld},\ and\ \citenamefont
  {Semenoff}}]{Carney2018need}%
  \BibitemOpen
  \bibfield  {author} {\bibinfo {author} {\bibfnamefont {D.}~\bibnamefont
  {Carney}}, \bibinfo {author} {\bibfnamefont {L.}~\bibnamefont {Chaurette}},
  \bibinfo {author} {\bibfnamefont {D.}~\bibnamefont {Neuenfeld}}, \ and\
  \bibinfo {author} {\bibfnamefont {G.}~\bibnamefont {Semenoff}},\ }\href@noop
  {} {\bibfield  {journal} {\bibinfo  {journal} {Journal of High Energy
  Physics}\ }\textbf {\bibinfo {volume} {2018}},\ \bibinfo {pages} {121}
  (\bibinfo {year} {2018}{\natexlab{b}})}\BibitemShut {NoStop}%
\bibitem [{\citenamefont {Boas}(1966)}]{Boas1966Mathematical}%
  \BibitemOpen
  \bibfield  {author} {\bibinfo {author} {\bibfnamefont {M.~L.}\ \bibnamefont
  {Boas}},\ }\href@noop {} {\emph {\bibinfo {title} {Mathematical Methods in
  the Physical Sciences}}}\ (\bibinfo  {publisher} {Wiley},\ \bibinfo {address}
  {Hoboken},\ \bibinfo {year} {1966})\BibitemShut {NoStop}%
\end{thebibliography}%

\end{document}